\documentclass[aps,prb,twocolumn,superscriptaddress]{revtex4-2}

\usepackage{amsmath}    

\usepackage{amsfonts}
\usepackage{amssymb}
\usepackage[colorlinks=true,citecolor=blue,linkcolor=red]{hyperref}
\usepackage{graphicx}
\usepackage{bbold}					
\usepackage[dvipsnames]{xcolor}

\usepackage{bm} 
\usepackage{dcolumn}
\usepackage{mathtools}

\usepackage{array,booktabs,ragged2e}	
\usepackage{soul}						
\usepackage{cancel}						
\usepackage{tabularx}
\usepackage{multirow}
\usepackage{hhline}
\usepackage{adjustbox}
\usepackage{sidecap}
\usepackage{threeparttable}
\usepackage{colortbl} 

\usepackage[utf8]{inputenc}
\usepackage[T1]{fontenc}

\def\bra#1{\left<{#1}\right|}					
\def\ket#1{\left|{#1}\right>}					
\newcommand{\angstrom}{\mbox{\normalfont\AA}}	

\def\degree {^{\circ}}

\newcolumntype{P}[1]{>{\raggedleft\arraybackslash}p{#1}}
\newcolumntype{R}[1]{>{\centering\arraybackslash}p{#1}}

\usepackage{bm,color,amsmath,amssymb,mathrsfs,latexsym,graphicx,psfrag,mathtools}
\usepackage{color}
\usepackage[dvipsnames]{xcolor}

\usepackage{empheq}

\newcommand{\bsl}[1]{\boldsymbol{#1}}

\renewcommand{\mod}{\,\mathrm{mod}\,}

\newcommand{\ii}{\mathrm{i}}

\newcommand{\U}{\mathrm{U}}

\newcommand{\eqnref}[1]{Eq.\,\eqref{#1}}
\newcommand{\figref}[1]{Fig.\,\ref{#1}}
\newcommand{\tabref}[1]{Tab.\,\ref{#1}}
\newcommand{\secref}[1]{Sec.\,\ref{#1}}
\newcommand{\appref}[1]{Appendix.\,\ref{#1}}
\newcommand{\refcite}[1]{Ref.\,\cite{#1}}

\newcommand{\mat}[1]{\left(\begin{matrix}#1\end{matrix}\right)}

\newcommand{\eq}[1]{\begin{equation} #1 \end{equation}}

\newcommand{\eqa}[1]{\begin{align}\begin{split} #1 \end{split}\end{align}}

\let\oldAA\AA
\renewcommand{\AA}{\text{\normalfont\oldAA}}

\newcommand{\ie}{{\emph{i.e.}}}
\newcommand{\eg}{{\emph{e.g.}}}
\newcommand{\TR}{\mathcal{T}}
\newcommand{\cc}{\mathcal{K}}

\renewcommand{\P}{\mathcal{P}}

\newcommand{\V}{\mathcal{V}}

\newcommand{\Q}{\mathcal{Q}}
\newcommand{\I}{\mathcal{I}}

\newcommand{\K}{\text{K}}

\newcommand{\tmt}{\text{$t$MoTe$_2$}}
\newcommand{\AAtmt}{\text{AA-$t$MoTe$_2$}}
\newcommand{\ABtmt}{\text{AB-$t$MoTe$_2$}}

\usepackage{comment}

\newcommand{\mt}{\text{MoTe}$_2$}

\newcommand{\la}{\lambda}

\newcommand{\bpm}{\begin{pmatrix}}
\newcommand{\epm}{\end{pmatrix}}

\renewcommand{\th}{\theta}

\newcommand{\mbf}[1]{\mathbf{#1}}
\renewcommand{\u}{\uparrow}
\renewcommand{\d}{\downarrow}
\newcommand{\bea}{\begin{equation} \begin{aligned}}
\newcommand{\eea}{\end{aligned} \end{equation} }

\begin{document}


\title{Moir\'e Fractional Chern Insulators I: First-principles calculations and Continuum Models of Twisted Bilayer MoTe$_2$}
 
\author{Yujin Jia}
\thanks{These authors contributed equally.}
\affiliation{Beijing National Laboratory for Condensed Matter Physics and Institute of physics,
Chinese academy of sciences, Beijing 100190, China}
\affiliation{University of Chinese academy of sciences, Beijing 100049, China}

\author{Jiabin Yu}
\thanks{These authors contributed equally.}
\affiliation{Department of Physics, Princeton University, Princeton, New Jersey 08544, USA}

\author{Jiaxuan Liu}
\thanks{These authors contributed equally.}
\affiliation{Beijing National Laboratory for Condensed Matter Physics and Institute of physics,
Chinese academy of sciences, Beijing 100190, China}
\affiliation{University of Chinese academy of sciences, Beijing 100049, China}

\author{Jonah Herzog-Arbeitman}
\affiliation{Department of Physics, Princeton University, Princeton, New Jersey 08544, USA}

\author{Ziyue Qi}
\affiliation{Beijing National Laboratory for Condensed Matter Physics and Institute of physics,
Chinese academy of sciences, Beijing 100190, China}
\affiliation{University of Chinese academy of sciences, Beijing 100049, China}

\author{Nicolas Regnault}
\affiliation{Department of Physics, Princeton University, Princeton, New Jersey 08544, USA}
\affiliation{Laboratoire de Physique de l’Ecole normale sup\'erieure,
ENS, Universit\'e PSL, CNRS, Sorbonne Universit\'e,
Universit\'e Paris-Diderot, Sorbonne Paris Cit\'e, 75005 Paris, France}

\author{Hongming Weng}
\affiliation{Beijing National Laboratory for Condensed Matter Physics and Institute of physics,
Chinese academy of sciences, Beijing 100190, China}
\affiliation{University of Chinese academy of sciences, Beijing 100049, China}
\affiliation{Songshan Lake Materials Laboratory, Dongguan, Guangdong 523808, China}
\date{\today}

\author{B. Andrei Bernevig}
\email{bernevig@princeton.edu}
\affiliation{Department of Physics, Princeton University, Princeton, New Jersey 08544, USA}
\affiliation{Donostia International Physics Center, P. Manuel de Lardizabal 4, 20018 Donostia-San Sebastian, Spain}
\affiliation{IKERBASQUE, Basque Foundation for Science, Bilbao, Spain}

\author{Quansheng Wu}
\email{quansheng.wu@iphy.ac.cn}
\affiliation{Beijing National Laboratory for Condensed Matter Physics and Institute of physics,
Chinese academy of sciences, Beijing 100190, China}
\affiliation{University of Chinese academy of sciences, Beijing 100049, China}
\date{\today}

\begin{abstract}
Recent experiments observed fractional Chern insulators (FCI) in twisted bilayer MoTe$_2$ at zero magnetic field, yet even the single-particle model of this material is controversial, leading to unreliable predictions of the experimental phase diagram as discussed in [Yu et al., 2023]. In this light, we revisit the single-particle model of twisted bilayer MoTe$_2$. Utilizing large-scale density functional theory, we calculate the band structure of twisted AA-stacked bilayer MoTe$_2$ at various twist angles relevant to experiment. We find that a band inversion occurs near $4.41^\circ$ between the second and third bands. Our \emph{ab initio} band structure is in qualitative agreement with [Wang et al., 2023], but shows important differences in the remote bands and in the $\Gamma$ valley.
We incorporate two higher harmonic terms into the continuum model to capture the highest 3 valence bands per valley. We confirm that the two highest valence bands per valley have opposite Chern numbers with $|C|=1$ for small angles, and also use our model to predict a variety of Chern states in the remote bands accessible by displacement field. Finally, we perform DFT calculations and build models for the AB stacking configuration. Our work serves as a foundation for accurate determination of the correlated phases in twisted bilayer MoTe$_2$. 
\end{abstract} 

\maketitle

\section{Introduction}

Fractional Chern insulators (FCI)~\cite{neupert, sheng, regnault} were show to naturally appear in zero magnetic field when nearly flat Chern bands~\cite{Tang11,Sun2011} are fractionally filled.
Over the last several years, there have been extensive theoretical~\cite{Bergholtz13, Parameswaran13,Abouelkomsan2020FCIMoire,Vishwanath2020FCITBG,RepellinFCITBG,Parker2021fieldtunedFCITBG,Wilhelm2021FCITBG,Stern2021FCITBG,Li2021FCItTMD,Crepel2022FCITMD,Fu2022FCIDirac,Abouelkomsan2023FCITBG,Cano2023FCItTMD,2023arXiv230907222W,reddy2023fractional,wang2023fractional,Wu2023IntegerFillingstMoTe2,Dong2023CFLtMoTe2,Goldman2023CFLtMoTe2,MacDonald2023MagicAngletTMD,Reddy2023GlobalPDFCI,Song2023tTMDFCI,Xu2023MLWOFCItTMD,Zaletel2023tMoTe2FCI,Liu2023HFtMoTe2num2,Yu2023FCI,Fu2023BandMixingFCItMoTe2} and experimental~\cite{cai2023signatures,zeng2023integer,park2023observation,Xu2023FCItMoTe2,Lu2023FCI} studies on zero-field FCIs in moir\'e materials~\cite{Cao2018TBGMott,Cao2018TBGSC}, as well as studies of fractional quantum Hall (FQH)-like states under nonzero external magnetic field~\cite{Young2018FCIBLGMoire} (which were called FCIs because the Chern bands where they appeared did not have flat Berry curvature and quantum geometry), and small B-field-induced FCIs \cite{Xie2021TBGFCIfiniteB}.
Remarkably, FCIs without any external magnetic fields have recently been observed in twisted bilayer MoTe$_2$ (\tmt)~\cite{cai2023signatures,zeng2023integer,park2023observation,Xu2023FCItMoTe2} and in the pentalayer-graphene/hBN moir\'e superlattice~\cite{LongJu2023FCIPentalayerGraphenehBN}. These \emph{moir\'e fractional Chern insulator} (mFCI) states provide the best experimental platform to date for the physics proposed in Refs. \cite{neupert, sheng, regnault}. 

In this work, we will focus on {\tmt}, where mFCIs are observed at fractional fillings $\nu=-2/3,-3/5$, as well as a Chern insulator (CI) at $\nu=-1$.  Throughout, $\nu$ is the electron filling measured from the charge neutrality point.
 FCIs  are expected to appear \cite{neupert, sheng, regnault} in the CI bands found in the {\tmt} model first proposed in \refcite{Wu2019TIintTMD}. However, an accurate model of the material is required to correctly reproduce the FCI states and competing non-topological states. Otherwise serious disagreement with experiment can occur, as has been somewhat overlooked in the flurry of recent literature. For example, the spin polarization at $\nu=-1/3,-4/3$ cannot be reproduced theoretically unless band-mixing is taken into account \cite{Yu2023FCI}, showing that a bands beyond the lowest valence manifold play an integral part in the physics. Recent theoretical work~\cite{reddy2023fractional,wang2023fractional,Dong2023CFLtMoTe2,Goldman2023CFLtMoTe2,Reddy2023GlobalPDFCI,Xu2023MLWOFCItTMD,Zaletel2023tMoTe2FCI,Yu2023FCI,Fu2023BandMixingFCItMoTe2,Fengcheng2023tMoTe2HFnum1}, some of which predicts the existence of fractional states not seen in the experiment, might then be subject to change when band-mixing is properly included. Even interaction-driven band mixing beyond the two-band Hilbert space considered in \refcite{Yu2023FCI} could be important if the single-particle bands are energetically close. Hence it is of the utmost importance to obtain a continuum model capturing the ab-initio band structure over the range of energies accessible by the Coulomb potential. 
  
 Most importantly, there are two different sets of ab-initio parameters \refcite{reddy2023fractional,wang2023fractional} used~\cite{reddy2023fractional,wang2023fractional,Dong2023CFLtMoTe2,Goldman2023CFLtMoTe2,Reddy2023GlobalPDFCI,Xu2023MLWOFCItTMD,Zaletel2023tMoTe2FCI,Yu2023FCI,Fu2023BandMixingFCItMoTe2,Fengcheng2023tMoTe2HFnum1} for interacting calculations in the continuum model of \refcite{Wu2019TIintTMD}. These sets of parameters give rise to different single particle phase diagrams, many-body spin polarizations, and stabilities of 2/3 and 1/3 mFCI states \refcite{Yu2023FCI}. Settling the single-particle model of {\tmt} is the subject of this first work in the mFCI series.

We perform large-scale density functional theory (DFT) calculations to study twisted bilayer MoTe$_2$ at various twist angles for AA stacking, which is the stacking configuration relevant to the experiments~\cite{cai2023signatures,zeng2023integer,park2023observation,Xu2023FCItMoTe2}.
To accurately capture the impact of crystal structure relaxation, we rigorously test 19 different van der Waals (vdW) exchange-correlation functionals, and determine that DFT-D2 yields the most reliable lattice parameters when compared with experimental results. 
With DFT-D2, we further use a highly efficient two-step method to obtain the structure.
Our first step is to use the DFT-D2 functional to train a machine learning algorithm that generates relaxed structures for a set of twist angles. 
The second step is to use these relaxed structures as the initial configurations, and further perform the full DFT relaxation to obtain final, accurate relaxed structures.
The moir\'e bands are eventually calculated with the final relaxed structures.
Our DFT results show that the $\pm\K$-valley valence bands dominate the low-energy physics while the $\Gamma$-valley valence bands are about 80meV away from the valence band maximum (VBM) for the experimentally relevant angle $3.89^\circ$. 
Furthermore, the second and the third valence bands in $\K$ valley (or $-\K$ valley related by time-reversal symmetry) undergo a gap closing around $4.41^\circ$, changing the Chern number of the second valence band from $-1$ at smaller angles to $1$ at larger angles. Our ab-initio results for the $\pm\K$ valley bands are closer to those of \refcite{wang2023fractional} than to those of \refcite{reddy2023fractional}, although our second and third band per valley are closer to each other than those of \refcite{wang2023fractional} at $3.89\degree$, which shifts their gap closing to smaller angles.  Our $\Gamma$-valley band is also lower in energy than that of \refcite{wang2023fractional}, and hence will not contribute to many-body physics near $\nu = -1$. The band structure in \refcite{reddy2023fractional} looks different from ours; we can reproduce similar band width of the topmost valence moir\'e bands in \refcite{reddy2023fractional} by using ISMEAR=1 parameters in the settings of VASP calculations. However, in our calculations all through this work we use ISMEAR=0~\cite{vaspmanual}, which leads to the narrow bandwidth.  

We then use the moir\'e model to capture the low-energy DFT bands.
If we only keep the first harmonics (FH) as in \refcite{Wu2019TIintTMD}, we can only manage to capture the top two valence bands in each valley.
To capture the top three valence bands in each valley, we add two extra second harmonic (SH) terms, and obtain a good match with the DFT band structure and symmetry representations (reps). We propose that this more accurate model be used in many-body calculations. 
In our model with SH terms, the gap closing between the second and third top valence bands in one valley happens around $4.2^\circ$ at zero displacement field, which is consistent with our DFT calculation.
We find that adding a displacement field can change the Chern number of the top valence band from $ 1$ to $0$ in $\K$ valley, and can also achieve a variety of Chern numbers (from -2 to 2) for the second and third top valence bands in one valley. Accessing these bands provides another route to integer Chern physics seen in twisted transition metal dichalcogenides\cite{2021Natur.600..641L,yankowitz2022moire,2022arXiv220702312Z,mak2022semiconductor,mai2023interaction}. 

In addition to the experimentally relevant AA-stacking configuration, we also study the AB-stacking configuration.
Our DFT results show that the $\Gamma$-valley bands in the AB stacking case is closer to the VBM (only about 30meV away) and are extremely flat.
We build a FH model capturing these $\Gamma$-valley bands, and show that they are extremely localized atomic bands whose flatness comes from zero hopping among atomic orbitals on the triangular lattice.
We also used the $\pm\K$-valley model in \refcite{Wu2019TIintTMD} to match the top two valence bands in each valley.

In the rest of this paper, we discuss the DFT calculations at a range of twist angles and stacking configurations in \secref{sec:DFT}, and the continuum models we employ to faithfully reproduce these calculations in \secref{sec:moire_model}.
We conclude the paper in \secref{sec:conclusion}, and provide more details in a series of appendices.

\section{DFT Calculations}
\label{sec:DFT}

In this section, we discuss the large-scale DFT calculations on the {\tmt} at various twist angles and different stacking configurations. We will mainly discuss the results for 3.89$\degree$, which is the closest one to the structure in recent experiments~\cite{cai2023signatures,zeng2023integer,park2023observation,Xu2023FCItMoTe2}, as well as the topological phase transition from large angle to 3.89$\degree$. A complete discussion can be found in \appref{app:DFT}.

\subsection{3.89$\degree$ {\tmt}}

Utilizing the coincidence lattice method \cite{coincidence1,coincidence2}, we construct twisted bilayer crystal structures of MoTe$_2$ at various commensurate angles: 13.2$\degree$, 9.43$\degree$, 7.34$\degree$, 5.09$\degree$, 4.41$\degree$, 3.89$\degree$ and 3.48$\degree$, considering both AA and AB stacking. Here, AA (AB) stacking implies that, without any twist, the Mo/Te atoms of the top layer respectively align with the Mo/Te (Te/Mo) atoms of the bottom layer. Subsequently, large-scale DFT calculations using the Vienna \emph{ab initio} simulation package (VASP) \cite{vasp1,vasp2,vasp3,vasp4,vasp5} are performed on these structures.

\begin{figure}
    \centering 
    \includegraphics[width=1.0\linewidth]{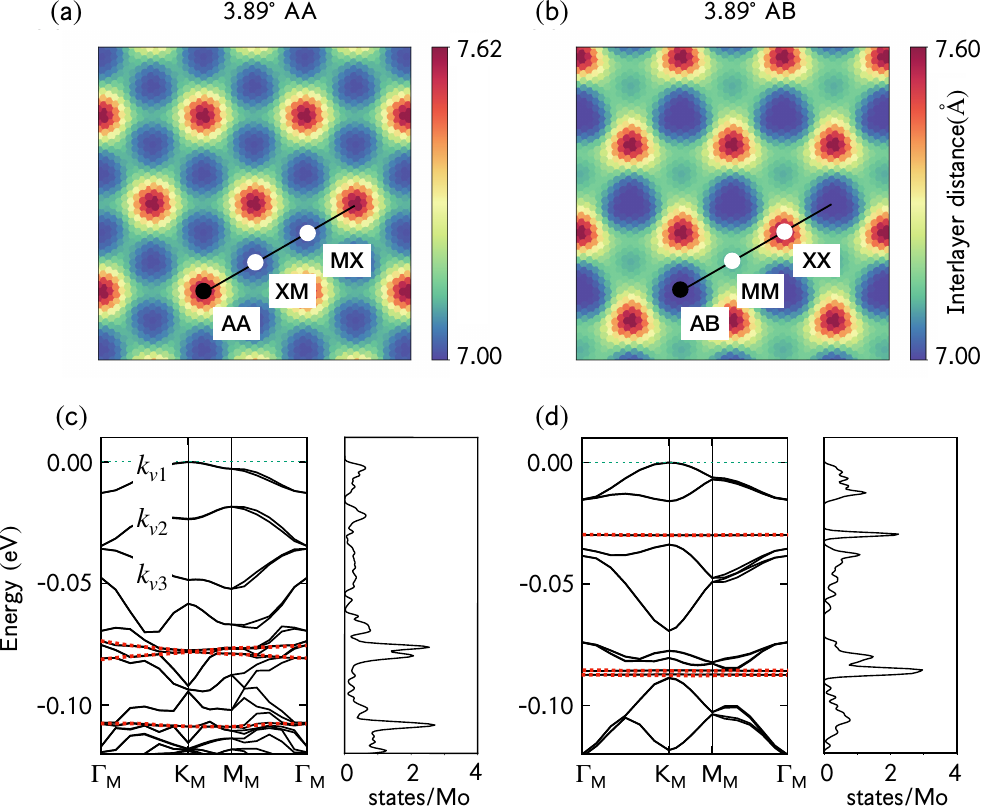}
    \caption{Relaxation and band structure of 3.89$\degree$ {\tmt}. $(a)$ Relaxation of the 3.89$\degree$ AA stacking. $(b)$ Relaxation of the 3.89$\degree$ AB stacking. $(c)$ and $(d)$ show the band structure and density of states (DOS) of the structures in $(a)$ and $(b)$ respectively, with red lines indicating bands from the $\Gamma$ valley and the green line marking the Fermi level.The notations $k_{v1}$, $k_{v2}$, and $k_{v3}$ represent the three highest pairs of valence bands of the AA-stacked configuration in the $\pm\K$ valleys. The DOS is normalized per molybdenum atom.
    }
    \label{fig:relaxband_main}
\end{figure}

To capture the van der Waals (vdW) interactions between top and bottom layers, we test 19 exchange-correlation functionals using the experimental bulk crystal structure \cite{1961_D.Puotinen} as a benchmark (see \tabref{vdw_functionals} in \appref{app:moireStructure}). Ultimately, we find that DFT-D2 gives the lattice parameter closest to the experimental value~\cite{1961_D.Puotinen}, and we use this functional throughout the work.
Secondly, different pseudo-potential (PP) combinations have also been tested. It has been found that the energy differences between different PPs are negligible (see \figref{pseudopotential-bs} of \appref{app:moireStructure}). Considering the computational cost, we choose the PAW pseudo-potential and PBE exchange-correlation functional.

Based on these funtionals, we develop a highly efficient two-step relaxation scheme which combines machine learning and DFT. First, we construct the machine-learned force field (MLFF) using the relaxation data generated by DFT-D2. The MLFF method has a much lower computational cost than the direct relaxation with DFT since the moir\'e unit cells are very large with 1302 aotms.
MLFF can produce a relaxed structure quickly. Using this structure as an initial guess, we further perform the full DFT relaxation. It turns out that the MLFF is reasonably good, making the DFT relaxation quite fast. To achieve total-force convergence with an accuracy of 5meV/$\AA$ and energy convergence of  $1\times 10^{-5}$eV, the DFT+MLFF method requires 17 ionic steps which take 7.5 hours using 16 NVIDIA A100 GPUs. In contrast, direct relaxation using DFT requires 178 ionic steps and take 55 hours when using 8 NVIDIA A100 GPUs, demanding approximately four times more resources than the DFT+MLFF method. 

The faster convergence of the DFT+MLFF method (compared to the DFT direct relaxation) does not sacrifice precision in the band structure.
To show it, we present the band structures obtained from three different approaches: MLFF relaxation, DFT+MLFF relaxation, and DFT direct relaxation, as shown in \figref{fig:3relax} of \appref{app:3.89band}. The band structures obtained from the DFT+MLFF and DFT relaxed methods have qualitatively the same shapes, which is consistent with the fact that the DFT relaxation upon the MLFF structures is quite fast.
In particular, the bands from the DFT+MLFF and the direct DFT relaxation methods are extremely similar, verifying the validity of the DFT+MLFF method.
Therefore, we use the DFT+MLFF method in this work instead of the DFT direct relaxation.
The stacking-dependent corrugated moir\'e structures of AA and AB configuration generated by the DFT+MLFF method are shown in \figref{fig:relaxband_main}(a,b). 

AA-stacked {\tmt} has a two-fold rotational symmetry axis along $y$ axis, $C_{2y}$, and the $C_3$ symmetry. The relaxed structure exhibits a maximum interlayer distance $d=7.62\angstrom$ in the AA region, where the metal atoms in top layer is aligned with metal atoms in bottom layer, while a minimum interlayer distance $d=7.0\angstrom$ in the MX (XM) region, where the top layer metal (chalcogen) atoms are aligned with chalcogen (metal) atoms of bottom layer, as shown in \figref{fig:AAandAB} of \appref{app:moireStructure}.

For AA {\tmt} at $3.89$ degree, the VBM is located at the $\K_M$ point in the moir\'e Brillouin zone (BZ), which is folded from the $\K$ point in the untwisted bilayer structure.
The top three pairs of valence bands, labeled as $k_{v1}$, $k_{v2}$ and $k_{v3}$, originate from the $\pm\K$ valleys, exhibiting bandwidths of 12.8 meV, 16.2 meV, and 16.5 meV, respectively. The combined effects of lattice relaxation and SOC lead to the $\Gamma$-valley bands shifting downward by about 80 meV from the VBM. The $\Gamma$-valley bands, illustrated in \figref{fig:relaxband_main} (c) and marked by red dashed lines, contribute to two distinct peaks in the density of states (DOS).

\begin{figure*}[htpb]
    \centering
    \includegraphics[width=1.0\linewidth]{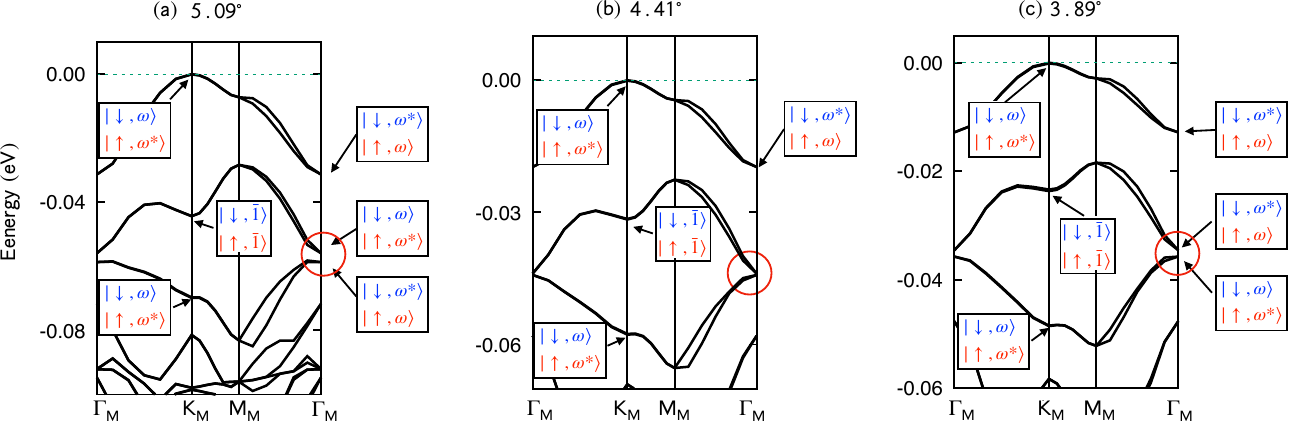}
    \caption{Irreducible representations at high-symmetry points $\Gamma$ and $K$ of valence bands of monolayer and twist MoTe$_2$ with twist angle 5.09$\degree$, 4.41$\degree$, and 3.89$\degree$ for relaxed AA stacking configuration with SOC. The eigenvalues of the $C_{3}$ are denoted as  $\omega=e^{i\pi/3}$, $\omega^*=e^{-i\pi/3}$ and $\bar{1}=e^{-i\pi}$. The critical point of band inversion between the 2nd and 3rd valence bands at $\gamma$ point of moire BZ happens at around twist angle 4.41$^{\circ}$. 
    } 
    \label{fig:irreps-AA-vbs}
\end{figure*}

The AB stacking configuration has a two-fold rotational symmetry $C_{2x}$ with the axis along $x$ as well as the three-fold rotation symmetry $C_{3}$ with axis along $z$. For the twist angle $3.89^\circ$, the top two pairs of valence bands also come from $\pm\K$ valley with band width of 16 meV, while the first pair of ultra flat bands from $\Gamma$ valley is only 14.6 meV below the VBM. The charge density of these ultra flat bands is highly localized in the AB region of the moir\'e lattice (see Fig.\ref{fig:fig-density} of \appref{app:3.89band}).

\subsection{Topological phase of valence bands from 5.09$\degree$ to 3.89$\degree$}

In this part, we will discuss the topological phase transition around $4.41^\circ$ of the valence bands in AA-stacked \tmt. Since the full DFT calculations are performed in huge unit cells, it is impractical to calculate the Chern number directly, and we turn to symmetry eigenvalues to efficiently deduce the topology\cite{PhysRevB.86.115112}.

\figref{fig:irreps-AA-vbs} shows the band structure and the $C_{3}$ eigenvalues at the high-symmetry points. Our convention is to use the spinful $C_{3}$ eigenvalues, which are labelled by $\omega=e^{i\pi/3}$, $\omega^*=e^{-i\pi/3}$ and $\bar{1}=e^{-i\pi}$.
Since $\pm\K$-valley bands are valley-spin locked, we distinguish two valleys by their spins (spin up/down for $\K$/$-\K$ valley).

For 5.09$\degree$, we observe that there is a gap of 2.8 meV between $k_{v2}$ and $k_{v3}$ bands at the moir\'e $\Gamma_M$ point. This gap is closed around 4.41$\degree$, and reopens at angle 3.89$\degree$. From the $C_{3}$ eigenvalues labeled in \figref{fig:irreps-AA-vbs}, we see that this band crossing exchanges the symmetry representations at $\Gamma_M$ for $k_{v2}$ and $k_{v3}$ bands. Explicitly, for 5.09$\degree$, the $k_{v2}$ band at $\Gamma_M$ has two spin-polarized states: $\ket{\uparrow, \omega^{*}}$ and $\ket{\downarrow, \omega}$, while the $k_{v3}$ band has $\ket{\uparrow,\omega}$ and $\ket{\downarrow,\omega^{*}}$.
When energy gap reopening occurs, the symmetry representations of $k_{v2}$ and $k_{v3}$ at 3.89$\degree$  switch in comparison to those at 5.09$\degree$.

The exchange of symmetry eigenvalues is proof of a band inversion and causes a change in the spin/valley Chern number. Specifically, recall that the Chern number can be determined from symmetry via $e^{i \frac{2\pi}{3}C} = - \xi_{\Gamma_M} \xi_{\K_M} \xi_{\K'_{M}}$ where $\xi_\mbf{k}$ is the spinful $C_{3}$ eigenvalue at high-symmetry point $\mbf{k}$~\cite{PhysRevB.86.115112}. Since $C_{2y} \mathcal{T}$ relates the moir\'e $\K_{M}$ and $\K_{M}'$ points and is anti-unitary, $\xi_{\K_{M}} =\xi_{\K_{M}'}$. Thus for $\th$ slightly larger than $4.41^\circ$, we find $e^{i \frac{2\pi}{3}C} = -(-1)(-1) \omega^*$ so that $C=1$ mod 3 for the second top spin-$\uparrow$ band, but for $\th$ slightly smaller than $4.41^\circ$, we find $e^{i \frac{2\pi}{3}C} = -(-1)(-1) \omega$ so that $C=-1$ mod 3. This topological phase transition is matched by the continuum model as we will show in \secref{sec:moire_model}.

\subsection{Comparison with \refcite{reddy2023fractional,wang2023fractional}}

\begin{figure*}
	\includegraphics[width=1\linewidth]{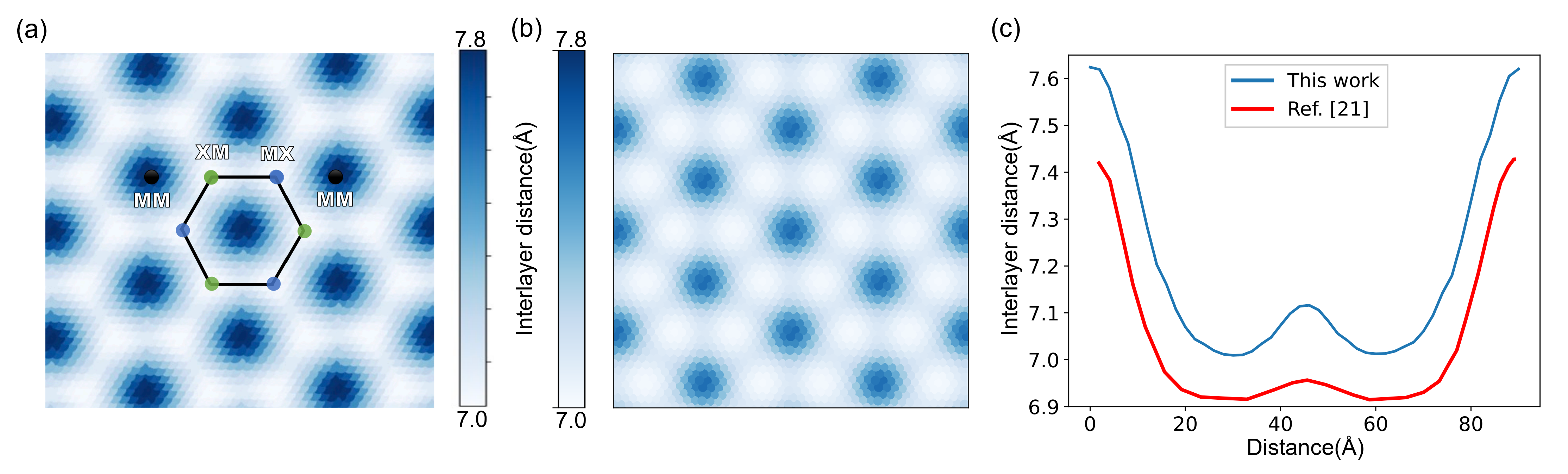}
\caption{Relaxation results of AA twisted structure. (a) Relaxation result of 4.41$\degree$ AA MoTe$_2$ from Fig.2 (a) of \refcite{reddy2023fractional}. (b) Our relaxation result of 4.41 $\degree$ AA MoTe$_2$. 
(c) The interlayer distance of 3.89$\degree$ AA MoTe$_2$ along the black arrow in \figref{fig:relaxband_main}(a). The red curve is the result from Fig.1(d) from \refcite{wang2023fractional}, and the blue curve is our result.}
\label{comparison}
\end{figure*}

Two recent works~\cite{reddy2023fractional,wang2023fractional} have also studied the relaxation and band structure of {\tmt}. The band structures in \refcite{reddy2023fractional,wang2023fractional} are different from our result, mainly due to the different relaxed structures.
\refcite{reddy2023fractional} employed the SCAN density functional with dDsC dispersion correction to perform crystal structure relaxation using VASP, while \refcite{wang2023fractional} used SIESTA with DFT-D2 functional to perform the relaxation.
The comparison of the relaxation results of {\AAtmt} between this work and \refcite{reddy2023fractional,wang2023fractional} is shown in \figref{comparison}.
As shown in \figref{comparison}, the interlayer distance in our relaxed structure has the qualitatively the same shape as those in \refcite{reddy2023fractional,wang2023fractional}---largest interlayer distance at MM and the smallest interlayer distance at MX/XM.
However, the interlayer distance in our relaxed structure has smaller spatial fluctuations than that in \refcite{reddy2023fractional} as shown in \figref{comparison}(a,b), while our relaxed structure has larger interlayer distance than that of \refcite{wang2023fractional} (\figref{comparison}(c)).
The maximum interlayer distances of this work, \refcite{reddy2023fractional}, and \refcite{wang2023fractional} are about 7.6$\angstrom$, 7.8$\angstrom$ and 7.4$\angstrom$, respectively, and the minimum interlayer distances are respectively about 7.0$\angstrom$, 7.0$\angstrom$ and 6.9$\angstrom$.

Our relaxation result is consistent with the AA and AB stacking untwisted bilayer structure.
In the AA region, the stacking configuration is close to that of AA untwisted bilayer structure, and thus the maximum interlayer distance should be close to but slightly smaller than (due to the corrugation effect due to the connection to other stacking configurations in the moir\'e structure) the interlayer distance of AA stacking untwisted bilayer (7.7$\angstrom$), which is consistent with our results but not with \refcite{reddy2023fractional}.
Furthermore, the MX region has the stacking configuration akin to that of AB untwisted bilayer structure. As a result, the smallest interlayer distance in {\tmt} structures should be close to but slightly larger than (due to the corrugation) 7$\angstrom$, which is consistent with our results but not with \refcite{wang2023fractional}.
More details about the relaxation and its influence on band structure are discussed in \appref{app:comparision}.

\begin{figure*}[t]
    \centering
    \includegraphics[width=1.8\columnwidth]{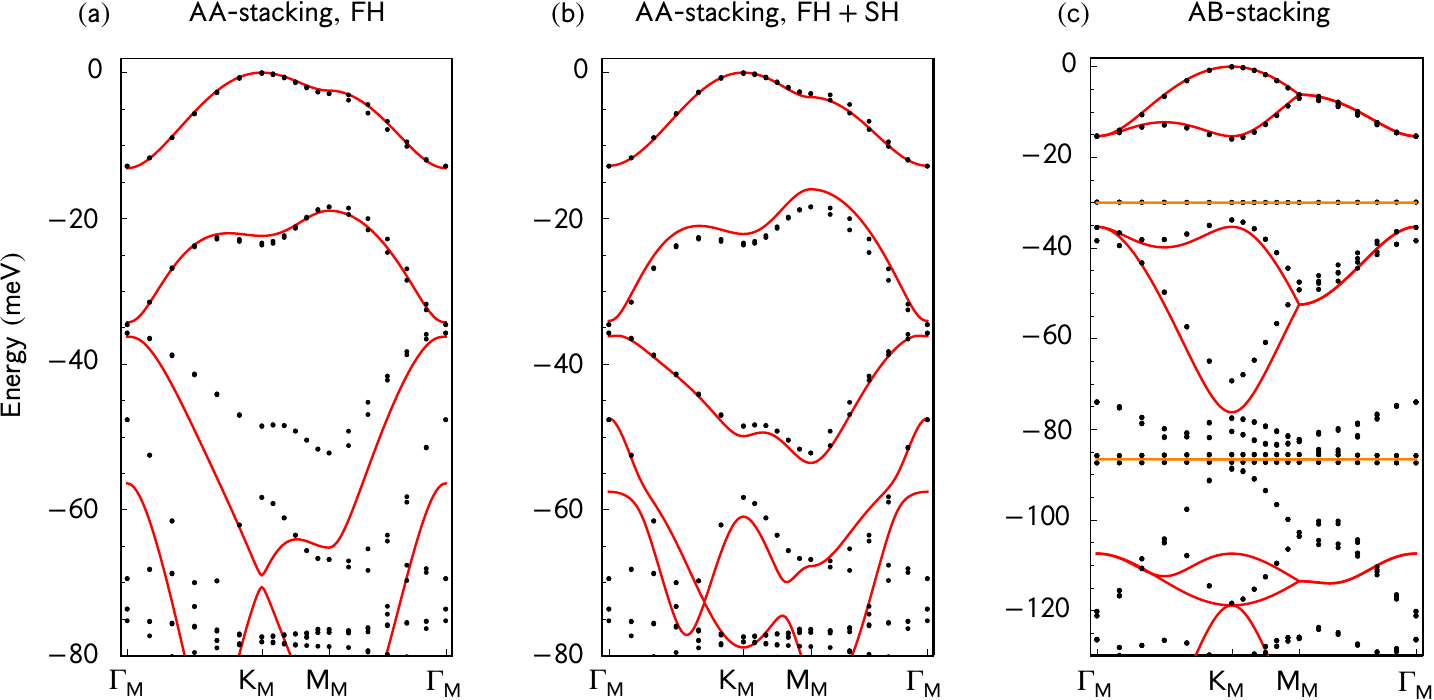}
    \caption{
    The comparison between the valence bands from the DFT calculation (black dots) and from the continuum model (red/orange line) for twist angle $3.89^\circ$ for (a,b) AA-stacking and (c) AB-stacking along the high-symmetry line.
    The red line comes from the $\pm\K$-valley model, while the orange line comes from the $\Gamma$-valley model.
    Each red line is doubly degenerate (except along $\Gamma_M-M_M$ in (c) where it is 4-fold degenerate).
    The orange line in (c) has double degeneracy around $-30$meV and 4-fold degeneracy around $-86$meV.
    The fitting in (a) is done with only FH terms, while SH terms are added in (b).
    }
    \label{fig:fitting_main}
\end{figure*}

\section{Continuum Models}

\label{sec:moire_model}

In this section, we use the continuum model to fit the DFT results.
Up to now, FCI states were only found for hole doping experimentally~\cite{cai2023signatures,zeng2023integer,park2023observation,Xu2023FCItMoTe2}; thus, we will focus on the model for the valence bands. 
We will first discuss the AA-stacking and then discuss the AB stacking.

\subsection{AA-Stacking}

According to the DFT results, the low-energy valence bands mainly originate from the $\pm\K$ valleys in the monolayer. 
The symmetry group of AA-stacking {\tmt} (\AAtmt) is generated by $C_3$, $C_{2y}$, and $\TR$, in addition to the moir\'e lattice translations~\cite{Wu2019TIintTMD}. We pick a convention where the top layer is rotated by $-\theta/2$ and the bottom layer by $\theta/2$.
The moir\'e lattice constant is
\eq{
a_M = \frac{a_0}{2 \sin\left( \frac{\theta}{2} \right)} \ ,
}
where $a_0 = 3.52\AA$ is the lattice constant of monolayer MoTe$_2$.

The continuum model for the monolayer $\pm\K$ valleys in general reads
\eq{
\label{main_eq:h_AA_Kvalley}
H_{\eta,0}^{AA} = \int d^2 r \left( c^\dagger_{\eta,b,\bsl{r}} , c^\dagger_{\eta,t,\bsl{r}} \right) h_{\eta,0}^{AA}(\bsl{r}) \mat{ c_{\eta,b,\bsl{r}} \\ c_{\eta,t,\bsl{r}} } \ ,
}
where $c^\dagger_{\eta,l,\bsl{r}}$ labels the basis of the continuum model, $\eta=\pm$ labels the $\pm \K$ valleys (or equivalently spins), $l=t,b$ labels the layer, and $\bsl{r}$ labels the position.
\refcite{Wu2019TIintTMD} proposed a model with only the first harmonics (FH); however, here we add the certain second harmonics (SH) terms in order to accurately match the higher bands. (See \appref{app:Kvalley_AA} for details.)
Expanding the potential terms and using symmetry, we find the form
\eqa{
\label{main_eq:Kvalley_V_t_FH_SH}
& V_{\eta,l}(\bsl{r}) = V e^{ - (-)^l \ii \psi} \sum_{i=1,2,3} e^{\ii \bsl{g}_i \cdot \bsl{r} } + V e^{ (-)^l \ii \psi} \sum_{i=1,2,3} e^{-\ii \bsl{g}_i \cdot \bsl{r}} \\
& \qquad + 2 V_2 \sum_{i=1}^3 \cos(\bsl{g}_{2i}\cdot \bsl{r} ) \\
& t_\eta(\bsl{r}) = w \sum_{i=1,2,3} e^{ - \eta \ii \bsl{q}_i \cdot \bsl{r} } + w_2 \sum_{i=1,2,3} e^{ - \eta \ii \bsl{q}_{2i} \cdot \bsl{r} }\ ,
}
where we take $(-)^t = 1$, $(-)^b = -1$, $\bsl{g}_{i} = C_3^{i-1} \bsl{b}_{M,1}$, $\bsl{b}_{M,1} = \frac{4 \pi}{\sqrt{3} a_{M}} (1,0)^T$ and $\bsl{b}_{M,2} = \frac{4 \pi}{\sqrt{3} a_{M}}$ are the basis moir\'e reciprocal lattice vectors, $\bsl{q}_1 = \frac{4 \pi}{3 a_0} 2 \sin\left(\frac{\theta}{2}\right) (0,1)^T$, $\bsl{q}_2 = C_3 \bsl{q}_1$, $\bsl{q}_3=C_3^2 \bsl{q}_1$,  $ \bsl{g}_{21} = \bsl{b}_{M,1} + \bsl{b}_{M,2}$, $\bsl{g}_{2i} = C_3^{i-1} \bsl{g}_{21} $, $ \bsl{q}_{21} = \bsl{b}_{M,1} + \bsl{q}_{_1}$, and $\bsl{q}_{2i} = C_3^{i-1} \bsl{q}_{21} $.
$V$, $\psi$ and $w$ characterize the FH terms, while $V_2$ and $w_2$ belong to the SH.
The AA-stacking FH $\pm\K$-valley model (\eqnref{main_eq:h_AA_Kvalley}) has effective inversion symmetry that makes the two bands from the two valleys identical (in accord with the DFT results which show small splitting about 1.2 meV 
The effective inversion symmetry is natural with only FH terms, and we only include the SH terms that preserve the effective inversion symmetry.
In total, the model (\eqnref{main_eq:h_AA_Kvalley}) has six real parameters $m^*$, $V$, $\psi$, $w$, $V_2$ and $w_2$.

We fit to the DFT band structure at $\theta=3.89^\circ$ in two ways.  (See \appref{app:Kvalley_AA} for details.)
First, we set $V_2=w_2=0$, which corresponds to the FH model.
In this case, we manage to fit the top 4 valence bands (2 in each valley) with the corresponding FH parameters in \tabref{tab:AA_parameters_DFT}, as shown in \figref{fig:fitting_main}(a).
Then, we allow nonzero $V_2$ and $w_2$, \ie, adding the SH terms.
We are now able to fit the top 6 valence bands (3 in each valley) with the corresponding FH+SH parameters in \tabref{tab:AA_parameters_DFT}, as shown in \figref{fig:fitting_main} (b).
The match is not only good along the high-symmetry line but also good in the full BZ as shown in \figref{fig:fitting_BZ} (a,b) in \appref{app:AA_fitting}.

As shown in \figref{fig:C3eigenvalues_highsymline} (a,b) in \appref{app:AA_fitting}, we can see that the $C_{3}$ eigenvalues for the top 6 valence bands match the DFT calculation in both FH and FH+SH cases.
Furthermore, the Chern numbers of the top 3 bands (in decreasing order of energy) in $\K$ valley are (1,-1,0) in both cases, which are consistent with the $C_3$ eigenvalues~\cite{PhysRevB.86.115112}. The $C_3$ eigenvalues and the Chern numbers of the top two bands per valley are the same as those in \refcite{wang2023fractional}.
It is clear that adding the SH terms improves the reliability of the model across a wider range of energies. We expect more remote bands to be accessible in future experiments, and hence the our FH+SH model is an essential improvement. 

\begin{table}[t]
    \centering
    \begin{tabular}{|c|c|c|c|c|c|c|}
    \hline
     Model & $m^*$ ($m_e$) &  $V$   & $\psi (^\circ)$ & $w$ & $V_2$  & $w_2$ \\ 
    \hline
     \text{FH} &  0.60  & 16.5 & -105.9 & -18.8 & 0 & 0 \\
     \hline
      \text{FH+SH}   &  0.62  & 7.94  & -88.43  & -10.77 & 20.00 & 10.21 \\ 
    \hline
    \end{tabular}
    \caption{Values of the parameters in the $\pm \K$-valley continuum model (\eqnref{main_eq:Kvalley_V_t_FH_SH}) for the AA-stacking {\tmt}.
    $V,w,V_2,w_2$ are in meV.
    ``FH'' means we only include the first harmonics, whereas ``FH+SH'' means that we include both the first harmonics and the effective-inversion-invariant second harmonics.
    }
    \label{tab:AA_parameters_DFT}
\end{table}

At last, we discuss the evolution of the bands of the AA-stacking moir\'e model in \eqnref{main_eq:h_AA_Kvalley} with the FH+SH parameters values in \tabref{tab:AA_parameters_DFT} as a funciton of the twist angle $\theta$ and the displacement field $\varepsilon$.
We will focus on the $\K$ valley.
As shown in \figref{fig:fitting_main} (a), the gap between second top and third top bands closes around $4.2^\circ$ for zero displacement field, which is close the the DFT's $4.41^\circ$ in \figref{fig:irreps-AA-vbs}; the gap closing will change the Chern numbers of the second and third top bands in $\K$ valley from (-1,0) to (1,-2).
Further increasing the angle at zero displacement field will cause a band inversion between the third and fourth top valence bands, which changes the Chern number of the third top band from -2 to -1 as shown in \figref{fig:fitting_main} (d).
\figref{fig:fitting_main} (b) shows that increasing the displacement field can trivialize the top valence band, while a variety of Chern numbers (ranging from -2 to 2) can arise for nonzero displacement field for the second and third top bands as shown in \figref{fig:fitting_main} (c,d).

\begin{figure}[t]
    \centering
    \includegraphics[width=1\columnwidth]{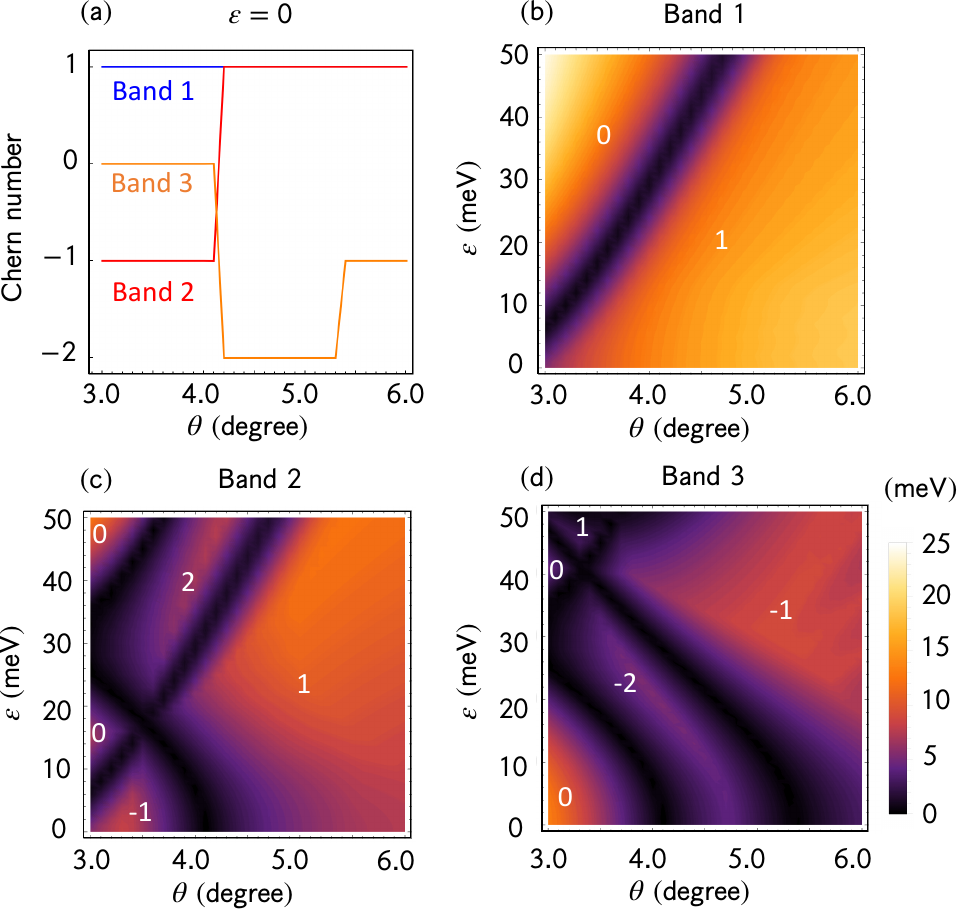}
    \caption{
    Phase diagram of AA-stacking moir\'e model in \eqnref{main_eq:h_AA_Kvalley} with the FH+SH parameters values in \tabref{tab:AA_parameters_DFT}.
    In this plot, band 1,2,3 refer to the top, second top and third top valence bands in the $\K$ valley.
    In (a), we show the Chern numbers of the three bands (blue for 1, red for 2 and orange for 3) for zero displacement field, which shows a band inversion between the second and third top bands around $4.2^\circ$. 
    In (b,c,d), we show the direct gaps and the Chern numbers of the top (a), second top (b) and third top (c) bands in the $\pm\K$ valley.
    Explicitly, the direct gaps are illustrated by the color (the color bar is the same across plots) and the Chern numbers are labeled in white.
    }
    \label{fig:AAEfieldThetaPhaseDiagram}
\end{figure}

\subsection{AB-Stacking}

The generators of the symmetry group of AB-stacking {\tmt} ({\ABtmt}), which can be thought of as twisting the top layer of AA-$t$MoTe$_2$ by another $180^\circ$, are $C_{3}, C_{2x}$, and $\mathcal{T}$. Note that $C_{2x}$ is local to the monolayer $\K$ point (unlike $C_{2y}$ in \AAtmt), and thus it preserves the valley quantum number in the moir\'e model. This difference in the valley symmetry group is important and, as we now show, leads to different behavior with the potential for interesting many-body spin and Hubbard physics \cite{2022NatNa..17..934X,xu2020correlated,kometter2023hofstadter, PhysRevB.107.235131,tan2023layer,PhysRevLett.128.217202,2023arXiv230409808F}. 

The DFT results show that the low-energy valence bands of {\ABtmt} come from both the $\pm\K$ valleys and the $\Gamma$ valleys in the monolayer. 
The $\pm \K$-valley model was proposed in \refcite{Wu2019TIintTMD}, which reads
\eqa{
H_{\eta,0}^{AB} & = \sum_l \int d^2 r  c^\dagger_{\eta,l,\bsl{r}}  \left[ \frac{\hbar^2 \nabla^2}{2 m^*} + V_{\eta,l}(\bsl{r}) +(-)^l \frac{\varepsilon}{2} \right]  c_{\eta,l,\bsl{r}} \\
& \ + \left[ \int d^2 r  c^\dagger_{\eta,b,\bsl{r}}  t_\eta(\bsl{r})  c_{\eta,t,\bsl{r}} + h.c. \right]\ ,
}
where $c^\dagger_{\eta,l,\bsl{r}}$ labels the basis of the continuum model with $\eta=\pm$ labeling the $\pm \K$ valleys (or equivalently spins), $l=t,b$ labels the layer, and $\bsl{r}$ labels the position. 
Since {\ABtmt} is given by rotating the top layer of {\AAtmt} by an extra $180^\circ$, the two layers in $\K$ or $-\K$ valley now have opposite spin.
With first harmonics, the forms of $V_{\eta,l}(\bsl{r})$ and $ t_\eta(\bsl{r}) $ derived from the symmetries read 
\eqa{
\label{main_eq:Kvalley_model_AB}
& V_{\eta,l}(\bsl{r}) = V e^{ - \ii \psi} \sum_{i=1,2,3} e^{\ii \bsl{g}_i \cdot \bsl{r} } + V e^{ \ii \psi} \sum_{i=1,2,3} e^{-\ii \bsl{g}_i \cdot \bsl{r}} \\
& t_\eta(\bsl{r}) = w \sum_{i=1,2,3} e^{\eta \ii  (i-1)\frac{2\pi}{3} } e^{ - \eta \ii \bsl{q}_i \cdot \bsl{r} }\ ,
}
where $V$ is real, and $w$ can made real by choosing the relative phase between the two layers.
For $\varepsilon = 0$, the minimal continuum model has the effective TR symmetry within each valley that flips layer (or equivalently spin), making the bands from the two valleys identical. 

On the other hand, the $\Gamma$-valley model has the following form
\eq{
H_{\Gamma} = \int d^2 r \mat{ \psi^\dagger_{\bsl{r},b} &  \psi^\dagger_{\bsl{r},t} } h_{\Gamma}^{AB}(\bsl{r})
\mat{ \psi^\dagger_{\bsl{r},b}  \\  \psi^\dagger_{\bsl{r},t} }\ ,
}
where we choose the kinetic term in the continuum model as the intralayer spin-independent $\nabla^2$ term,
\eqa{
\label{main_eq:Gammavalley_model_AB} 
 & h_{\Gamma}^{AB}(\bsl{r}) =
\frac{\hbar^2 \nabla^2 }{2 m_\Gamma^*} +E_\Gamma  + \mat{
 V_{\Gamma,b}(\bsl{r}) -\epsilon/2 & t_{\Gamma}(\bsl{r})\\
t_{\Gamma}^\dagger (\bsl{r}) &  V_{\Gamma,t}(\bsl{r})+ \epsilon/2
}\ ,
}
\eq{
\psi^\dagger_{\bsl{r},l} = ( \psi^\dagger_{\bsl{r},l,\uparrow} , \psi^\dagger_{\bsl{r},l,\downarrow} )\ ,
}
$t$ and $b$ correspond to the top and bottom layers, respectively, $E_\Gamma$ accounts for the energy difference between the $\Gamma$-valley and $\pm\K$-valley bands, and $V_{\Gamma,l}(\bsl{r})$ and $t_{\Gamma}(\bsl{r})$ are $2\times 2$ matrix functions.
For zero displacement field, the $\Gamma$-valley model has effective TR symmetry with in each spin-subspace, which makes the bands from the two spins identical (resulting in at least double degeneracy of each band).

We fit the low-energy bands at $\theta=3.89^\circ$, and the resulting parameter values are summarized in \tabref{tab:AB_parameters_DFT}.
As the illustration, we show the good match between the DFT bands and those from the models along the high-symmetry line in \figref{fig:fitting_main} (b).
The bands match well also in the full BZ, and the $C_{3}$ eigenvalues also match the DFT calculation as discussed in \appref{app:ABFitting}.

For the $\pm\K$-valley model, we notice that the interlayer coupling can be set to zero $w=0$ in the $\pm\K$-valley model while keeping the match of the bands good, which can be understood as the follows.
The two layers in one valley now have opposite spins; owing to the spin U(1) symmetry for the low-energy states near $\pm \K$ valleys in monolayer {\mt}, we expect the spin $U(1)$ symmetry is approximately preserved in {\tmt}, which means the interlayer coupling is very small for the AB-stacking.
The zero interlayer coupling makes the eigenstates have well-defined valley and layer/spin.
As a result, the two states with the same spin in the $\pm\K$-valley model are degenerate at $\K_M$, since the combination of the effective TR symmetry and the TR symmetry leaves the spin invariant; this is consistent with DFT calcualtions as shown in \figref{fig:irreps-2H-vbs} of \appref{app:ABFitting}.
The top four valence bands is in the $A_1@1a$ atomic limit, and the band structure can be thought of as arising from hopping on this moir\'e triangular lattice. 

The $\Gamma$-valley bands are extremely flat as shown in \figref{fig:fitting_main} (b), owing to the large $m^*_{\Gamma} = 10 m_e$ ($-\frac{\hbar^2 |\bsl{q}_1|^2}{2 m^*_{\Gamma}} = -2.49$meV) compared to the potential $V_{\Gamma}=72$meV.
As a result, those flat bands are extremely localized atomic states localized around the minima of the intralayer potential (\ie, 1a positions due to $\psi=0$), and their energies can be approximately calculated from an array of decoupled harmonic oscillators related by moir\'e translations. (See \appref{app:ABFitting}.)

At last, we discuss the effect of the displacement field.
As shown in \figref{fig:AB_Efield} in \appref{app:ABFitting}, the effect of the displacement field on the $\pm \K$-valley bands are just to shift the bands from different layers relative to each other, since the valley is a good quantum number due to the zero interlayer coupling (see \tabref{tab:AB_parameters_DFT}).
On the other hand, the effect of the displacement field on the low-energy $\Gamma$-valley bands is negligible, which is consistent with the fact that the very large interlayer coupling makes the eigenstates equally distributed between the two layers (see \tabref{tab:AB_parameters_DFT}).

\begin{table}[t]
    \centering
    \begin{tabular}{|c|c|c|c|c|}
    \hline
      $m^*$ ($m_e$) &  $V$ (meV)  & $\psi (^\circ)$ & $w$ (meV) \\ 
    \hline
       0.62  & 53 & -56 & 0  \\
    \hline\hline
      $m_{\Gamma}^*$ ($m_e$) &  $V_\Gamma$ (meV)  & $\psi_\Gamma (^\circ)$ & $w_\Gamma$ (meV) \\ 
    \hline
     10  & 72 & 0 & 300  \\
    \hline
    \end{tabular}
    \caption{Values of the parameters in the $\pm \K$-valley continuum model in \eqnref{main_eq:Kvalley_model_AB} (first and second rows) and the $\Gamma$-valley continuum model in \eqnref{main_eq:Gammavalley_model_AB} (third and fourth rows) for the AB-stacking {\tmt}.
    }
    \label{tab:AB_parameters_DFT}
\end{table}

\section{Conclusions}
\label{sec:conclusion}

Our extensive DFT study, accelerated by machine learning, has confirmed that, and our continuum model analysis show that the lowest two bands in the $\K$ valley have Chern numbers $+1$ and $-1$ respectively, at a twist angle of $3.89\degree$ for the experimentally relevant AA stacking. These Chern numbers mod 3 are accessible directly from the DFT data, and we compute the exact value from the continuum model fit to the DFT bands. The Chern numbers are consistent with \refcite{wang2023fractional} but do not agree with those computed by \refcite{reddy2023fractional} at larger twist angles and extrapolated to $3.89\degree$. 

Unlike the fitting in \refcite{reddy2023fractional,wang2023fractional} which captures the top two valence bands per valley, we succeeded in matching the top three valence bands (as well as much of the fourth band) per valley by adding only two higher harmonic terms to the moir\'e Hamiltonian (see Eq. \eqref{main_eq:h_AA_Kvalley}). This Hamiltonian still preserves the effective intra-valley inversion symmetry, and will serve as a faithful model of the dispersion and topology for a wide range of electron fillings. Excitingly, our model predicts a rich topological phase diagram accessible through displacement fields in the remote bands. Our forthcoming work will study the many-body physics of this model, contributing to the study of correlations and topology \cite{2023arXiv230914340S,PhysRevB.103.205415,2022PhRvL.129d7601S,Bultinck2019GroundSA,2023PhRvB.107x5145S,2022arXiv221200030H,Phillips2023DQMCHaldaneModel,2023arXiv230105588W,Phillips2023DQMC, ding2023particle,mai2023interaction} now accessible in experiment.

\section{Acknowledgments}
The authors thank  Allan H. MacDonald, Kin-Fai Mak, Jie Shan, Zhijun Wang and Xiaodong Xu for helpful discussions, with special thanks to D. Xiao, Y. Zhang and Jianpeng Liu for their helpful cross-checking of the details in the DFT calculations. J.Y. and J. H.-A. are grateful for conversations with Pok Man Tam. 
This work was supported by the Ministry of Science and Technology of China (Grant No. 2022YFA1403800), the Science Center of the National Natural Science Foundation of China (Grant No. 12188101) and the National Natural Science Foundation of China (Grant No.12274436). H.W. acknowledge support from the Informatization Plan of the Chinese Academy of Sciences (CASWX2021SF-0102). 
B. A. B.’s work was primarily supported by the DOE Grant No. DE-SC0016239 and the Simons Investigator Grant No. 404513. N.R. also acknowledges support from the QuantERA II Programme that has received funding from the European Union’s Horizon 2020 research and innovation programme under Grant Agreement No 101017733 and from the European Research Council (ERC) under the European Union’s Horizon 2020 Research and Innovation Programme (Grant Agreement No. 101020833). 
. O.V. was funded by the Gordon and Betty Moore Foundation’s EPiQS Initiative Grant GBMF11070, National High Magnetic Field Laboratory
through NSF Grant No. DMR-1157490 and the State of Florida. J. H.-A. is supported by a Hertz Fellowship, with additional support from DOE Grant No. DE-SC0016239 by the Gordon and Betty Moore Foundation through Grant No. GBMF8685 towards the Princeton theory program, the Gordon and Betty Moore Foundation’s EPiQS Initiative (Grant No. GBMF11070), Office of Naval Research (ONR Grant No. N00014-20-1-2303), BSF Israel US foundation No. 2018226 and NSF-MERSEC DMR. J. Y. is supported by the Gordon and Betty Moore Foundation through Grant No. GBMF8685 towards the Princeton theory program.

\appendix

\section{DFT Results}
\label{app:DFT}

\subsection{Atomic structures of {\tmt}}
\label{app:moireStructure}

\begin{figure*}[!htpb]
    \centering
    \includegraphics[width=1.0\linewidth]{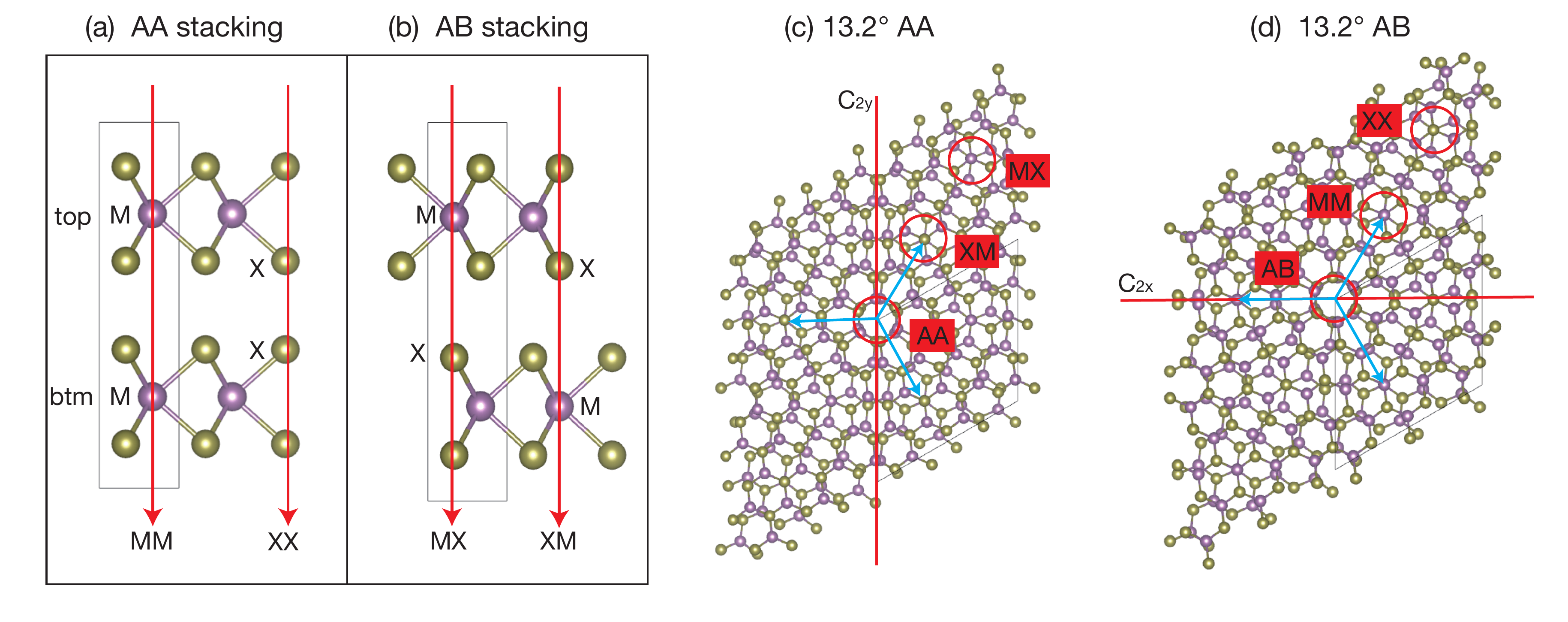}
    \caption{Rigid atomic structures of 13.2$\degree$ {\tmt}. The purple one is Mo atom and yellow-brown one is Te atom. (a) Bilayer AA stacking configuration. (b) Bilayer AB stacking configuration. (c) 13.2$\degree$ {\tmt} AA configuration with in-plane 2-fold rotational symmetry axis along the y-axis. (b) 13.2$\degree$ {\tmt} AB configuration with in-plane 2-fold rotational symmetry axis along the x-axis. Both AA and AB have the $C_3$ 3-fold rotation symmetry axis perpendicular to the plane. Mo atom is labeled by M, while chalcogen atom is labeled by X. MX represents the top layer Mo atom is aligned with bottom layer chalcogen atom. The similar with XM and MM. }
    \label{fig:AAandAB}
\end{figure*}
Bulk 2H-MoTe$_2$, as sketched in \figref{fig:AAandAB} (b),  has a hexagonal structure with space group 194 generated by the inversion $\P$, a three-fold symmetry $C_3$ with axis along z axis, a two-fold symmetry $C_{2x}$ with axis along x axis, and lattice translations. 
Experimentally, the lattice constant is found to be a=3.519$\AA$  and c=13.976$\AA$~\cite{1961_D.Puotinen}.

Before performing DFT calculations for band structure, we need to choose the suitable exchange-correlation functionals to describe the electron-electron interaction and the suitable pseudopotentials to deal with the interactions between electrons and nucleus. 
To pick the best exchange-correlation functionals, we try 19 different ones to calculate the relaxed bulk crystal structure, and show the results in \tabref{vdw_functionals}.
We find that the DFT-D2 functional (IVDW=10) provides a lattice constant of $a=b=3.518\angstrom$ and $c=13.976\angstrom$ for the bulk {\mt} primitive crystal structure, which is closest to the experimental values in \refcite{1961_D.Puotinen}.
Therefore, we conclude that the DFT-D2 functional (IVDW=10) is the most appropriate choice for the MoTe$_2$ structural relaxation among the 19 functionals.

\begin{table*}
\begin{tabular}{|c|c|c|c|c|c|}
\hline
\textbf{}       & \textbf{experiment}  & \textbf{optB86}     & \textbf{optB88}   & \textbf{vdw-DF}     & \textbf{vdw-DF-cx}       \\ \hline
\textbf{a,b(Å)} & 3.519                 & 3.527                & 3.567              & 3.631                & 3.502                     \\ \hline
\textbf{c(Å)}   & 13.964                & 14.032               & 14.213             & 15.007               & 13.867                    \\ \hline
\textbf{}       & \textbf{PBE}         & \textbf{optPBE-vdw} & \textbf{rVV10}    & \textbf{SCAN+rVV10} & \textbf{r$^2$SCAN+rVV10} \\ \hline
\textbf{a,b(Å)} & 3.551                 & 3.580                & 3.546              & 3.503                & 3.542                     \\ \hline
\textbf{c(Å)}   & 15.095                & 14.475               & 13.949             & 14.223               & 14.187                    \\ \hline
\textbf{}       & \textbf{rev-vdw-DF2} & \textbf{vdw-DF2}    & \cellcolor{lightgray}\textbf{IVDW=10}  & \textbf{IVDW=11}    & \textbf{IVDW=12}         \\ \hline
\textbf{a,b(Å)} & 3.529                 & 3.711                & \cellcolor{lightgray}3.519              & 3.512                & 3.490                     \\ \hline
\textbf{c(Å)}   & 14.028                & 14.891               & \cellcolor{lightgray}13.976             & 13.984               & 13.649                    \\ \hline
\textbf{}       & \textbf{IVDW=20}     & \textbf{IVDW=21}    & \textbf{IVDW=263} & \textbf{IVDW=4}     & \textbf{IVDW=3}          \\ \hline
\textbf{a,b(Å)} & 3.514                 & 3.516                & 3.490              & 3.515                & 3.531                     \\ \hline
\textbf{c(Å)}   & 13.923                & 13.826               & 13.709             & 14.047               & 14.222                    \\ \hline
\end{tabular}
\caption{Relaxed lattice constant of bulk MoTe$_2$ using different vdw functionals. The IVDW number corresponds to different vdw functionals provided in VASP. Here IVDW=10 is the DFT-D2 method of Grimme (marked by gray). IVDW=11 is the DFT-D3 method of Grimme with zero-damping function. IVDW=13 is the DFT-D4 method. IVDW=20 is the Tkatchenko-Scheffler method. IVDW=21 is the Tkatchenko-Scheffler method with iterative Hirshfeld partitioning. IVDW=263 is the Many-body dispersion energy with fractionally ionic model for polarizability method. IVDW=4 is the dDsC dispersion correction method. IVDW=3 is the DFT-ulg method. Experimental data is from \cite{1961_D.Puotinen}. }
\label{vdw_functionals}
\end{table*}

\begin{table}
\begin{tabular}{|c|c|c|c|c|}
\hline
                & \textbf{experiment} & \textbf{Mo-Te} & \textbf{Mo\_pv-Te} & \textbf{Mo\_sv-Te} \\ \hline
\textbf{a,b(Å)} & 3.519                & 3.519           & 3.521               & 3.523               \\ \hline
\textbf{c(Å)}   & 13.964               & 13.976          & 13.985              & 13.997              \\ \hline
\end{tabular}
\caption{Relaxed lattice constant of bulk MoTe$_2$ using different pseudopotentials. Three pseudopotentials combinations show little difference. Experimental data is from \cite{1961_D.Puotinen}. }
\label{pseudopotentials-vdw}
\end{table}

\begin{figure*}
    \centering
    \includegraphics[width=1.0\linewidth]{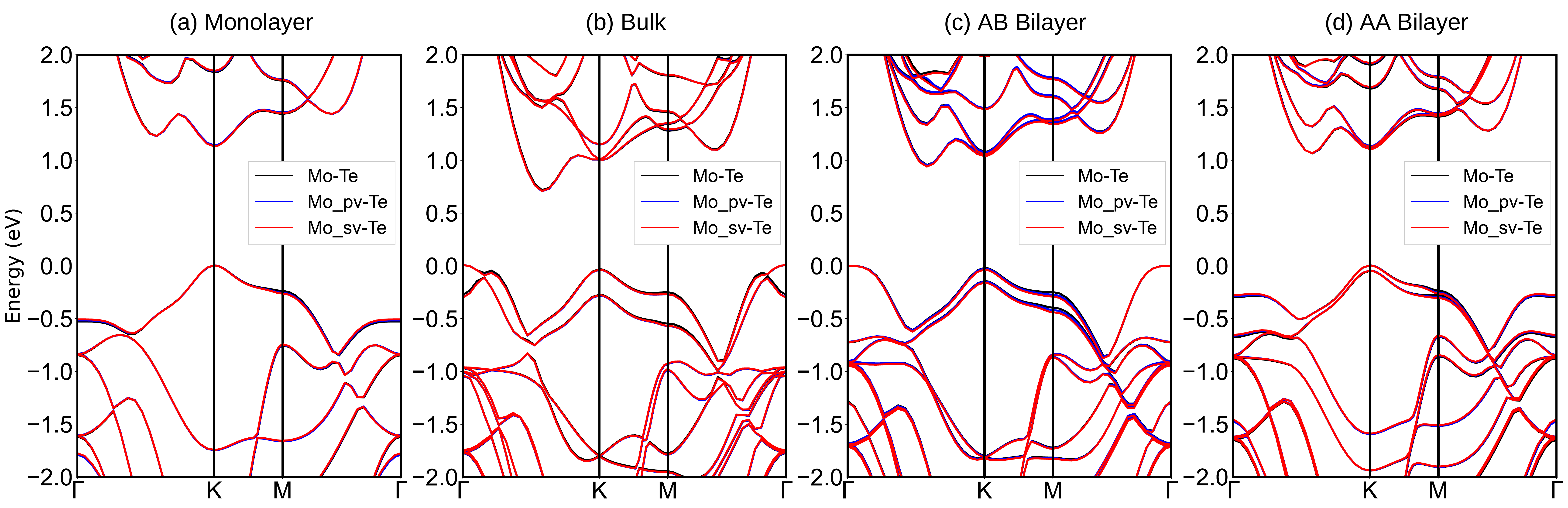}
    \caption{Band structure of bulk, monolayer, AA bilayer and AB bilayer MoTe$_2$ calculated with different pseudopotentials. The crystal structures are relaxed with corresponding pseudopotentials and IVDW=10(DFT-D2) functional.  }
    \label{pseudopotential-bs}
\end{figure*}

After selecting the exchange-correlation functionals, we test the different pseudopotentials---Mo-Te, Mo\_pv-Te, Mo\_sv-Te---where the pseudopotentials without suffices are general pseudopotential, and the suffix `\_sv' (`\_pv') means that the inner s (p) electrons are considered as valence electrons.
These combinations show little difference in the relaxed lattice parameters, and is consistent with experimental data. 
In addition, the band structure calculated by different pseudopotentials coincides with each other in different MoTe$_2$ structures.
Therefore, it is legitimate to choose any of them.
Considering the balance of accuracy and computational cost, here we choose the "Mo-Te" combination.

Before studing the electronic structures of the {\tmt} system, we first generated their crystal structures.
We start from both the AA and AB stacking\cite{olin2023ab}. As illustrated in \figref{fig:AAandAB}, AA stacking means that, when there is no twist, the Mo/Te atoms of the top layer
aligns with the Mo/Te atoms of the bottom layer.  AB stacking means that, when there is no twist, the Mo/Te atoms of the top layer
is directly above the Te/Mo atoms of the bottom layer (see \figref{fig:AAandAB} (a)). 
The bilayer untwisted crystal structure relaxed by DFT-D2 functional
gives lattice parameter $a=3.5228\angstrom$, and thus two primitive lattice vectors are $a_1=a(1,0,0)$ and 
$a_2=\frac{a}{2}(-\frac{1}{2},\frac{\sqrt{3}}{2}$,0). 
For twisted homobilayer with both top and bottom layer being MoTe$_2$, a commensurate structure occurs
when the moir\'e lattice vector of top layer and bottom layer satisfy the commensurate lattice condition $\bsl{a}_{Mb}=n_1 \bsl{a}_1+n_2\bsl{a}_2=m_1\bsl{a}_1^r+m_2\bsl{a}_2^r=\bsl{a}_{Mt}$ 
for certain integers $n_1,n_2,m_1,m_2$, where $\bsl{a}_1^r$ is the primitive lattice vectors rotated by an angle $\theta$.
In this way, we can obtain the rigid {\tmt} structures with different angles. All the rigid structures were generated using a homemade software 2DTwist.

The moir\'e structures twisted from AA stacking has a hexagonal structure in space group 150 , with $C_{3}$, $C_{2y}$ (see \figref{fig:AAandAB} (a)) as well as the time-reversal (TR) symmetry. Thus, the little group in the $\K$ valley is generated by $C_{3}, C_{2y}\mathcal{T}$, and $\pm\K$  valleys are exchanged by $C_{2y}$ and $\mathcal{T}$. 
In contrast, \figref{fig:AAandAB} (b) shows the moir\'e structures twisted from AB stacking which has a hexagonal structure of space group 149, with $C_{3}$, $C_{2x}$  and $\TR$. For the AB configuration, the little group in the $\K$ valley is generated by $C_{3}$ and $C_{2x}$, and the two valleys are exchanged by $\mathcal{T}$. 

As the twist angle becomes smaller but nonzero, a $\U(1)$ valley symmetry emerges due to the exponential suppression of inter-valley scattering off the moir\'e potential. This will enable us to build continuum models around the monolayer $\pm\K$ points using the little group symmetries to constrain the low-order terms.

\subsection{Relaxation of {\tmt}}

Since the relaxation will greatly affect the band structure, it is necessary to perform relaxation on {\tmt}. However, there are 1302 atoms for 3.89$^\circ$ and 1626 atoms for 3.48$^\circ$ in the moire cell of twisted structures, making the relaxation process difficult to converge. We decided to construct a Machine Learning Force Field (MLFF) to get relaxed structures in an efficient way. MLFF is a machine learning algorithm that will ``learn'' energies and forces of atoms from ab-initio calculation and can be applied to predict forces and energies for similar systems. 
We note that MLFF is not a relaxed structure---it is a function that maps structures to forces/energies; it can be used to efficiently generate the relaxed structures.
We firstly constructed a MLFF and applied it to get an MLFF-relaxed structure. Then, we performed DFT relaxation on the MLFF-relaxed structures. It only took around 20 DFT steps in relaxing the largest moire structure to converge in this strategy. In comparison, the relaxation from rigid 3.89$\degree$ {\AAtmt} takes 178 steps to converge.

\subsubsection{Construction of Machine Learning Force field}

During the construction of MLFF, two software packages are used. One is the VASP together with its integrated MLFF module~\cite{jinnouchi_--fly_2019}, and the second is NequIP\cite{batzner_e3-equivariant_2022}, which is an MLFF built on an E(3) equivariant neural network. 

We use the VASP MLFF module to generate the \emph{ab-initio} data needed for training a high-precision MLFF.
The VASP MLFF module itself is a way to generate MLFF and accelerate the Molecular Dynamics (MD) simulation.
It firstly runs several \emph{ab initio} MD steps, collects all the energies, forces and structure data into a dataset, then trains a MLFF, and finally estimates the error by doing MLFF-based MD calculations.
Here the MLFF-based MD calculation is based on Bayesian linear regression, and the error is directly estimated by the spread of the Gaussian distribution. 
If the estimated error of MLFF is large, an additional \emph{ab initio} MD step will be performed to enlarge dataset, the MLFF is retrained with the new dataset, and estimate the error again.
The procedure will be repeated until the total number of MD steps exceeds the preset.

However, the MLFF algorithm in VASP is lightweight. It helps accelerate the MD simulation, but the MLFF generated by the algorithm is not accurate enough.
Therefore, we will not directly use the generated MLFF; instead, we just run the VASP MLFF module for tens of thousands of steps to generate a set of \emph{ab-initio} MD data.

With the set of \emph{ab-initio} MD data generated from the VASP MLFF module, we use the NequIP software to train an accurate MLFF. NequIP is based on an E(3) equivariant neural network, meaning that the input and output of each neural network layer are equivariant (in other words, covariant) under the rotation, reflection and translation in 3D space. The NequIP software is reported to outperform several other MLFF algorithms in both data efficiency and accuracy \cite{batzner_e3-equivariant_2022}.

The training process is summarized as the follows. 
We started from small supercells of untwisted AA and AB bilayer MoTe$_2$ with different in-plane shift between top and bottom layer. We ran MD simulation using VASP MLFF module on those structures and collected the data from all ab-initio MD steps. We merged all the collected data and trained a NequIP MLFF, which is then used in Atomic Simulation Environment \cite{HjorthLarsen2017} to relax the $t$MoTe$_2$ at various angles.

\subsubsection{Relaxation results}

\begin{figure*}
	\includegraphics[width=1\linewidth]{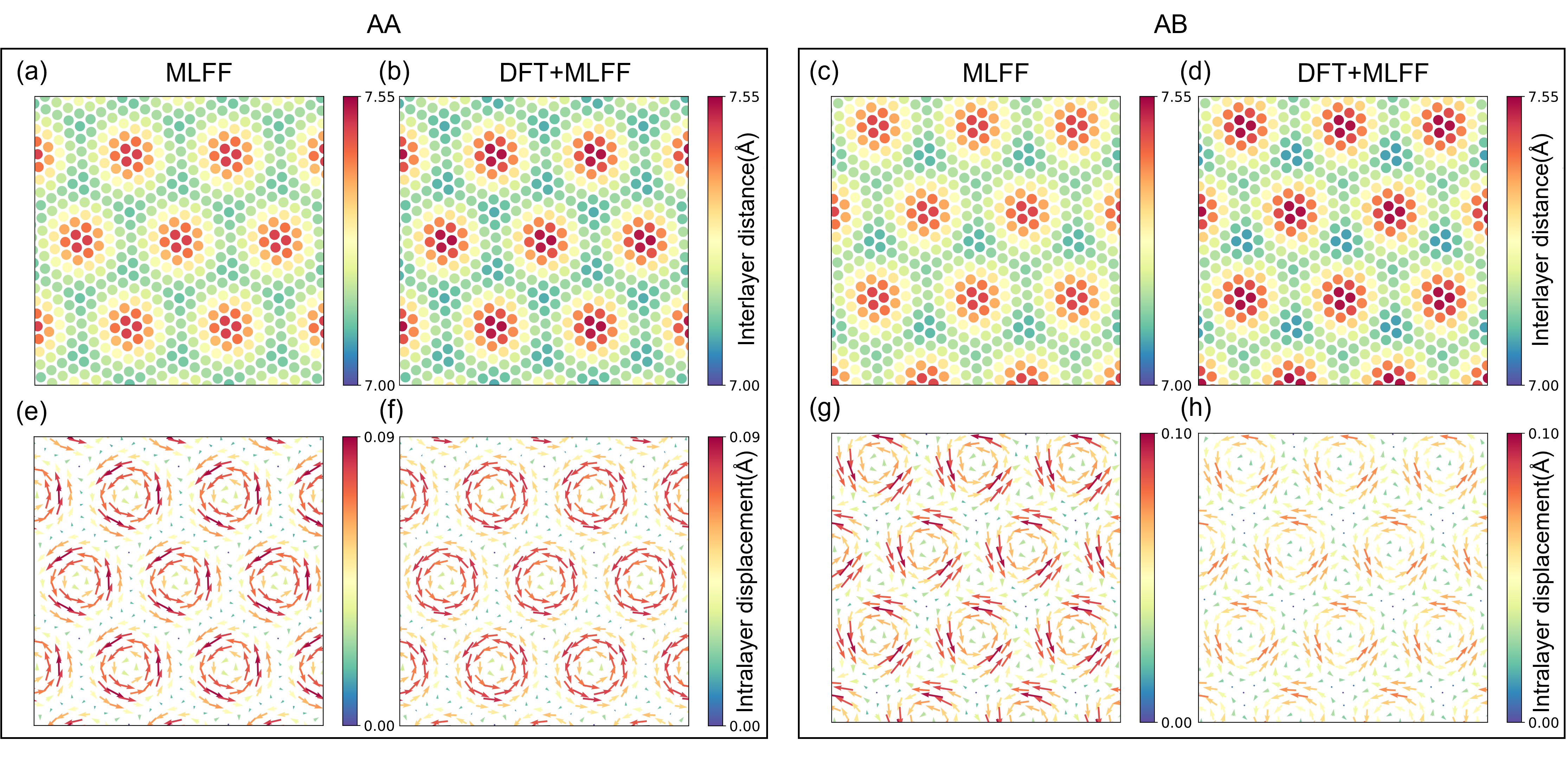}
\caption{Relaxation results of 7.34$\degree$ AA and AB {\tmt}. (a), (b), (c) and (d) are interlayer distances of MLFF-relaxed AA structure, DFT+MLFF-relaxed AA structure, MLFF-relaxed AB structure and DFT+MLFF-relaxed AB structure respectively. (e), (f), (g) and (h) are intralayer displacements of MLFF-relaxed AA structure, DFT+MLFF-relaxed AA structure, MLFF-relaxed AB structure and DFT+MLFF-relaxed AB structure. Interlayer distance is the distance between the top and bottom layer, while the intralayer displacement indicates the in-plane displacement from rigid positions to relaxed positions of a Mo atom in top layer. In the interlayer distance plots, rigid structures are selected to have the same lattice constant as relaxed structures. }
\label{relaxation_MLFF-DFT_7}
\end{figure*}

\begin{figure*}
	\includegraphics[width=1\linewidth]{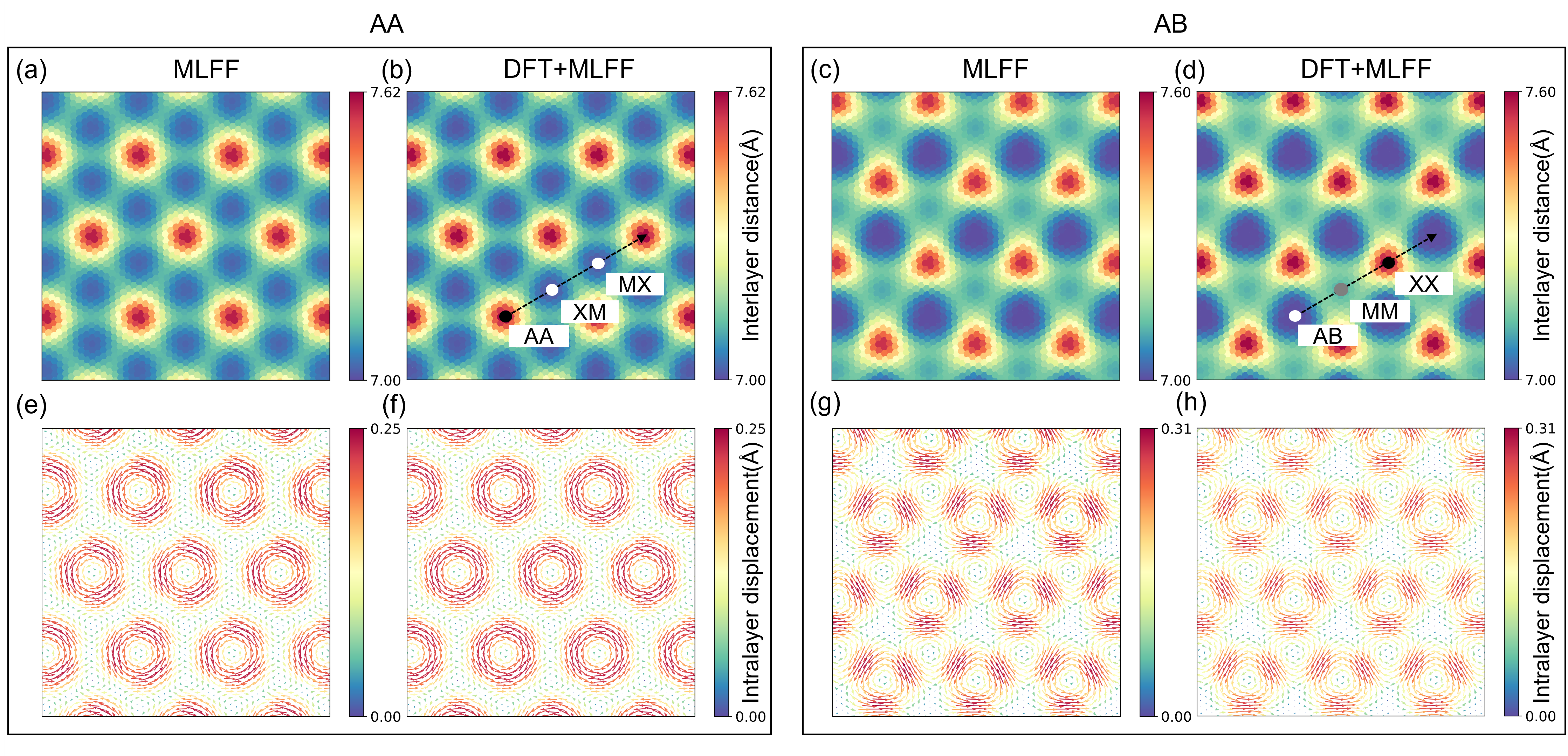}
\caption{Relaxation results of 3.98$\degree$ AA and AB {\tmt}. (a), (b), (c) and (d) are interlayer distances of MLFF-relaxed AA structure, DFT+MLFF-relaxed AA structure, MLFF-relaxed AB structure and DFT+MLFF-relaxed AB structure respectively. (e), (f), (g) and (h) are intralayer displacements of MLFF-relaxed AA structure, DFT+MLFF-relaxed AA structure, MLFF-relaxed AB structure and DFT+MLFF-relaxed AB structure respectively. }
\label{relaxation_MLFF-DFT_3}
\end{figure*}

\begin{figure}
	\includegraphics[width=1\linewidth]{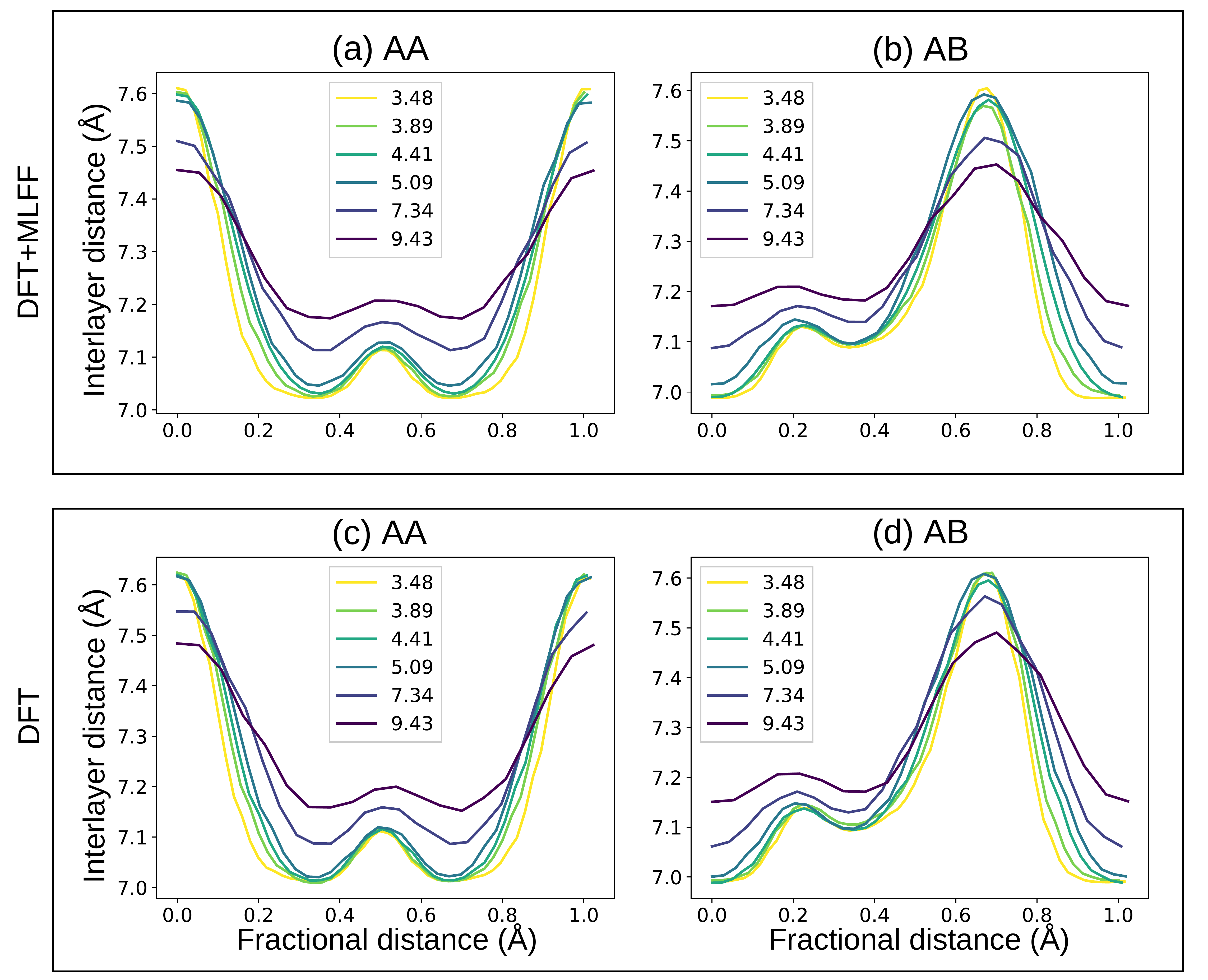}
\caption{Relaxation results of different angles along black arrows in \figref{relaxation_MLFF-DFT_3} (b, d). (a) and (b), interlayer distance of AA structures given by MLFF and DFT+MLFF, respectively. (c) and (d), interlayer distance of AB structures given by MLFF and DFT+MLFF.}
\label{ILD_result}
\end{figure}

The relaxation results of AA and AB MoTe$_2$ in different twist angles are listed in \figref{relaxation_MLFF-DFT_7}, \figref{relaxation_MLFF-DFT_3} and \figref{ILD_result}.  
As we mentioned before, the DFT+MLFF-relaxed results are obtained by further DFT relaxation based on the MLFF-relaxed structures. In \figref{relaxation_MLFF-DFT_7} and \figref{relaxation_MLFF-DFT_3}, the MLFF-relaxed structures are close to the DFT+MLFF-relaxed structure. The extra DFT relaxation in the DFT+MLFF method only modified the quantitative details, keeping the qualitative shape of the MLFF-relaxed structures unchanged. It indicates that the MLFF can reproduce the main part of DFT relaxation, making the DFT relaxation easier to converge.

In structures of smaller twist angles, the local conformation is more similar to untwisted structures. Because our MLFF are constructed from untwisted structures, more accurate results in smaller angles are expected. Compare \figref{relaxation_MLFF-DFT_3} with \figref{relaxation_MLFF-DFT_7}, the MLFF-relaxed results are indeed better at 3.89$\degree$ than that at $7.34\degree$. 

\figref{ILD_result} shows the relaxation results of AA and AB {\tmt} in different twist angles. 
In both AA and AB stacking, interlayer distance comes to the lowest point in MX/XM/AB region and becomes higher in MM/XX/AA region. In AA stacking, the interlayer distance of MX region is the same as XM region since MX and XM configurations are related by $C_{2}$ symmetry. In the case of AB stacking, however, the MM region is different from XX region, as they are not symmetry-related. The interlayer distance of XX region is always higher than MM region, which results in the asymmetry in \figref{ILD_result} (b) and \figref{ILD_result} (d).

\subsection{Electronic structures of {\tmt}}
\label{app:BandtMT}

In this part, we discuss the electron band structures of {\tmt} obtained from the DFT+MLFF relaxed structure. 

\subsubsection{Energy bands of monolayer, AA bilayer and AB bilayer MoTe$_2$}

\begin{figure*}
    \centering
    \includegraphics[width=0.98\linewidth]{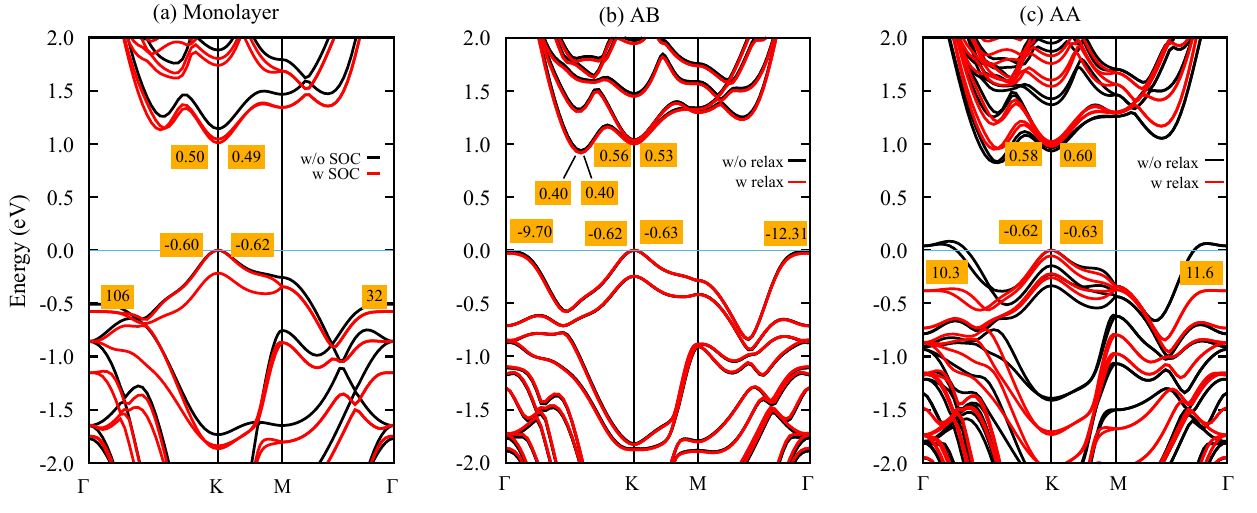}
    \caption{Band structure of monolayer, AB bilayer and AA bilayer MoTe$_2$ with and without relaxation. 
    (a), band structure of monolayer, the effective mass of the monolayer MoTe$_2$ of the VBM at the $\Gamma$ and $\K$ points, and the CBM at the $\K$ point, along two different directions, are indicated in the orange boxes with units $m_e$. 
   (b), band structure of AB bilayer, where the rigid and relaxed) bands of AA and AB bilayer MoTe$_2$ are indicated by black and red lines, respectively. (c, band structure of AA bilayer. All the effective masses are calculated for relaxed structures with SOC.
    }
    \label{fig:bs-mono-AA-2H}
\end{figure*}

Before diving into electronic structure of the {\tmt}, we first study the monolayer and bilayer structures. For the monolayer structure, the inclusion of SOC brings a clear splitting for top valence band at $\K$ point. Additionally, At the $\Gamma$ point, SOC causes a downward shift of the top valence band by 70 meV in comparison to the scenario without SOC, as depicted in \figref{fig:bs-mono-AA-2H} (a).

\figref{fig:bs-mono-AA-2H} (b) shows that lattice relaxation has negligible influence on the band structure of AB stacking. The reason is that the functional we used is optimized for the bulk MoTe$_2$ which is also AB stacked. 
Conversely, \figref{fig:bs-mono-AA-2H} (c) demonstrates that lattice relaxation significantly alters the band structure of AA stacking. This is due to the fact that the relaxed interlayer distance measures approximately 7.7 \AA, which is around 10\% greater than the interlayer distance of the rigid structure at 7 \AA. Relaxation causes a downward shift of the first valence band at the $\Gamma$ point, resulting in an energy about 380meV lower at $\Gamma$ compared to the $\K$ point. For conduction bands, upon relaxation, the bottom bands along $\Gamma$-$\K$ path are elevated to a similar energy level as those at the $\K$ point.

\subsubsection{13.2$\degree$ {\tmt}}
\label{app:13.2degree_bands}

\begin{figure*}
    \centering
    \includegraphics[width=1\linewidth]{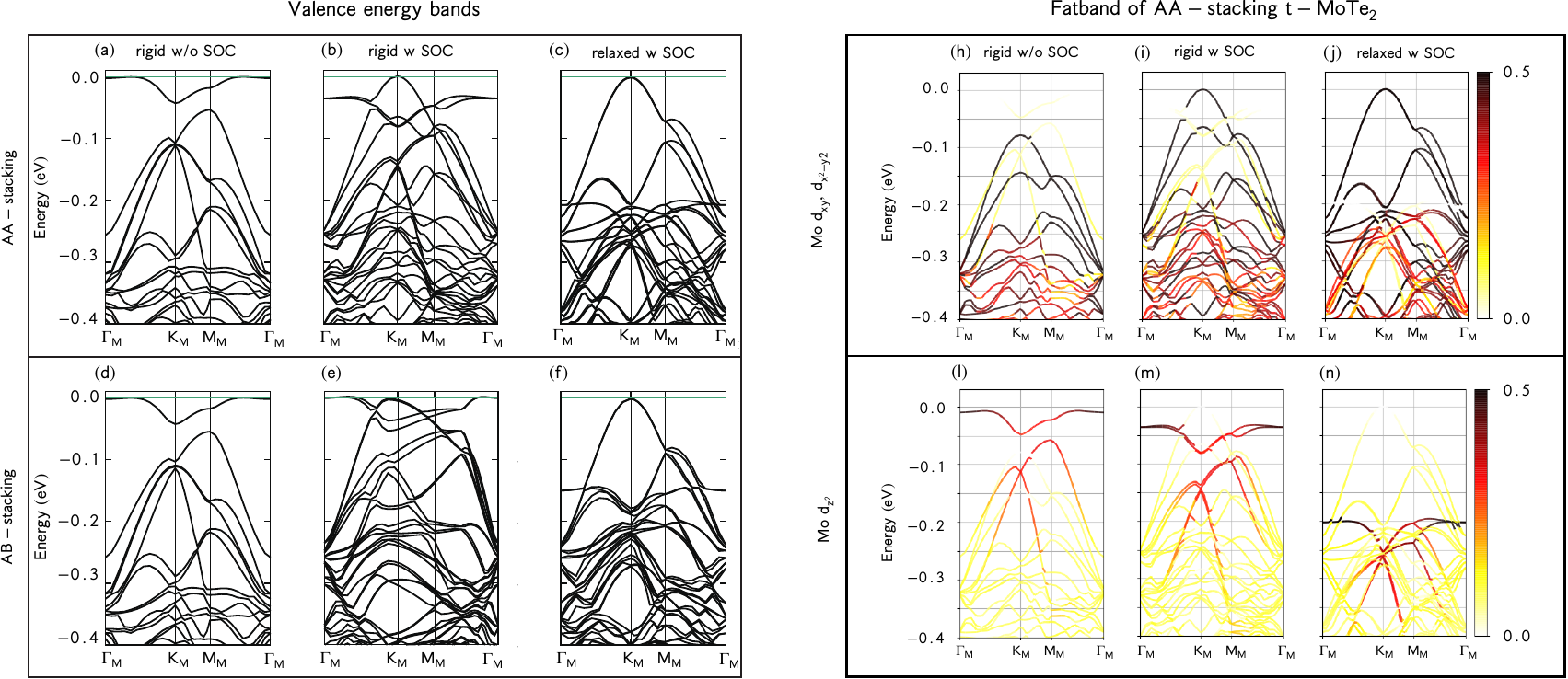}
    \caption{Band structure and band character analysis (fatband) of 13.2$\degree$ {\tmt}. (a) is the band structure for rigid AA without SOC, showing negligible difference with rigid AB
    without SOC in (d). Considering SOC, the top isolated valence band tangles with other valence bands both for rigid AA and AB. SOC brings larger splitting
    in $\Gamma-M$ k-path for rigid AA in (b) while larger splitting around $K$ point for rigid AB in (e). After the relaxation, the bands from $\pm\K$ valley are lifted
    up to the top of valence bands, both for relaxed AA in (c) and relaxed AB in (f). 
    } 
    \label{fig:13.2band}
\end{figure*}

We start from the $13.2\degree$ {\tmt} and gradually decrease the twist angle to follow the evolution of the band structure. In \figref{fig:13.2band} (a), when spin orbit coupling (SOC) is not considered, there is an isolated narrow valence band (NVB) located at the top of the valence bands for AA stacking $13.2\degree$ {\tmt} without relaxation. 
A very similar band is also present for the AB stacking with negligible difference (\figref{fig:13.2band} (d)).
After considering the SOC, the NVB start to go down and entangle with other bands. (See \figref{fig:13.2band} (b,e).)
Further including the relaxations, we obtain stacking-dependent corrugated moir\'e structures. Relaxation pushes the NVB further below as shown in \figref{fig:13.2band} (c,f). 

By projecting the moir\'e bands into atoms' orbitals as shown in \figref{fig:13.2band} (l)-(n), it is clear that the NVB is consist of Mo atoms $d_{z^2}$ orbitals which is the same band character of the VBM at $\Gamma$ point of monolayer MoTe$_2$ (see \figref{fig:fatband} (a)). So we can say that the NVB is the $\Gamma$ valley band. 
Using the orbital nature, we can see that upon considering both SOC and lattice relaxation effects, the $\Gamma$-valley bands are pushed down by 250 meV below the valence band maximum (VBM) for AA configuration, while about 180 meV for AB configuration. 
Then, the state around the VBM are composed of molybdenum (Mo) $d_{x^2-y^2}$ and $d_{xy}$ orbitals which have the same orbital characteristics as the VBM at the $\pm\K$ points in a monolayer MoTe$_2$ (see \figref{fig:fatband} (a)). 
Therefore, the VBM of the moiré energy bands primarily originates from the $\pm\K$ valley. 
We note that the downward shift of the $\Gamma$ bands, caused by the SOC and lattice relaxation effects, is consistent with the lower $\Gamma$-valley bands after including the relaxation in the AA-stacking untwisted case in \figref{fig:bs-mono-AA-2H}(c).

\subsubsection{AA stacking: Evolution of band structure from 9.43$\degree$ to 3.89$\degree$}

\begin{figure*}
    \centering
    \includegraphics[width=1\linewidth]{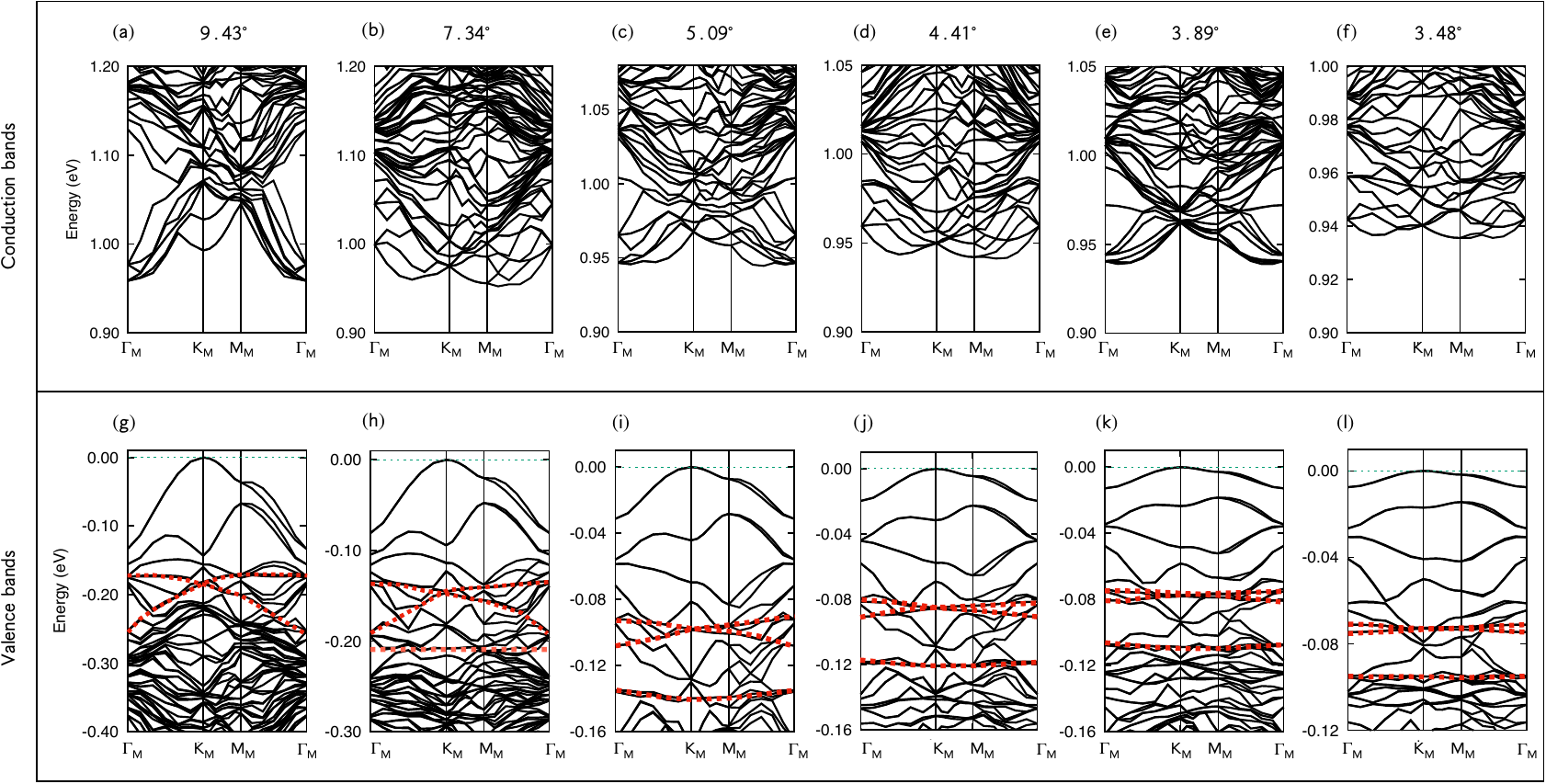}
    \caption{Evolution of band structures of AA stacking configuration {\tmt} with twist angles ranging from 9.43$\degree$ to 3.48$\degree$ . The six band structures are calculated from DFT+MLFF relaxed moir\'e structure with the consideration of SOC effects. The dashed red lines show the flat bands coming from $\Gamma$ valley.}
    \label{fig:AArelaxevolution}
\end{figure*}

We also calculate the band structures of several relaxed AA stacking {\tmt} structure at 9.43$\degree$, 7.34$\degree$, 5.09$\degree$, 4.41$\degree$, 3.89$\degree$ and 3.48$\degree$ as shown  in \figref{fig:AArelaxevolution}. 

\begin{figure*}
    \centering
    \includegraphics[width=0.95\linewidth]{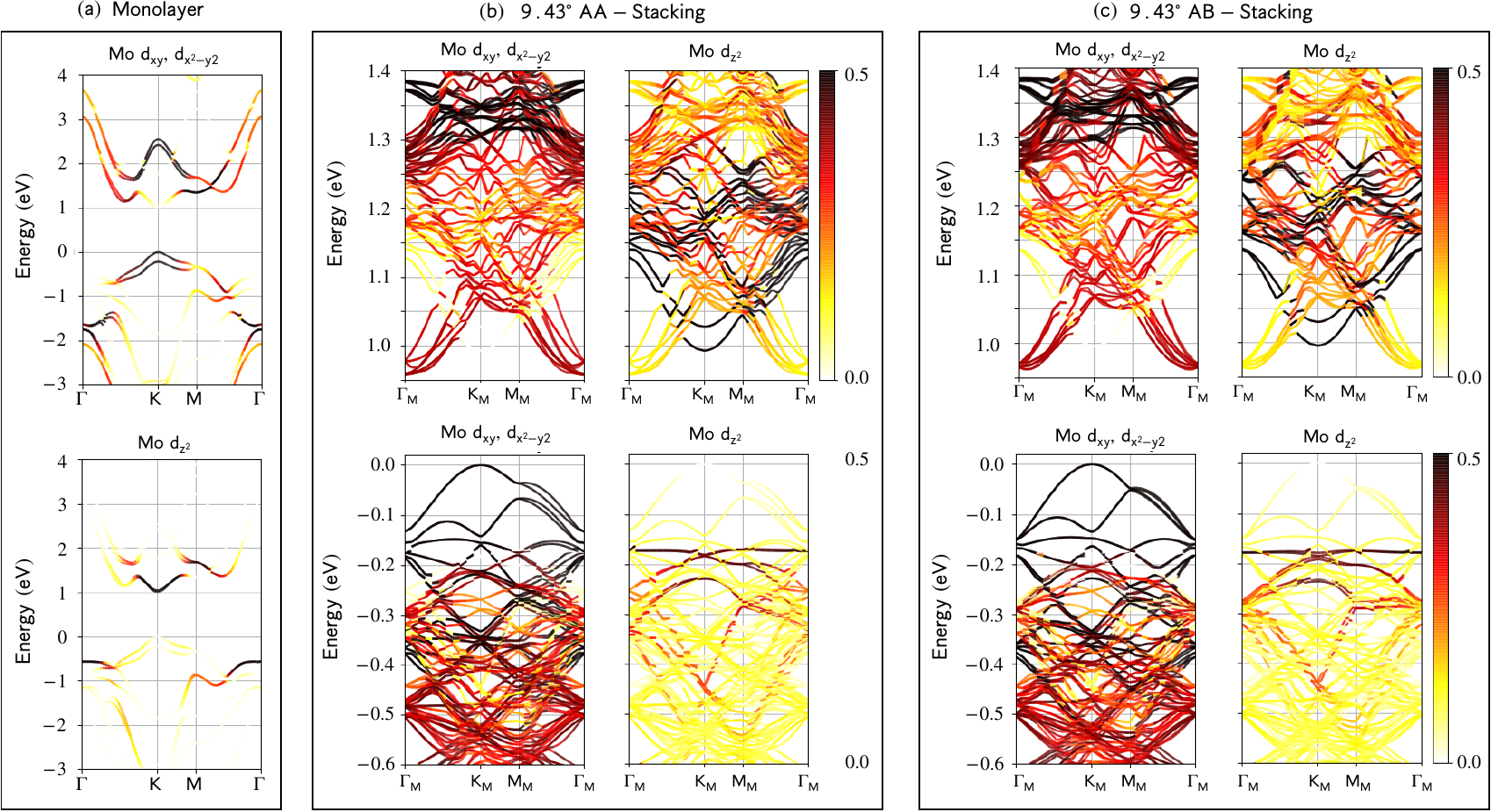}    
    \caption{Orbital-projected band structure of monolayer layer MoTe$_2$ and AA (AB) {\tmt} with twisted angle 9.43$^\circ$. Bands are projected on $d_{xy}$, $d_{x^2-y^2}$ and $d_{z^2}$ of Mo atoms. Figure obtained with the open-source code PYPROCAR\cite{pyprocar}.}
    \label{fig:fatband}
\end{figure*}

We discuss the valence bands first.
The valence band maximum is at the $\K_M$ points of the moire BZ. The top pair of valence bands remains quasi degenerate in $\Gamma_M-K_M-M_M$ high symmetry line, suggesting the presence of an extra symmetry than TR symmetry, which would map $-\bsl{k}$ to $\bsl{k}$. 
As shown in \figref{fig:fatband} on the orbital natures of the bands at 9.43$\degree$, the valence bands near the VBM mainly come from $\pm\K$ valleys, owing to their $d_{x^2-y^2}$ and $d_{xy}$ nature, similar to the discussion of \appref{app:13.2degree_bands}; the expectation is the bands labeled by red dashed lines in \figref{fig:AArelaxevolution}, which comes form the $\Gamma$ valley owing to its $d_{z^2}$ orbital nature.

Analyzing the evolution of the valence bands in \figref{fig:AArelaxevolution} from panel (g) to (l), it becomes evident that the bandwidth of the top two pairs of valence bands decreases.  narrows as the twist angle decreases—--a characteristic commonly observed in twisted systems. Moreover, there is a discernible indication of band inversion between $k_{v2}$ and $k_{v3}$ at $\Gamma_M$, occurring as the twist angle decreases from 5.09$^\circ$ and 3.48$^\circ$. Specifically, at the $\Gamma_M$ point for 5.09$^\circ$, there is a gap of approximately 2.8 meV separating the second and third pairs of top valence bands. This gap closes at a twist angle of 4.41$^\circ$ and reopens to about 1.2 meV at 3.89$^\circ$. This cycle of gap closing and reopening suggests the possibility of a band inversion. To investigate this, we calculated the irreducible representations of the six highest valence bands at the $\K_M$ and $\Gamma_M$ points for twist angles ranging from 7.34$^\circ$ to 3.89$^\circ$ and have presented the findings in \figref{fig:irreps-AA-vbs}.

As shown in \figref{fig:AArelaxevolution} from panel (a) to (f), the conduction bands are quite messy.
Based on the decomposition of orbital content depicted in \figref{fig:fatband}, we determine that the conduction band minimum (CBM) at the $\Gamma_M$ point for a 9.43° twist exhibits the same orbital characteristics—specifically Mo $d_xy$  and $d_{x^2-y^2}$—as those found in the electron pocket along the $\Gamma$-$\K$  path in the monolayer band structure, also illustrated in \figref{fig:fatband}.
The conduction band minimum is around the $\Gamma_M$ point rather than $\K_M$ point can be understood from the AB and AA band structures in \figref{fig:bs-mono-AA-2H}.

The presence of the conduction band minimum around the $\Gamma_M$ point, as opposed to the $\K_M$ point, can be elucidated by examining the band structures for AB and AA configurations in \figref{fig:bs-mono-AA-2H}. In the AB bilayer band structure, the dip along the  $\Gamma-K$ path is lower than at the $\K$ point itself. Conversely, in the AA bilayer, the $\Gamma-K$ path's minimum aligns with the level at the $\K$  point. Given that the moir\'e unit cell's stacking transitions between AA and AB types, the moir\'e conduction band minimum consequently shifts away from the $\K_M$ point.

We discuss the erratic evolution of the electron bands in \appref{app:electronpocket}, which we can attribute to the lowest energy states coming from an electron-like pocket. This pocket occurs at generic momentum $\mbf{p}_i$ in the untwisted BZ, so that the bands are folded around a generic point  $\mbf{p}_i \mod \mbf{G}_M$ in the moir\'e BZ 
that depends extremely sensitively on the angle through $\mbf{G}_M$.

\subsubsection{3.89$\degree$ AA stacking {\tmt} electronic structures}
\label{app:3.89band}
\begin{figure}
    \centering
    \includegraphics[width=1\linewidth]{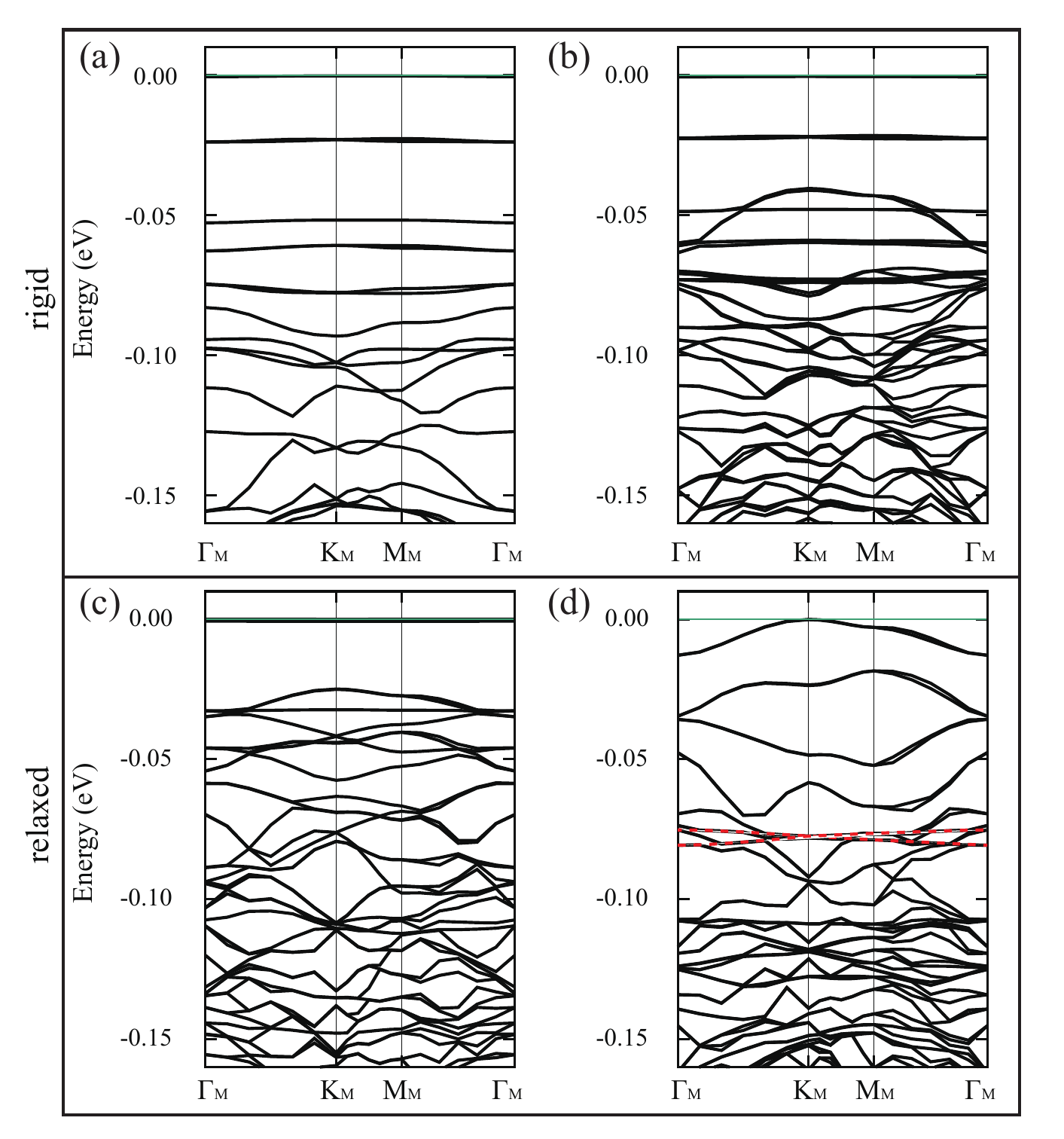}
    \caption{AA stacking configuration band structure of 3.89$\degree$ {\tmt}. (a) and (b) show the band structures without SOC and with SOC bands for rigid
    structure, respectively. (c) and (d) show the band structures without SOC and with SOC bands for relaxed
    structure, respectively}
    \label{fig:3.89AA4pics}
\end{figure}

Let us focus on the 3.89° twisted AA stacking. In \figref{fig:3.89AA4pics}, we present the band structure calculations for both the rigid and relaxed configurations, with and without SOC. Before relaxation,
the $\Gamma$ valley ultra flat bands are located at the top of the valence bands when SOC is omitted. The introduction of SOC raises some dispersive bands; however, even with SOC, the rigid structure maintains flat $\Gamma$ valley bands at the top. Relaxation dependent on stacking has a significant impact on the band structure. When SOC is absent, the relaxation raises the $\pm\K$ bands. Further incorporating SOC,  $\pm\K$ bands new becomes the VBM, and the red dashed bands originating from the $\Gamma$ valley, as denoted in \figref{fig:3.89AA4pics} (d), are markedly diminished, sinking approximately 80 meV below the VBM.
Our results are different from \refcite{wang2023fractional}, where the $\Gamma$ bands are close to the top two valence bands from $\pm\K$ valley, with only 30 meV away from the VBM. The top valence band in \refcite{wang2023fractional} has the bandwidth of about 9 meV, while the width of the same band in our results is about 12.8 meV.

\begin{figure*}
    \centering
    \includegraphics[width=1\linewidth]{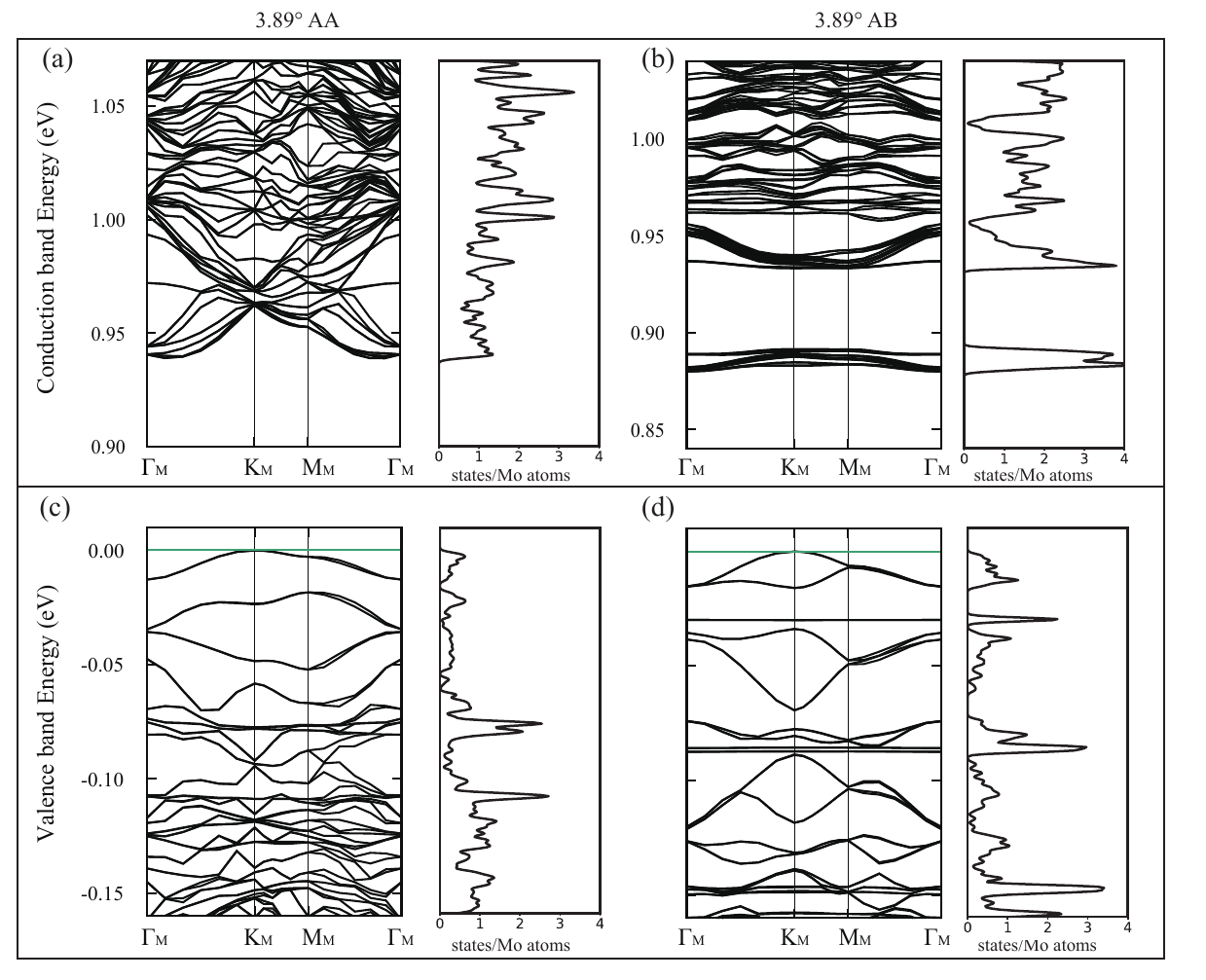}
    \caption{Conduction bands and valence bands with their density of states (DOS) in unit of states per Mo atom of 3.89$\degree$ for relaxed AA stacking and AB stacking configuration with SOC. a and b are the conduction bands of AA and AB stacking configuration with DOS. c and d are the valence bands of AA and AB stacking configuration. The green line represents the Fermi level.}
    \label{fig:3.89cvbands}
\end{figure*}

The density of states (DOS) calculations depicted in \figref{fig:3.89cvbands} (c) reveal two distinct peaks within the valence bands around -75meV and -105meV, which correspond to the $\Gamma$ valley band embedded in the backdrop of dispersive bands. Additionally, several minor peaks around 0meV, 20meV, are observed, which originate from the valence bands of the $\pm\K$ valleys.

\begin{figure*}
    \centering
    \includegraphics[width=1\linewidth]{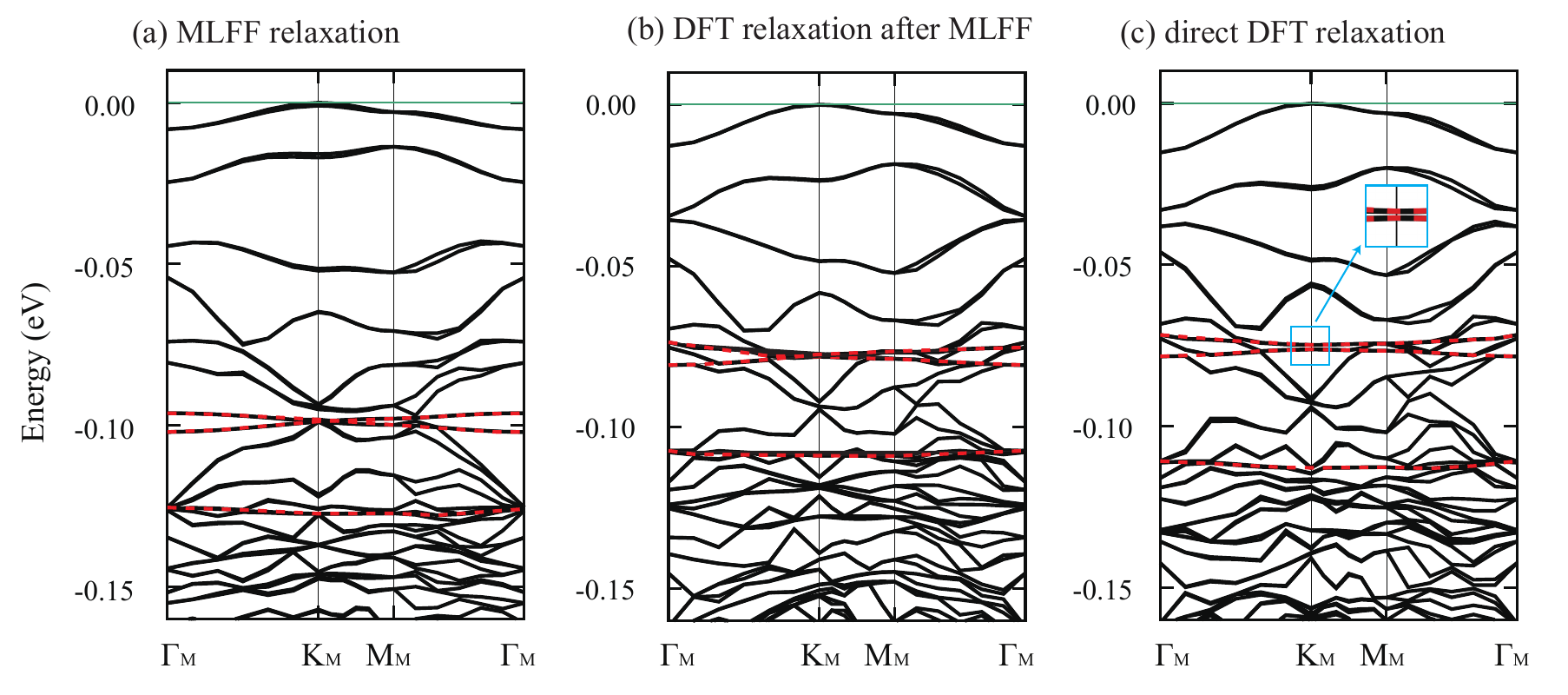}
    \caption{Comparision of different relaxation band structures. a, band structure of MLff relaxation. b, band structure of MLFF+DFT relaxation. c, band structure of direct DFT relaxation from rigid structure.}
    \label{fig:3relax}
\end{figure*}

In \figref{fig:3relax}, we show the band structure given by three different relaxation methods: a, only using MLFF, b, two-step MLFF+DFT relaxation  and c, relaxation directly from rigid structure using DFT. MLFF can an accuracy considerably close to the DFT calculations, especially the top three pairs of $\pm\K$-valley valence bands and the highest $\Gamma$-valley bands as shown in \figref{fig:3relax}(a). However, we find the forces of MLFF relaxed structure is not small enough (with a mean absolute force of $7.8 \times 10^{-2}eV/\angstrom$), meaning that the MLFF result can be relaxed further using the more accurate DFT relaxation method. After the further DFT relaxation, the mean absolute force reached $2.1\times 10^{-3}eV/\angstrom$, and the band structure changes quantitatively (see in \figref{fig:3relax} (a) and (b)). If the relaxation is performed from the rigid structure using DFT only, we will get almost the same band structure as the two-step relaxation band.

\begin{figure*}
    \centering
    \includegraphics[width=1\linewidth]{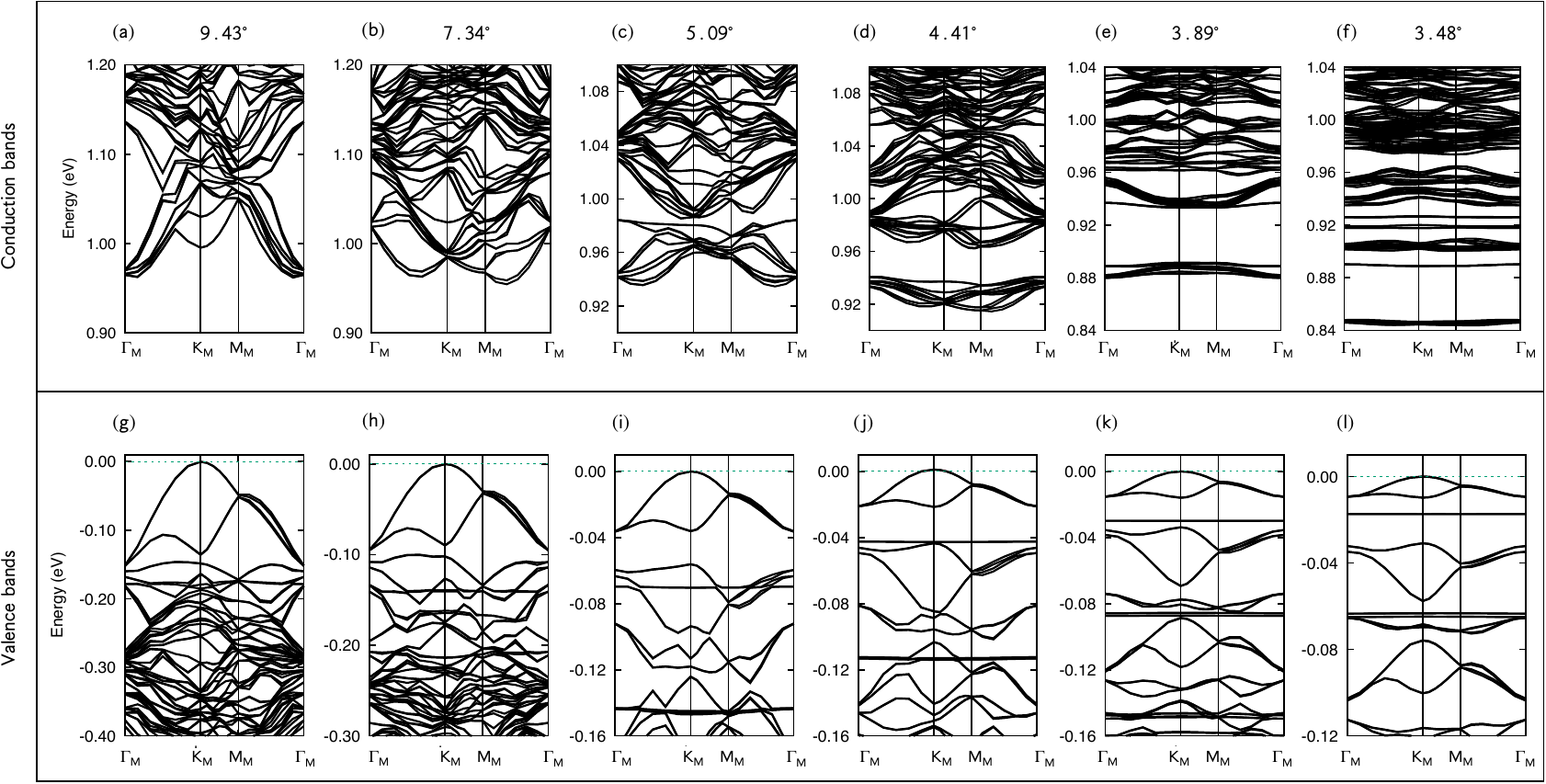}
    \caption{Evolution of band structures for various angles range from 9.43$\degree$ to 3.48$\degree$ AB stacking configuration {\tmt}. Top(bottom) row is for conduction(valence) bands. The lattice was fully relaxed with DFT-D2 vdW functional. SOC are considered in the calculation.}
    \label{fig:ABrelaxevolution}
\end{figure*}

\begin{figure}
    \centering
    \includegraphics[width=0.8\linewidth]{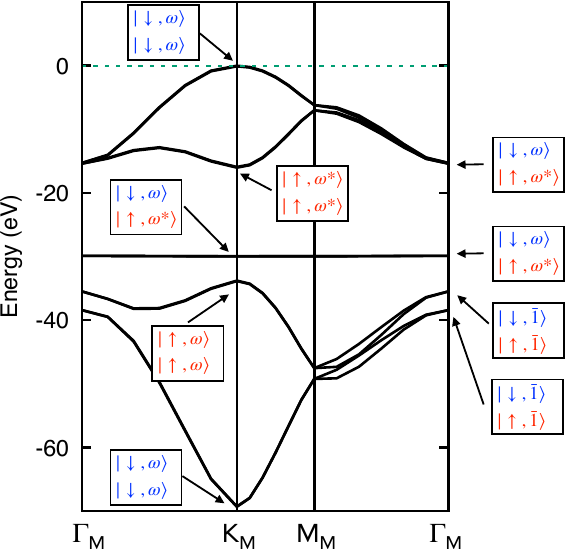}
    \caption{Irreducible representations at high-symmetry points $\Gamma$ and $K$ of valence bands of  3.89$\degree$ for relaxed AB stacking configuration with SOC. The eigenvalues of the $C_{3}$ are denoted as  $\omega=e^{i\pi/3}$, $\omega^*=e^{-i\pi/3}$ and $\bar{1}=e^{-i\pi}$}
    \label{fig:irreps-2H-vbs}
\end{figure}

The squared wavefunctions $|\psi_{\Gamma_M}|^2$ of the top four valence bands at the $\Gamma_M$ point in the $\K$ valley is illustrated in \figref{fig:fig-density}(a). The wavefunctions of the first and second topmost valence bands in the $\K$ valley are predominantly localized in the XM and MX regions, collectively manifesting a hexagonal lattice pattern in real space. In contrast, the wavefunction of the third topmost valence band primarily occupies the AA region, delineating a triangular lattice. Meanwhile, the fourth band's wavefunction is concentrated at the XM, MX, and AA regions, constituting another hexagonal lattice configuration.

\subsubsection{Comparision with \refcite{reddy2023fractional,wang2023fractional}}
\label{app:comparision}

\refcite{reddy2023fractional} and \refcite{wang2023fractional} have studied the relaxation and band structures of {\tmt}. \refcite{reddy2023fractional} employed the SCAN density functional with dDsC dispersion correction to perform crystal structure relaxation. In our assessment of 19 different functionals, as shown in \tabref{vdw_functionals}, we find that the SCAN functional and dDsC dispersion correction (IVDW=4 in \tabref{vdw_functionals}) yielded a larger c-axis lattice parameter and smaller a(b)-axis lattice parameters, whereas the DFT-D2 functional that we use (IVDW=10 in \tabref{vdw_functionals}) provide the lattice parameters closest to the experimental results. After crystal structure relaxation, \refcite{reddy2023fractional} showed a larger interlayer distance compared to our relaxation.

In contrast to both our approach and \refcite{reddy2023fractional}, \refcite{wang2023fractional} used SIESTA with DFT-D2 functional to perform the DFT calculations. After the relaxation, \refcite{wang2023fractional} gets a smaller inter-layer distance (ILD) of which the minimum is about 6.9 \angstrom. This Mo-Mo ILD is smaller than our relaxation result shown in \figref{comparison}, 
even smaller than ILD of bulk MoTe$_2$ crystal structure with 7.0 \angstrom.  

Our relaxation results shows that, in the AA region, where the stacking configuration is close to that of AA untwisted bilayer structure, and thus the maximum interlayer distance should be close to but slightly smaller than (due to the drag of surrounding environment which is not aligned in AA form) the interlayer distance of AA stacking untwisted bilayer (7.7$\angstrom$), which is not satisfied by \refcite{reddy2023fractional}.
Furthermore, the MX region has the stacking configuration akin to that of AB untwisted bilayer structure. As a result, the smallest interlayer distance in {\tmt} structures should be close to but slightly larger than (due to the surrounding environment's influence) 7$\angstrom$, which is violated by \refcite{wang2023fractional}.

 From \figref{fig:3.89AA4pics}, it is clear that relaxation will push the $\Gamma$-valley bands down as well as lift the bands from $\pm\K$ valleys up. The relaxation in \refcite{wang2023fractional} gets the smaller ILD, resulting that their $\Gamma$-valley bands are not pushed down significantly and close to the $\pm\K$-valley bands. The relaxed $\Gamma$-valley bands in \refcite{wang2023fractional} are only 30 meV below the valence band maximum, while our $\Gamma$-valley bands are 80 meV below the VBM.
Meanwhile, the relaxation in \refcite{reddy2023fractional} gives the maximal ILD of 7.8 \angstrom which is larger that us, due to the overestimated c-axis lattice parameter by SCAN functional.
\refcite{reddy2023fractional} calculated the band structure of 4.4$\degree$ moir\'e structure (1014 atoms per unit cell). Due to \refcite{reddy2023fractional} only calculated 4 k point along the high-symmetry line, the band width is estimated from their fitted continuum model. From the model, the top valence band from $\pm \K$ valleys has band width of about 30 meV at 4.41$\degree$, while our 4.41$\degree$ structure (1014 atoms per unit cell) $\pm\K$-valley band has a smaller bandwidth. For 3.89$\degree$ structure in \refcite{wang2023fractional}, the top pair of the valence bands have band width of about 9 meV, which is smaller than our result of 12.8 meV.

\subsubsection{AB stacking: Evolution of band structure from 9.43$\degree$ to 3.89$\degree$}

The symmetry group of the AB stacking can be given by replacing $C_{2y}$ in the group for $AA$ stacking by $C_{2x}$. At large twist angle, such as 13.2$\degree$, the electronic structures of the two stacking configurations appear similar, as illustrated in \figref{fig:13.2band}. However, they exhibit notable differences at smaller twist angles (most notable for angles smaller than $5\degree$, as seen in \figref{fig:ABrelaxevolution} and \figref{fig:AArelaxevolution}. 

As listed in \tabref{ABBandWidth} ,  the bandwidth of the top two pairs of bands narrows with decreasing twist angle, specifically to 16 meV for 3.89$^\circ$ and 9.8 meV for 3.48$^\circ$. As the twist angle decreases, a pair of ultra flat valence bands gradually moves up, and  become isolated for $\theta<4.41^\circ$ as shown in \figref{fig:ABrelaxevolution}. When the twist angle decreases to 3.89$\degree$, the two set of ultra flat valence bands are separated with about 56.0 meV, bringing two distinct peak in the DOS as shown in \figref{fig:3.89cvbands}. 

Similar with that of AA, the top two pairs of valence bands of AB configuration consist of $d_{x^2-y^2}$ and $d_{xy}$ orbitals of Mo atoms as shown in \figref{fig:fatband}, indicating these bands come from the $\pm\K$ valleys. Besides, the flat valence bands of AB come from $\Gamma$ valley because they consist of Mo $d_{z^2}$ orbitals.

\begin{figure*}
    \centering
    \includegraphics[width=0.95\linewidth]{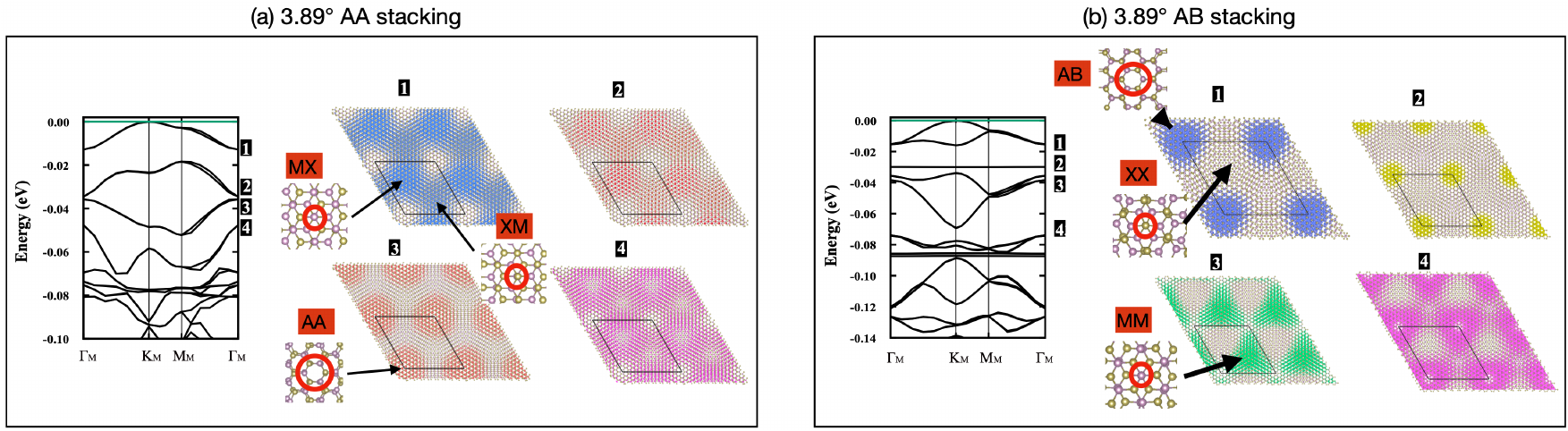}
\caption{Charge densities of the valence-band wavefunctions at $\gamma$ point in relaxed 3.89$^\circ$ AA-twist and AB-twist  MoTe$_2$, The isosurface value for these plots is $2\times10^{-5}e/\AA^3$}
    \label{fig:fig-density}
\end{figure*}

Distinct from from AA stacking, there are isolated moiré bands on the conduction band side for the AB stacking. As the twist angle decreases to 4.41$^\circ$ (see \figref{fig:ABrelaxevolution}(d)), there are 12 bands at the bottom of the conduction bands that are isolated from the higher-energy bands. The degeneracy of these bands suggests they likely originate from the pockets along the $\Gamma$-$\K$ line of the conduction bands, as depicted in \figref{fig:bs-mono-AA-2H}(b) and (c).  Compared with AA stacking, the valence bands from the $\Gamma$ valley in AB stacking are extremely flat (see \figref{fig:ABrelaxevolution}(k) and (l)), resulting in pronounced peaks in the density of states (DOS) (see \figref{fig:3.89cvbands}). To faciliate further comprehensive analysis in \appref{app:SP_Continuum_AA_Stacking} and \appref{app:SP_Continuum_AB_Stacking}, we have also calculated the irreducible representations of the top 4 valence bands for AB stacking, which can be found in \figref{fig:irreps-2H-vbs}.
In \figref{fig:fig-density}(b), we plot the charge densities of first four pairs of valence bands' wavefunctions at $\Gamma_M$. The top $\pm\K$ valley bands wavefunctions are localized at MX region, similar with the top $\Gamma$ valley bands. The third pair valence bands' wavefunctions are localized at MM region, forming triangular lattice, while the forth pair valence bands' wavefunctions form triangular rings with minimum at MX region.

\begin{table*}
\begin{tabular}{|c|c|c|c|c|c|c|}
\hline
\textbf{Twist angle}        & \textbf{9.43$\degree$}      & \textbf{7.34$\degree$} & \textbf{5.09$\degree$} & \textbf{4.41$\degree$} & \textbf{3.89$\degree$} & \textbf{3.48$\degree$} \\ \hline
\textbf{Bandwidth (meV) } & 135.4   &     89.8         &   36.1         &    22.5            &   16.0  & 9.8           \\ \hline
\end{tabular}
\caption{Band width of the top two pairs of valence band of 3.89 AB stacking $\tmt$.}
\label{ABBandWidth}
\end{table*}

\section{Single-Particle Continuum Model and Fitting: AA-Stacking}
\label{app:SP_Continuum_AA_Stacking}

The symmetry group of AA-stacked {\tmt} (\AAtmt) is generated by $C_3$, $C_{2y}$, and $\TR$, in addition to the moir\'e lattice translations $T_{\bsl{R}_M}$~\cite{Wu2019TIintTMD}, where $\bsl{R}_M$ labels the moir\'e lattice vectors. 
We label the moir\'e lattice basis vectors as $\bsl{a}_{M,1} = a_{M} (\frac{\sqrt{3}}{2},-\frac{1}{2})^T$ and $\bsl{a}_{M,2} = C_3 \bsl{a}_{M,1}$, and the moir\'e reciprocal lattice basis vectors as $\bsl{b}_{M,1} = \frac{4 \pi}{\sqrt{3} a_{M}} (1,0)^T$ and $\bsl{b}_{M,2} = \frac{4 \pi}{\sqrt{3} a_{M}} (\frac{1}{2},\frac{\sqrt{3}}{2})^T$,
where
\eq{
a_M = \frac{a_0}{2 \sin\left( \frac{\theta}{2} \right)} \ ,
}
and $a_0 = 3.52\AA$ is the lattice constant of the monolayer MoTe$_2$.

Up to now, FCI states were only found for hole doping experimentally~\cite{cai2023signatures,zeng2023integer,park2023observation,Xu2023FCItMoTe2}. From \figref{fig:AArelaxevolution}, it is immediately apparent that isolated, nearly flat bands (understood to be an important precursor to the FCI phase \cite{regnault,sheng,neupert}) appear only in the valence bands, which are accessible through hole doping. In contrast the conduction bands do not have well-separated bands. As such, we focus on building a model for the valence bands in this section.

We will disucss the continuum models for both $\pm\K$ valleys and the $\Gamma$ valley.

\subsection{AA-Stacking: $\pm\K$ Valleys}
\label{app:Kvalley_AA}

\subsubsection{Microscopic Basis, Symmetries, and Inter-layer coupling}

We now derive the moir\'e states that make up the continuum model basis for the AA-stacked twisted heterostructure. This derivation follows \cite{2011PNAS..10812233B,2018arXiv180710676S} by expanding the tight-binding states around the monolayer $\K$ point. ($\K$ can be straightforwardly obtained via time-reversal symmetry.) We consider a two-layer system where the $l = t,b$ layer is twisted via the linear transformation $M_\ell = 1 - \ii \frac{(-)^l \th}{2} \sigma_y$ to leading order $\th \ll 1$, where $(-)^t=-(-)^b = 1$. We will derive the symmetry representations as well as the form of the interlayer coupling within the two-center approximation. 

From our first principles calculations in \figref{fig:fatband}, we see the valence band maximum around $\K$ is spanned by $d_{x^2-y^2} + i d_{xy}$ orbitals on the Mo atoms, which we will refer to as $d$ for brevity in this part. The $d_{z^2}$ Mo orbitals and Te orbitals do not contribute significantly to the density of states near the active bands. We write $M_\ell\mbf{R}$ as the positions of the Mo atoms on the $\ell$th layer, where $\mbf{R}$ is an untwisted lattice vector. The states carried by these orbitals are
\bea
\label{eq:envelope}
 \ket{\mbf{p},\ell}= \frac{1}{\sqrt{N}} \sum_{\mbf{R}} e^{i M_\ell \mbf{R} \cdot \mbf{p}} \ket{M_\ell \mbf{R}', \ell}
\eea
for $\ell = 0,1$ corresponding to top/bottom, and $\mbf{p} = \mbf{K}+\delta \mbf{p}$ is a momentum near the $\mbf{K}$ point. The intra-layer $C_3$ symmetry acts as
\bea
\label{eq:c3micro}
& C_3 \ket{\mbf{K}+\delta\mbf{p},\ell} \\
&= \frac{1}{\sqrt{N}} \sum_{\mbf{R}} e^{i M_\ell \mbf{R} \cdot C_3 (\mbf{K}+\delta\mbf{p})} \ket{M_\ell \mbf{R}, \ell} e^{i \la_d} \\
&= \frac{1}{\sqrt{N}} \sum_{\mbf{R}} e^{i M_\ell \mbf{R} \cdot (\mbf{K} - \mbf{G} +C_3 \delta\mbf{p})} \ket{M_\ell \mbf{R}, \ell} e^{i \la_d} \\
&= \ket{\mbf{K}+ C_3 \delta\mbf{p},\ell} e^{i \lambda_d}
\eea
where $e^{i \lambda_d} = e^{i \frac{2\pi}{3}}$ is the (spin-less) $C_3$ eigenvalue of $d$ orbital, and $\mbf{G} = C_3 \mbf{K}-\mbf{K}$ is a \tmt reciprocal lattice vector obeying $\mbf{G} \cdot \mbf{R} = 0 \mod 2\pi$. We also used that the rotation matrix $C_3$ commutes with $M_\ell$, since both are rotations. Note that $e^{i \lambda_d}$ is simply an overall phase of the rotation representation. Secondly, the $C_{2y}\mathcal{T}$ representation is
\bea
\label{eq:c2ytmicro}
& C_{2y}\mathcal{T} \ket{\mbf{K}+\delta\mbf{p},\ell} \\
&= \frac{1}{\sqrt{N}} \sum_{\mbf{R}} e^{-i M_\ell \mbf{R} \cdot (\mbf{K}+\delta\mbf{p})} \ket{C_{2y} M_\ell \mbf{R}, -\ell} \\
&= \frac{1}{\sqrt{N}} \sum_{\mbf{R}} e^{-i M_\ell \mbf{R} \cdot (\mbf{K}+\delta\mbf{p})} \ket{M_{-\ell} C_{2y} \mbf{R}, -\ell} \\
&= \frac{1}{\sqrt{N}} \sum_{\mbf{R}} e^{-i M_\ell C_{2y} \mbf{R} \cdot (\mbf{K}+\delta\mbf{p})} \ket{M_{-\ell}  \mbf{R}, -\ell} \\
&= \sum_{\ell'} \ket{\mbf{K}+C_{2y} (-\delta\mbf{p}),\ell'} [\sigma_x]_{\ell \ell'}
\eea
using the fact that $\mathcal{T}$ is anti-unitary and $C_{2y}$ flips layer, the reflection property $C_{2y} M_\ell C_{2y}^{-1} = M_{-\ell}$, and the trivial transformation of $d$ orbital under $C_{2y} \mathcal{T}$ \cite{Yu2023FCI}. In this discussion, we have neglected the spin degree of freedom, which can be easily re-introduced because of the spin-valley locking in {\mt}.
Lastly, we note that the emergent intra-valley inversion (which will be discussed in \appref{app:continuum_model_K_AA}) has no representation on the microscopic basis since it is not a true symmetry of the model, much like the emergent particle-hole symmetry in twisted bilayer graphene \cite{2018arXiv180710676S,2022PhRvB.106h5140H}.

We now derive the form of the interlayer Hamiltonian. Formally, we compute the overlap
\bea
H^{inter}_{l,-l}(\mbf{p}, \mbf{p}') &= \frac{1}{N} \sum_{\mbf{R},\mbf{R}'} e^{-i M_l \mbf{R}  \cdot \mbf{p} + i M_{-l} \mbf{R}' \cdot \mbf{p}'} \\
&\qquad \bra{M_{l}\mbf{R},l}H\ket{M_{-l}\mbf{R}',-l} \ . \\
\eea
It is convenient to shift the sum so that the bottom layer is unrotated and the top layer is rotated by $-\th$:
\eqa{
& H^{inter}_{l,-l}(\mbf{p}, \mbf{p}') \\
& = \frac{1}{N} \sum_{\mbf{R},\mbf{R}'} e^{-i M \mbf{R}  \cdot \mbf{p} + i \mbf{R}' \cdot \mbf{p}'} \bra{M \mbf{R},l}H\ket{\mbf{R}',-l} \\
}
where $M = 1 - i \th \sigma_2$. To proceed, we assume that the matrix element of the Hamiltonian is only dependent on the distance between orbitals (the ``two-center" approximation) leading to
\eqa{
& \bra{M\mbf{R},l}H\ket{\mbf{R}',-l} \\
&= \frac{1}{N\Omega} \sum_{\mbf{q} \in BZ} \sum_{\mbf{G}} t_{\mbf{q} + \mbf{G}} e^{i (\mbf{q} + \mbf{G}) \cdot( M \mbf{R} -   \mbf{R}')} \ . 
}
Plugging this expression into the interlayer Hamiltonian and using $M \mbf{r} \cdot \mbf{k} = (M\mbf{r})^T \mbf{k} = \mbf{r} \cdot M^T \mbf{k}$ gives
gives
\bea
H^{inter}_{l,-l}(\mbf{p},\mbf{p}') 
&=  \sum_{\mbf{G}_1,\mbf{G}_2} \frac{t_{\mbf{p} + \mbf{G}_1} }{\Omega} \delta_{\mbf{p}+\mbf{G}_1,\mbf{p}'+M \mbf{G}_2} \ . \\
\eea
Using $\mbf{p} = \mbf{K} + \delta \mbf{p}$ (and similarly for $\mbf{p}'$), we see that $H^{inter}_{l,-l}(\mbf{p},\mbf{p}') $ connects momenta apart by $M\mbf{K}-\mbf{K} = \mbf{q}_1$. We now keep only the lowest order $\mbf{G}_1$ since $t_\mbf{p}$ is rapidly decaying. Thus in real space, this Hamiltonian can be written 
\bea
H^{inter}(\mbf{K} + \delta \mbf{p}, M \mbf{K} + \delta \mbf{p}') &= \sum_{j=1}^3 w \delta_{\delta \mbf{p}, \delta \mbf{p}' + \mbf{q}_i}
\label{eq:hintermicro}
\eea
where $w = t_{\bsl{K}}$. We will find agreement with this term and the symmetry-based approach in the following section. 

The intra-layer Hamiltonian, in the two-center approximation, is given by expanding the monolayer dispersion $h_{mono}(\mbf{K} + \delta \mbf{p}) = -\frac{\hbar^2 \delta \mbf{p}^2}{2 m^*}$ to leading order. Note that the two-center approximation cannot capture the intra-layer moir\'e potential which arises due to relaxation within the moir\'e unit cell.

\begin{figure*}[t]
    \centering
    \includegraphics[width=1.9\columnwidth]{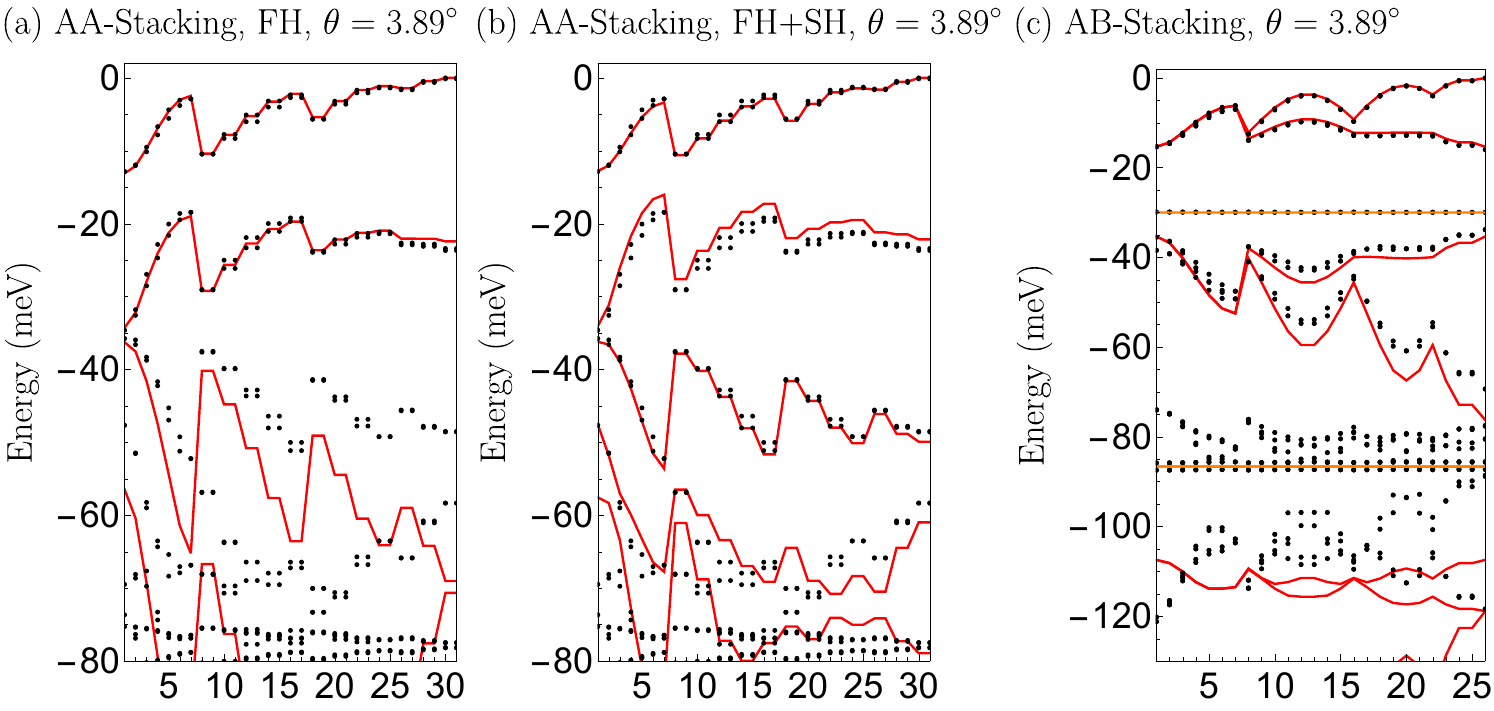}
    \caption{The comparison between the valence bands from the DFT calculation (black dots) and from the continuum model (red/orange line) for twist angle $3.89^\circ$ in the fundamental zone for (a,b) AA-stacking and (c) AB-stacking .
    The red line comes from the $\pm\K$-valley model, while the orange line comes fro mthe $\Gamma$-valley model.
    The horizontal axies labels the 31 momenta in the $k$-mesh of the fundamental zone that we choose. 
    Each red line is at least doubly degenerate.
    The orange line in (c) has double degeneracy around $-30$meV and 4-fold degeneracy around $-86$meV.
    The fitting in (a) is done with FH terms, while SH terms are added in (b).
    }
    \label{fig:fitting_BZ}
\end{figure*}

\subsubsection{Continuum Model}
\label{app:continuum_model_K_AA}

In this part, we discuss the continuum model for the low-energy states around the $\pm\K$ valleys based on symmetries. 
\refcite{Wu2019TIintTMD} proposed a model with first harmonics, and we will introduce more terms into it.
In monolayer MoTe$_2$, the strong spin-orbit coupling locks the spin degree of freedom to the valley degree of freedom. 
Explicitly, the highest valence band around the $\K$ valley is made up of $d_{x^2-y^2} + \ii d_{xy}$ Mo orbitals with spin $\uparrow$~\cite{Xiao2012TMD}.
The highest and second highest valence bands are seperated by the large energy 200meV~\cite{Wu2019TIintTMD}, and thus we only consider the highest electron valence band around $\K$ and $-\K$ (in both layers) to construct the moir\'e model.
The basis of the continuum model is labelled by $c^\dagger_{\eta,l,\bsl{r}}$ with $\eta=\pm$ labels the $\pm \K$ valleys (or equivalently spins), $l=t,b$ labels the layer, and $\bsl{r}$ labels the position. The wavefunction of $c^\dagger_{\eta,l,\bsl{r}}$ is a continuous approximation to the microscopic basis in \eqnref{eq:envelope}. Using Eqs.\,\ref{eq:c2ytmicro} and \ref{eq:c3micro} and after accounting for the electron spin, the \emph{spinful} symmetry representations furnished by $c^\dagger_{\eta,l,\bsl{r}}$ read 
\eqa{
\label{eq:sym_rep_moire_Kvalley}
 & C_3 c^\dagger_{\eta,l,\bsl{r}} C_3^{-1} = c^\dagger_{\eta,l,C_3 \bsl{r}} e^{\ii \eta \frac{\pi}{3} } \\
 & (C_{2y}\TR) c^\dagger_{\eta,l,\bsl{r}} (C_{2y}\TR)^{-1} = c^\dagger_{\eta,\bar{l}, C_{2y} \bsl{r}} \\
 & \TR c^\dagger_{\eta,l,\bsl{r}} \TR^{-1} = c^\dagger_{-\eta,l, \bsl{r}} (-\eta)\\
 & T_{\bsl{R}_M} c^\dagger_{\eta,l,\bsl{r}} T_{\bsl{R}_M}^{-1} = c^\dagger_{\eta,l, \bsl{r}+\bsl{R}_M} e^{-\ii \eta \bsl{R}_M\cdot \K_l}\ ,
}
where $\bar{l}=b,t$ for $l=t,b$ respectively, and $\K_l$ is the rotated $\K$ point of the $l$th layer.
For {\AAtmt}, we rotate the top layer by $-\theta/2$ and the bottom layer by $\theta/2$, and thus
\eqa{
& \K_b= \frac{4 \pi}{3 a_0} \left( \cos\left(\frac{\theta}{2}\right) , \sin\left(\frac{\theta}{2}\right) \right)^T \\
& \K_t= \frac{4 \pi}{3 a_0} \left( \cos\left(\frac{\theta}{2}\right) , - \sin\left(\frac{\theta}{2}\right) \right)^T\ .
}
For the convenience of the discussions in the rest of this part, we define 
\eqa{
& \bsl{q}_1 = \K_b - \K_t = \frac{4 \pi}{3 a_0} 2 \sin\left(\frac{\theta}{2}\right) \mat{ 0 \\ 1} \\
& \bsl{q}_2 = C_3 \bsl{q}_1 \ ,\ \bsl{q}_3=C_3^2 \bsl{q}_1 \ .
}

The general form of the single-particle Hamiltonian in the $\eta$ valley reads
\eq{
H_{\eta,0}^{AA} = \int d^2 r \left( c^\dagger_{\eta,b,\bsl{r}} , c^\dagger_{\eta,t,\bsl{r}} \right) \mat{ h_{\eta,b}(\bsl{r}) & t_\eta(\bsl{r}) \\ t_\eta^*(\bsl{r}) & h_{\eta,t}(\bsl{r})  } \mat{ c_{\eta,b,\bsl{r}} \\ c_{\eta,t,\bsl{r}} } \ ,
}
where $h_{\eta,l}(\bsl{r})$ is hermitian.
The symmetry requirements on the terms in the Hamiltonian read
\eqa{
\label{eq:symmetry_requirements_Kvalley}
& C_3:\ h_{\eta,l}(C_3\bsl{r}) =  h_{\eta,l}(\bsl{r})\ ,\ t_\eta(C_3\bsl{r}) = t_\eta(\bsl{r}) \\ 
& C_{2y} \TR: \ h_{\eta,\bar{l}}(C_{2y}\bsl{r}) =  h_{\eta,l}^*(\bsl{r})\ ,\ t_\eta(C_{2y}\bsl{r}) = t_\eta(\bsl{r}) \\ 
& \TR: \ h_{-\eta,l}(\bsl{r}) =  h_{\eta,l}^*(\bsl{r})\ ,\ t_{-\eta}(\bsl{r}) = t_{\eta}^*(\bsl{r}) \\ 
& T_{\bsl{R}_M}: h_{\eta,l}(\bsl{r}+\bsl{R}_M) =  h_{\eta,l}(\bsl{r})\ ,\\
& \qquad \ \ t_\eta(\bsl{r}+\bsl{R}_M) = t_\eta(\bsl{r}) e^{-\ii \eta \bsl{q}_1 \cdot \bsl{R}_M }
}

Owing to $C_3$ symmetry, the dispersion around $\pm\K$ in the monolayer MoTe$_2$ is quadratic.
Thus, the spatial derivatives are kept to the second order in the intra-layer term:
\eq{
h_{\eta,l}(\bsl{r}) = \frac{\hbar^2 \nabla^2}{2 m^*} + V_{\eta,l}(\bsl{r}) \ ,
}
where $m^*$ is real and positive as we are considering the valence electron bands, the $C_{2y}\TR$ and $\TR$ symmetries require $m^*$ to be the same for all values of $\eta,l$, and hermiticity requires $V_{\eta,l}(\bsl{r})$ to be real. The kinetic term follows from the two-center approximation (see the discussion after Eq. \ref{eq:hintermicro}). Owing to the constraints imposed by moir\'e lattice translations on the potential terms (see \eqnref{eq:symmetry_requirements_Kvalley}), $V_{\eta,l}(\bsl{r})$ and $e^{\ii \eta \bsl{q}_1 \cdot \bsl{r} } t_{\eta}(\bsl{r})$ can be expanded in series of the moir\'e lattice vector $\bsl{G}_M$:
\eqa{
 &  V_{\eta,l}(\bsl{r}) = \sum_{\bsl{G}_M} e^{ - \eta \ii \bsl{G}_M \cdot \bsl{r} } V_{\eta ,l,\bsl{G}_M} \\
 & t_\eta(\bsl{r}) = \sum_{\bsl{G}_M} e^{ - \eta  \ii (\bsl{q}_1+\bsl{G}_M) \cdot \bsl{r} } t_{\eta ,\bsl{q}_1+\bsl{G}_M}\ ,
}
where the TR symmetry requires $V_{\eta ,l,\bsl{G}_M} = V_{-\eta ,l,\bsl{G}_M}^*$ and $t_{\eta ,l,\bsl{q}_1+\bsl{G}_M} = t_{-\eta ,l,\bsl{q}_1+\bsl{G}_M}^*$.

In \refcite{Wu2019TIintTMD}, only the first harmonics of $V_{\eta,l}(\bsl{r})$ and $t_{\eta}(\bsl{r})$ is kept, \ie, only including $\bsl{G}_M \in \{ \pm \bsl{g}_i | i=1,2,3 \}$ with $\bsl{g}_{i} = C_3^{i-1} \bsl{b}_{M,1}$ for the intralayer terms, and only include $\bsl{q}_1+\bsl{G}_M \in \{ - \bsl{q}_i | i=1,2,3 \}$ for the interlayer terms.
Combined with $C_{3}$, $C_{2y}\TR$ and TR symmetries, $V_{+,l}(\bsl{r})$ and $t_+(\bsl{r})$ take the forms of 
\eqa{
& V_{\eta,l}^{FH}(\bsl{r}) = V e^{ - (-)^l \ii \psi} \sum_{i=1,2,3} e^{\ii \bsl{g}_i \cdot \bsl{r} } + V e^{ (-)^l \ii \psi} \sum_{i=1,2,3} e^{-\ii \bsl{g}_i \cdot \bsl{r}} \\
& t_\eta^{FH}(\bsl{r}) = w \sum_{i=1,2,3} e^{ - \eta \ii \bsl{q}_i \cdot \bsl{r} }\ ,
}
where ``$FH$'' labels the first harmonics, $(-)^t = 1$, $(-)^b = -1$, $V$ and $\psi$ are real, and $w$ is chosen to be real non-positive by tuning the relative phase between the two layers.
The lowest-harmonics model has effective inversion symmetry
\eq{
\I c^\dagger_{\eta,l,\bsl{r}} \I^{-1} = c^\dagger_{\eta,\bar{l},-\bsl{r}}\ .
}
Adding displacement field would induce an energy difference between the two layers, \ie,
\eq{
H_{\eta,\varepsilon} =  \int d^2 r \sum_{l} c^\dagger_{\eta,l,\bsl{r}}c_{\eta,l,\bsl{r}} (-)^l \frac{\varepsilon}{2}\ ,
}
which breaks the effective inversion symmetry, as well as the $C_{2y}$ symmetry.

As discussed in the previous work (\eg, \refcite{wang2023fractional}) and in \appref{app:AA_fitting}, the lowest-harmonics model can only well describe the highest two valence bands per valley and the gap between 2nd and 3rd valence bands at $\Gamma$.
In our work, we want to capture the 3rd band in each valley by including one more harmonics.
In general, including higher harmonics may break the effective inversion symmetry.
However, according to the DFT band structure in \figref{fig:AArelaxevolution}, the 3rd pairs of valence bands are still approximately degenerate for angles below 5.09 degrees; therefore, we will keep the effective inversion symmetry when adding extra harmonics. 
The second harmonics (SH) have $|\bsl{G}_M|=|\bsl{b}_{M,1}+\bsl{b}_{M,2}|$ for the intra-layer potential and have $|\bsl{q}_1+\bsl{G}_M |$ = $\bsl{q}_1 + \bsl{b}_{M,1}$ for the inter-layer potential; combined with $C_{3}$, $C_{2y}\TR$, TR and the effective inversion symmetries, the form of the SH for the intra-layer potential read
\eqa{
& V_{\eta,l}^{SH}(\bsl{r}) = 2 V_2 \sum_{i=1}^3 \cos(\bsl{g}_{2i}\cdot \bsl{r} ) \\
& t_\eta^{SH}(\bsl{r}) = w_2 \sum_{i=1,2,3} e^{ - \eta \ii \bsl{q}_{2i} \cdot \bsl{r} }\ ,
}
where $ \bsl{g}_{21} = \bsl{b}_{M,1} + \bsl{b}_{M,2}$, $\bsl{g}_{2i} = C_3^{i-1} \bsl{g}_{21} $, $ \bsl{q}_{21} = \bsl{b}_{M,1} + \bsl{q}_{_1}$, $\bsl{q}_{2i} = C_3^{i-1} \bsl{q}_{21} $, and we have enforced the effective inversion symmetry, which makes $V_2$ and $w_2$ real.
With these extra terms, the final form of $V_\eta(\bsl{r})$ and $t_\eta(\bsl{r})$ reads
\eqa{
\label{eq:Kvalley_V_t_FH_SH}
& V_{\eta,l}(\bsl{r}) = V e^{ - (-)^l \ii \psi} \sum_{i=1,2,3} e^{\ii \bsl{g}_i \cdot \bsl{r} } + V e^{ (-)^l \ii \psi} \sum_{i=1,2,3} e^{-\ii \bsl{g}_i \cdot \bsl{r}} \\
& \qquad + 2 V_2 \sum_{i=1}^3 \cos(\bsl{g}_{2i}\cdot \bsl{r} ) \\
& t_\eta(\bsl{r}) = w \sum_{i=1,2,3} e^{ - \eta \ii \bsl{q}_i \cdot \bsl{r} } + w_2 \sum_{i=1,2,3} e^{ - \eta \ii \bsl{q}_{2i} \cdot \bsl{r} }\ ,
}

The moir\'e translations allow us to express the Hamiltonian in the momentum space.
The Fourier transformation of the basis reads
\eq{
\label{eq:Kvalley_FT}
c^\dagger_{\eta,l,\bsl{r}} = \frac{1}{\sqrt{\V}} \sum_{\bsl{k}} \sum_{\bsl{Q}\in\Q_l^{\eta} } e^{-\ii (\bsl{k} - \bsl{Q}) \cdot \bsl{r} } c^\dagger_{\eta,l,\bsl{k}-\bsl{Q}}\ ,
}
where
\eq{
\Q_l^\eta = \{ \bsl{G}_M + \eta (-)^l \bsl{q}_1 \}  \ .
}
As a result, $H_{\eta,0}$ reads
\eq{
\label{eq:Kvalley_H_momentum_space_general_AA}
H_{\eta,0}^{AA} = \sum_{\bsl{k}} \sum_{\bsl{Q},\bsl{Q}'\in \Q} c^\dagger_{\eta,\bsl{k},\bsl{Q}} h_{\eta,\bsl{Q}\bsl{Q}'}^{AA}(\bsl{k})c_{\eta,\bsl{k},\bsl{Q}'}\ ,
}
where we have included the displacement field, $\Q= \{ \bsl{G}_M + \bsl{q}_1 \} \cup \{ \bsl{G}_M - \bsl{q}_1 \} $,
\eq{
\label{eq:Kvalley_moire_basis_momentum_space}
c^\dagger_{\eta,\bsl{k},\bsl{Q}} = c^\dagger_{\eta,l_{\bsl{Q}}^{\eta},\bsl{k}-\bsl{Q}}\ ,
}
and $l_{\bsl{Q}}^{\eta} = l $ for $\bsl{Q}\in \Q_l^{\eta}$.%
The general form of $h_{\eta,\bsl{Q}\bsl{Q}'}^{AA}(\bsl{k})$ reads
\eqa{
\label{eq:Kvalley_model_AA_General}
& h_{\eta,\bsl{Q}\bsl{Q}'}^{AA}(\bsl{k})  =  \delta_{\bsl{Q}\bsl{Q}'} \left( \frac{- \hbar^2 \left( \bsl{k} -\bsl{Q} \right)^2}{ 2 m^*} + \frac{\varepsilon}{2} \eta (-)^{\bsl{Q}} \right) \\
& \quad + \sum_{\bsl{G}_M} \delta_{\bsl{Q}_1,\bsl{Q}_2 + \eta \bsl{G}_M} V_{\eta,l_{\bsl{Q}_1}^{\eta}, \bsl{G}_M} \\
& \quad + \sum_{\bsl{G}_M} \delta_{\bsl{Q}_1,\bsl{Q}_2 + \eta (\bsl{q}_1+\bsl{G}_M) } t_{\eta,l_{\bsl{Q}_1}^{\eta}, \bsl{q}_1+\bsl{G}_M } \\
& \quad + \sum_{\bsl{G}_M} \delta_{\bsl{Q}_1,\bsl{Q}_2 - \eta (\bsl{q}_1+\bsl{G}_M) } t_{\eta,l_{\bsl{Q}_1}^{\eta}, \bsl{q}_1+\bsl{G}_M }^* \ .
}
With the FH and SH terms in \eqnref{eq:Kvalley_V_t_FH_SH}, the matrix Hamiltonian $h_{\eta,\bsl{Q}\bsl{Q}'}^{AA}(\bsl{k})$ has the form
\eqa{
\label{eq:Kvalley_model_AA}
& h_{\eta,\bsl{Q}\bsl{Q}'}^{AA}(\bsl{k})  =  \delta_{\bsl{Q}\bsl{Q}'} \left( \frac{- \hbar^2 \left( \bsl{k} -\bsl{Q} \right)^2}{ 2 m^*} + \frac{\varepsilon}{2} \eta (-)^{\bsl{Q}} \right) \\
& \quad + V \sum_{i=1}^3 \left[  e^{-\eta (-)^{\bsl{Q}}\ii \psi} \delta_{\bsl{Q}+\bsl{g}_i,\bsl{Q}'} +  e^{\eta (-)^{\bsl{Q}}\ii \psi} \delta_{\bsl{Q}-\bsl{g}_i,\bsl{Q}'}\right]  \\
& \quad + V_2 \sum_{i=1}^3 \left[   \delta_{\bsl{Q}+\bsl{g}_{2i},\bsl{Q}'} +  \delta_{\bsl{Q},\bsl{Q}'+\bsl{g}_{2i}}\right]  \\
& \quad + w \sum_{i=1}^{3} \left[  \delta_{\bsl{Q}+\bsl{q}_i,\bsl{Q}'} + \delta_{\bsl{Q},\bsl{Q}'+\bsl{q}_i} \right] \\
& \quad + w_2 \sum_{i=1}^{3} \left[  \delta_{\bsl{Q}+\bsl{q}_{2i},\bsl{Q}'} + \delta_{\bsl{Q},\bsl{Q}'+\bsl{q}_{2i}} \right] \ ,
}
where $(-)^{\bsl{Q}} =\pm1$ for $\bsl{Q}\in\{ \bsl{G}_M \pm \bsl{q}_1 \}$, and we used $ \eta (-)^{\bsl{Q}} =  (-)^{l^{\eta}_{\bsl{Q}}}$
\subsection{AA-Stacking: $\Gamma$ Valley}

In this part, we discuss the continuum model for the $\Gamma$ valley.
In the monolayer MoTe$_2$, the bands around $\Gamma$ mainly come from the $d_{z^2}$  orbital, and time-reversal symmetry at $\Gamma$ leads to a Kramers degeneracy between spin up and spin down states. 
Thus, the basis of the moir\'e model at $\Gamma$ valley is labelled as $\psi^\dagger_{\bsl{r},l,s}$, where $l$ is the layer index, and $s=\uparrow/\downarrow$ labels the spin.
Then, the continuum model in general has the following form
\eq{
\label{eq:Gamma_model_general}
H_{\Gamma}^{AA} = \int d^2 r \mat{ \psi^\dagger_{\bsl{r},b} &  \psi^\dagger_{\bsl{r},t} } 
\mat{
h_{\Gamma,b}(\bsl{r}) & t_{\Gamma}(\bsl{r})\\
t_{\Gamma}^\dagger (\bsl{r}) & h_{\Gamma,t}(\bsl{r})
}
\mat{ \psi^\dagger_{\bsl{r},b}  \\  \psi^\dagger_{\bsl{r},t} }\ ,
}
where 
\eq{
\psi^\dagger_{\bsl{r},l} = ( \psi^\dagger_{\bsl{r},l,\uparrow} , \psi^\dagger_{\bsl{r},l,\downarrow} )\ ,
}
$t$ and $b$ correspond to the top and bottom layers, respectively, and $h_{\Gamma,l}(\bsl{r})$ and $t_{\Gamma}(\bsl{r})$ are $2\times 2$ moir\'e-periodic matrix functions.

The spinful representations of the $d_{z^2}$ orbital symmetries of the moir\'e model are
\eqa{
& C_3 \psi^\dagger_{\bsl{r},l} C_3^{-1} = \psi^\dagger_{C_3 \bsl{r},l} e^{-\ii s_z \frac{\pi}{3} }\\
& C_{2y} \TR \psi^\dagger_{\bsl{r},l} (C_{2y}\TR)^{-1} = \psi^\dagger_{C_{2y} \bsl{r},\bar{l} }\\
& \TR \psi^\dagger_{\bsl{r},l} \TR^{-1} = \psi^\dagger_{ \bsl{r},l} \ii s_y\\
& T_{\bsl{R}_M} \psi^\dagger_{\bsl{r},l} T_{\bsl{R}_M}^{-1} = \psi^\dagger_{ \bsl{r}+\bsl{R}_M,l} \ ,
}
where $s_x,s_y,s_z$ are the Pauli matrices for the spin index, and $\bar{l} = t,b$ if $l=b,t$.
Based on symmetries, we know 
\eqa{
& e^{-\ii s_z \frac{\pi}{3} } h_{\Gamma,l}(\bsl{r}) e^{\ii s_z \frac{\pi}{3} } = h_{\Gamma,l}(C_3 \bsl{r}) \\
&e^{-\ii s_z \frac{\pi}{3} } t_{\Gamma}(\bsl{r}) e^{\ii s_z \frac{\pi}{3} } = t_{\Gamma}(C_3 \bsl{r})\\ 
& h_{\Gamma,l}^*(\bsl{r})  = h_{\Gamma,\bar{l}}(C_{2y} \bsl{r}) \ ,\ t_{\Gamma}^*(\bsl{r})  = t_{\Gamma}^\dagger (C_{2y}\bsl{r})  \\ 
&  \sigma_y h_{\Gamma,l}^*(\bsl{r}) \sigma_y = h_{\Gamma,l}( \bsl{r}) \ ,\ \sigma_y t_{\Gamma}^*(\bsl{r}) \sigma_y = t_{\Gamma} (\bsl{r})  \\ 
&  h_{\Gamma,l}(\bsl{r}) = h_{\Gamma,l}( \bsl{r} + \bsl{R}_M) \ ,\  t_{\Gamma}(\bsl{r}) = t_{\Gamma} (\bsl{r}+ \bsl{R}_M)  \ .
}

The intra-layer kinetic energy term must take the form
\eqa{
\label{eq:h_AA_Gamma_Valley}
h_{\Gamma,l}(\bsl{r}) = \frac{\hbar^2 \nabla^2 }{2 m_\Gamma^*} s_0 + V_{\Gamma,l}(\bsl{r}) 
}
since a linear Rashba-like kinetic term is forbidden by $C_3$, $\mathcal{T}$, and $M_z$ symmetry in the monolayer. The effective mass $m_\Gamma^*$ is the same in both layers because of the $C_{2y}$ symmetry.
We note that the value of $m_\Gamma^*$ in \eqnref{eq:h_AA_Gamma_Valley} might not be equal to that for monolayer {\mt} due to the inter-layer coupling renormalizing the bands, due to the fact that the bands near $\Gamma$ in the monolayer is quite flat as shown in \figref{fig:bs-mono-AA-2H} (a). The potentials are periodic and can be expanded as
\eqa{
& V_{\Gamma,l}(\bsl{r}) =  \sum_{\bsl{G}} V_{\Gamma,l,\bsl{G}} e^{\ii \bsl{G}\cdot\bsl{r} }\\
& t_{\Gamma}(\bsl{r}) =  \sum_{\bsl{G}}t_{\Gamma,\bsl{G}} e^{\ii \bsl{G}\cdot\bsl{r} } 
}
whose components are restricted by symmetry to obey
\eqa{
\label{eq:sym_rep_AA_Gamma_Valley}
& e^{-\ii s_z \frac{\pi}{3} } V_{\Gamma,l,\bsl{G}} e^{\ii s_z \frac{\pi}{3} } = V_{\Gamma,l,C_3 \bsl{G}} \ ,\ e^{-\ii s_z \frac{\pi}{3} } t_{\Gamma,\bsl{G}} e^{\ii s_z \frac{\pi}{3} } = t_{\Gamma,C_3 \bsl{G}} \\ 
&  V_{\Gamma,l,\bsl{G}}^* = V_{\Gamma,\bar{l},-C_{2y}\bsl{G}}  \ ,\ t_{\Gamma,\bsl{G}}^* = t^\dagger_{\Gamma,-C_{2y}\bsl{G}}  \\ 
&  s_y V_{\Gamma,l,\bsl{G}}^* s_y = V_{\Gamma,l,-\bsl{G}} \ ,\ s_y t_{\Gamma,\bsl{G}}^* s_y = t_{\Gamma,-\bsl{G}}  \\
& V_{\Gamma,l,\bsl{G}}^\dagger = V_{\Gamma,l,-\bsl{G}} \ .
}

To capture the dominant contribution, we only include the first harmonics in $V_{\Gamma,l}(\bsl{r})$, and the zeroth harmonic in the $t_{\Gamma}(\bsl{r})$, \ie,
\eqa{
\label{eq:Vt_AA_Gamma_Valley}
& V_{\Gamma,l}(\bsl{r}) =  \sum_{\bsl{G}\in \{ \pm \bsl{g}_i | i=1,2,3 \}} V_{\Gamma,l,\bsl{G}} e^{\ii \bsl{G}\cdot\bsl{r} } \\
& t_{\Gamma}(\bsl{r}) = t_{\Gamma,0} \ .
}
We neglect the zeroth harmonic in $V_{\Gamma,l}(\bsl{r})$ since it is just a total shift of energy (although it must be included to compare the energies of the in-active $\Gamma$ valley and active $\pm\K$ valleys). 
According to \eqnref{eq:sym_rep_AA_Gamma_Valley}, the first harmonic terms in $V_{\Gamma,l}(\bsl{r})$ can be written
\eqa{
& V_{\Gamma,b,\pm \bsl{g}_i} = V_{\Gamma} e^{\pm \ii \psi_{\Gamma}} s_0 \\
& V_{\Gamma,t,\pm \bsl{g}_i} = V_{\Gamma} e^{\mp \ii \psi_{\Gamma}} s_0 \ ,
}
and similarly, $t_{\Gamma,0}$ has the form
\eq{
t_{\Gamma,0} = w_{\Gamma} e^{\ii s_z \phi_\Gamma}\ .
}
Here $V_{\Gamma}$, $\psi_{\Gamma}$, $w_{\Gamma}$ and $\phi_\Gamma$ are real. We set $\phi_\Gamma = 0$ as a gauge choice of the relative phase between the layers. As a result, \eqnref{eq:Vt_AA_Gamma_Valley} is further simplified to
\eqa{
\label{eq:Vt_AA_Gamma_Valley_sim}
& V_{\Gamma,l}(\bsl{r}) =  \sum_{i=1}^3 V_{\Gamma} e^{\ii (-)^l \psi_{\Gamma}}  e^{\ii \bsl{g}_i\cdot\bsl{r} } +  \sum_{i=1}^3 V_{\Gamma} e^{-\ii (-)^l  \psi_{\Gamma}}  e^{-\ii \bsl{g}_i\cdot\bsl{r} } \\
& t_{\Gamma}(\bsl{r}) = w_{\Gamma} s_0 \ .
}

With the simplification in \eqnref{eq:h_AA_Gamma_Valley} and \eqnref{eq:Vt_AA_Gamma_Valley_sim}, the $\Gamma$-valley continuum model in \eqnref{eq:Gamma_model_general} has effective inversion symmetry, \ie,
\eq{
\I \psi^\dagger_{\bsl{r},l} \I^{-1} = \psi^\dagger_{-\bsl{r}, \bar{l} }\ .
}
Similar to the $\pm\K$-valley case, adding displacement breaks the effective inversion symmetry, as well as the $C_{2y}$ symmetry, where the displacement field term reads
\eq{
H_{\Gamma,\varepsilon} =  \int d^2 r \sum_{l} \psi^\dagger_{\bsl{r},l}\psi_{\bsl{r},l} (-)^l \frac{\varepsilon}{2}\ .
}
Including the first harmonics in $t_{\Gamma}(\bsl{r})$ and the higher harmonics in $V_{\Gamma,l}(\bsl{r})$ may also break the effective inversion symmetry.

\begin{figure*}[t]
    \includegraphics[width=1.9\columnwidth]{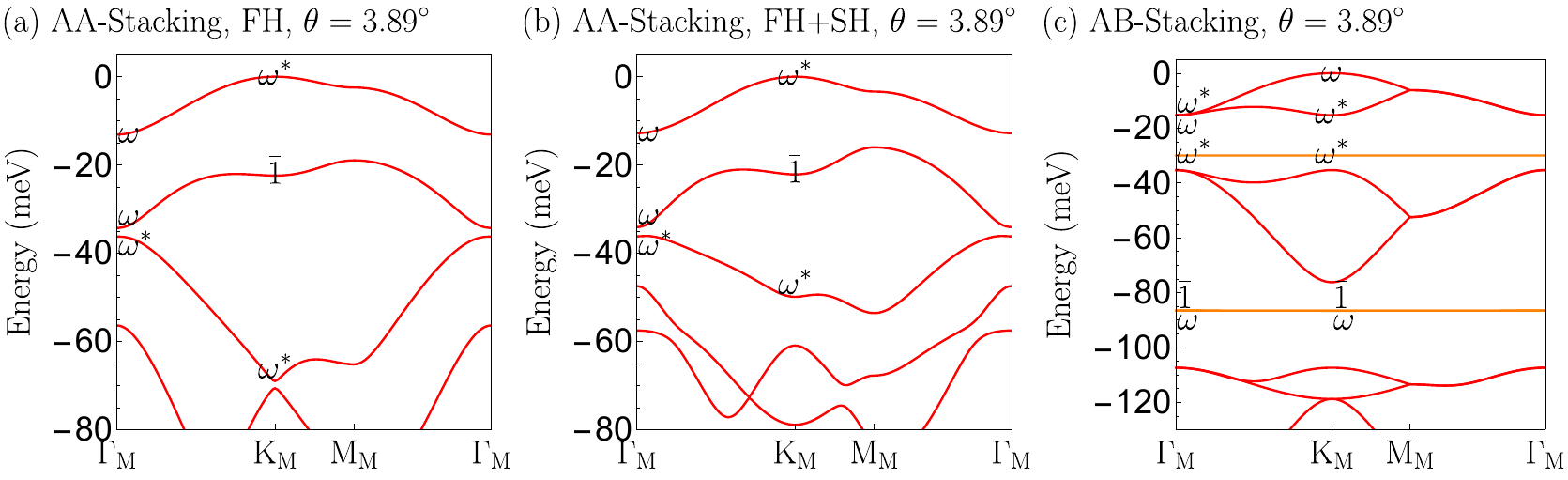}
    \caption{The $C_3$ eigenvalues of the bands from the continuum model for twist angle $3.89^\circ$ for (a,b) AA-stacking and (c) AB-stacking .
    The plot in (a) is done with FH terms, while SH terms are added in (b). 
    In (a,b), we only include $C_3$ eigenvalues for the $\K$-valley bands (red) at $\Gamma_M$ and $\K_M$, since $K_M'$ point is related to the $K_M$ point by $C_{2y} \TR$ and the $-\K$ valley can be obtained from the TR symmetry.
    In (c), we only include $C_3$ eigenvalues for the $\K$-valley bands (red) and the $\Gamma$-valley spin-up bands at $\Gamma_M$ and $\K_M$, since $K_M'$ point is related to the $K_M$ point by the effective TR symmetry, and those for the $-\K$ valley and the $\Gamma$-valley spin down can be obtained from the TR symmetry.
    Like in \figref{fig:irreps-AA-vbs}, $\omega=e^{\ii \pi/3}$ and $\bar{1}=-1$.
    }
    \label{fig:C3eigenvalues_highsymline}
\end{figure*}

\subsection{AA-Stacking: Fitting to the DFT Data}
\label{app:AA_fitting}

As shown in \figref{fig:AArelaxevolution}, the $\Gamma$-valley valence bands are below the 6th highest valence bands for $\theta\leq 5.09^\circ$ and zero displacement field $\varepsilon=0$.
Therefore, we only use the $\pm \K$-valley model (\eqnref{eq:Kvalley_model_AA}) in the fitting.

We fit the DFT bands at $3.89^\circ$ in two ways.
First, we set $V_2=w_2=0$, which corresponds to the FH model.
In this case, we manage to fit the top 4 valence bands (2 in each valley) with the corresponding FH parameters in \tabref{tab:AA_parameters_DFT}, as shown in \figref{fig:fitting_main} (a) along the high-symmetry lines and in \figref{fig:fitting_BZ} (a) for the $\bsl{k}$ points in the fundamental zone.
Then, we allow nonzero $V_2$ and $w_2$, which means we add the effective-inversion-symmetric SH terms.
We are now able to fit the top 6 valence bands (3 in each valley) with the corresponding FH+SH parameters in \tabref{tab:AA_parameters_DFT}, as shown in \figref{fig:fitting_main} (b) along the high-symmetry lines and in \figref{fig:fitting_BZ} (b) for the $\bsl{k}$ points in the fundamental zone.

\begin{figure*}
    \centering
    \includegraphics[width=17cm]{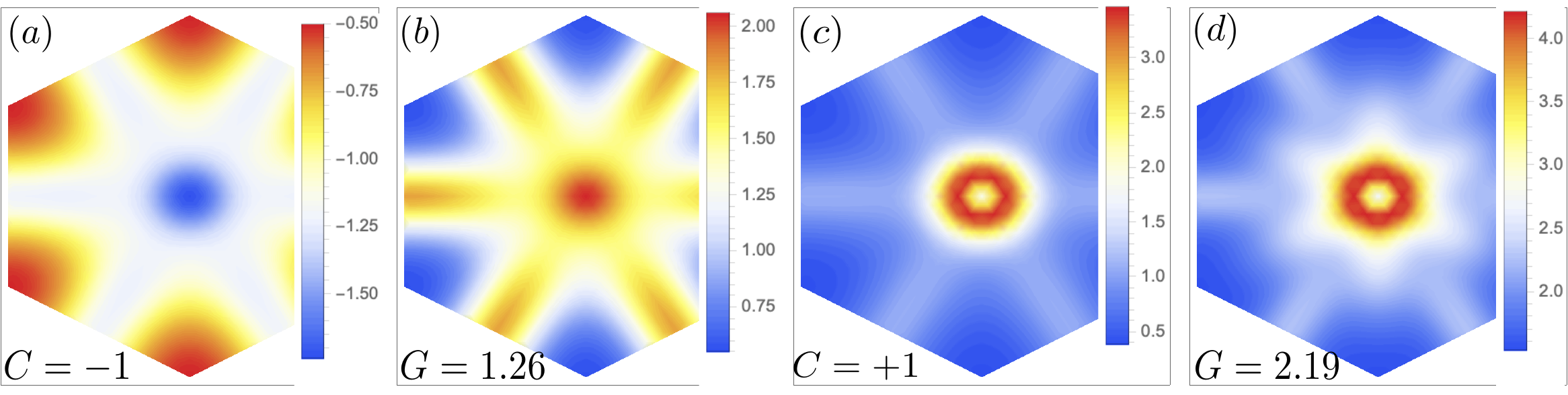}
    \caption{Quantum geometry of the AA-stacked continuum FH model ($\th = 3.89^\circ$) in \eqref{eq:Kvalley_model_AA} (without higher harmonics). $(a,b)$ show the Berry curvature and Fubini-Study metric of the highest valence band, respectively, and $(c,d)$ show the corresponding plots for the second highest valence band. The Chern numbers are $C = -1,+$ for $(a)$ and $(c)$ respectively, and integrated Fubini-Study metrics are marked.
    }
    \label{fig:continuummodelAAQG_FH}
\end{figure*}

\begin{figure*}
    \centering
    \includegraphics[width=17cm]{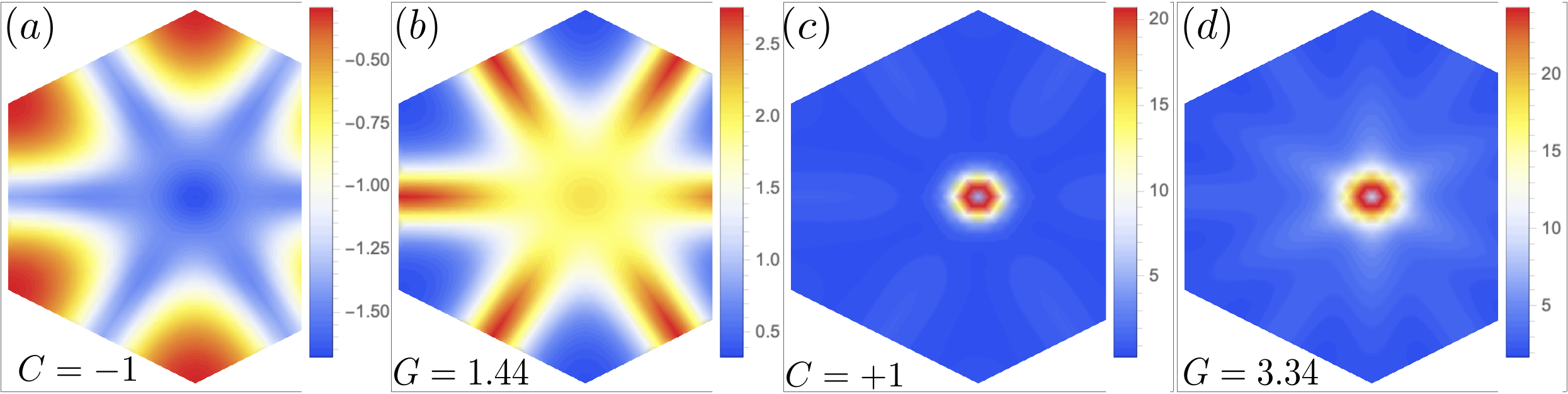}
    \caption{Quantum geometry of the AA-stacked continuum model ($\th = 3.89^\circ$) including second-shell harmonics in \eqref{eq:Kvalley_model_AA}. $(a,b)$ show the Berry curvature and Fubini-Study metric of the highest valence band, respectively, and $(c,d)$ show the corresponding plots for the second highest valence band. The Chern numbers are $C = -1,+$ for $(a)$ and $(c)$ respectively.
    }
    \label{fig:continuummodelAAQG}
\end{figure*}

As shown in \tabref{tab:AA_parameters_DFT}, the value of the effective mass $m^*$ is similar to the monolayer and untwisted AA-stacking bilayer structures masses, as shown in \figref{fig:bs-mono-AA-2H}.
We note that all bands from the $-\K$-valley continuum model are the same as those from the $\K$-valley continuum model, due to the combination of the effective inversion and the TR symmetry. 
By comparing \figref{fig:C3eigenvalues_highsymline} (a,b) to \figref{fig:irreps-AA-vbs} (c), we can see that the $C_3$ eigenvalues for the top 6 valence bands from the model are the same as those from the DFT calculation in both FH and FH+SH cases at $3.89^\circ$.
At $3.89^\circ$, The Chern numbers of the top 3 bands in $\K$ valley are (1,-1,0) for the highest, the second highest and the third highest valence bands in both FH and FH+SH  cases, respectively, which are consistent with the $C_3$ eigenvalues. 
The Chern numbers and the $C_3$ eigenvalues of the top two bands per valley are the same as those in \refcite{wang2023fractional} at $3.89^\circ$.
The quantum geometry on the moir\'e Brillouin zone is shown in \figref{fig:continuummodelAAQG_FH}-\ref{fig:continuummodelAAQG}, with Chern number $C$ and integrated Fubini-Study metric $G/2\pi$ (see Ref. \cite{Yu2023FCI}). We observe that the Berry curvature and Fubini-Study metric are more uniform in the first band than in the second (remote) band, with the second remote band showing a peak around $\Gamma$ where the gap to the third band is smallest. Note also that the effect of second harmonic terms is weak on the first ($C=-1$) valence band, whereas the second valence band becomes significantly more strongly peaked at the $\Gamma$ point and flatter elsewhere on the BZ. This can be understood from the decrease in the gap at $\Gamma$ between the second and third valence bands due to the improved accuracy of the higher-harmonic model. 

Since the FH+SH model captures more bands than the FH model, we use the FH+SH model to discuss the phase diagram under the twist angle $\theta$ and the displacement field $\varepsilon$. 
\figref{fig:AAEfieldThetaPhaseDiagram} in the Main Text shows the single-particle phase diagram for the top three valence bands in $\K$ valley. The highest valence band -- most relevant to the filling factors where FCIs have been observed.
The top valence $\K$-valley band displays a Chern number $C=1$ throughout the phase diagram at $\varepsilon = 0$, and shows a phase transition into a trivial insulator as $\varepsilon$ is turned on. 
The phase diagram of the remote bands is richer, showing all Chern numbers $C= -2,\dots, 2$ throughout the $\th, \varepsilon$ plane. 
In particular, at $\varepsilon=0$, the second and third top bands have a gap closing around $\theta=4.2^\circ$, which is close to the gap closing around $4.41^\circ$ shown in \figref{fig:irreps-AA-vbs}.

\section{Single-Particle Continuum Model and Fitting: AB-Stacking (2H-Stacking)}
\label{app:SP_Continuum_AB_Stacking}

Replacing $C_{2y}$ by $C_{2x}$ in the generators of the symmetry group of {\AAtmt} gives the generators of the symmetry group of AB-stacking {\tmt} ({\ABtmt}), which can be thought of as twisting the top layer of AA-$t$MoTe$_2$ by another $180^\circ$. The change of symmetry has a dramatic effect on the model. Since $C_{2x}$ is local to the monolayer $\K$ point (unlike $C_{2y}$), it preserves the valley quantum in the moir\'e model.

Different from the {\AAtmt}, the DFT results show that the low-energy valence bands of {\ABtmt} come from both the $\pm\K$ valleys and the $\Gamma$ valleys. 
Therefore, in the following, we will review the continuum model for the $\pm\K$ valleys proposed in \refcite{Wu2019TIintTMD}, and will also discuss the continuum model for the $\Gamma$ valley.

\subsection{AB-Stacking: $\pm\K$ Valleys}

In this part, we review the continuum model of {\ABtmt} for the $\pm\K$ valleys proposed in \refcite{Wu2019TIintTMD}.
The basis of the continuum model is still labelled by $c^\dagger_{\eta,l,\bsl{r}}$ with $\eta=\pm$ labeling the $\pm \K$ valleys (or equivalently spins), $l=t,b$ labels the layer, and $\bsl{r}$ labels the position.
Since {\ABtmt} is given by rotating the top layer of {\AAtmt} by an extra $180^\circ$, the two layers in one valley now have opposite spin. 
Then, the symmetry reps furnished by $c^\dagger_{\eta,l,\bsl{r}}$ now read
\eqa{
\label{eq:sym_rep_moire_Kvalley_AB}
 & C_3 c^\dagger_{\eta,l,\bsl{r}} C_3^{-1} = c^\dagger_{\eta,l,C_3 \bsl{r}} e^{ - \ii \eta (-)^l \frac{\pi}{3} } \\
 & C_{2x} c^\dagger_{\eta,l,\bsl{r}} C_{2x}^{-1} = c^\dagger_{\eta,\bar{l}, C_{2x} \bsl{r}} (-\ii)  \\
 & \TR c^\dagger_{\eta,l,\bsl{r}} \TR^{-1} = c^\dagger_{-\eta,l, \bsl{r}} (-\eta)\\
 & T_{\bsl{R}_M} c^\dagger_{\eta,l,\bsl{r}} T_{\bsl{R}_M}^{-1} = c^\dagger_{\eta,l, \bsl{r}+\bsl{R}_M} e^{-\ii \eta \bsl{R}_M\cdot \K_l}\ ,
}
where $\bar{l}=b,t$ for $l=t,b$ respectively, $\K_l$ is the rotated $\K$ point of the $l$th layer, and $(-)^t = -(-)^b =1$.

The general form of the single-particle Hamiltonian in the $\eta$ valley reads
\eqa{
H_{\eta,0}^{AB} & = \sum_l \int d^2 r  c^\dagger_{\eta,l,\bsl{r}}  \left[ \frac{\hbar^2 \nabla^2}{2 m^*} + V_{\eta,l}(\bsl{r}) \right]  c_{\eta,l,\bsl{r}} \\
& \ + \left[ \int d^2 r  c^\dagger_{\eta,b,\bsl{r}}  t_\eta(\bsl{r})  c_{\eta,t,\bsl{r}} + h.c. \right]\ ,
}
where $V_{\eta,l}(\bsl{r})$ is real, and we have used the fact that the dispersion around $\pm\K$ in the monolayer MoTe$_2$ is quadratic and the fact that $m^*$ is the same for two valleys and two layers owing to $C_{2x}$ and $\TR$ symmetries.
The symmetry requirements on $V_{\eta,l}(\bsl{r})$ and $t_\eta(\bsl{r})$ read
\eqa{
\label{eq:symmetry_requirements_Kvalley_AB}
& C_3:\ V_{\eta,l}(C_3\bsl{r}) =  V_{\eta,l}(\bsl{r}) \ ,\ t_\eta(C_3\bsl{r}) = t_\eta(\bsl{r}) e^{-\ii \eta \frac{2\pi}{3}} \\ 
& C_{2x}: \ V_{\eta,l}(C_{2x}\bsl{r}) =  V_{\eta,\bar{l}}(\bsl{r}) \ ,\ t_\eta^\dagger(C_{2x}\bsl{r}) = t_\eta(\bsl{r}) \\ 
& \TR: \ V_{-\eta,l}(\bsl{r}) =  V_{\eta,l}^*(\bsl{r}) \ ,\ t_{-\eta}(\bsl{r}) = t_{\eta}^*(\bsl{r}) \\ 
& T_{\bsl{R}_M}: V_{\eta,l}(\bsl{r}+\bsl{R}_M) =  V_{\eta,l}(\bsl{r}) \\
& \qquad \ \ t_\eta(\bsl{r}+\bsl{R}_M) = t_\eta(\bsl{r}) e^{\ii \eta \bsl{q}_1 \cdot \bsl{R}_M }
}
By keeping the first harmonics, the symmetry requirements lead to 
\bea
\label{eq:2HVr}
& V_{\eta,l}(\bsl{r}) = V e^{ - \ii \psi} \sum_{i=1,2,3} e^{\ii \bsl{g}_i \cdot \bsl{r} } + V e^{ \ii \psi} \sum_{i=1,2,3} e^{-\ii \bsl{g}_i \cdot \bsl{r}} \\
& t_\eta(\bsl{r}) = w \sum_{i=1,2,3} e^{\eta \ii  (i-1)\frac{2\pi}{3} } e^{ - \eta \ii \bsl{q}_i \cdot \bsl{r} } \ .
\eea
As can be checked from the expressions above, this lowest-harmonics model has effective anti-unitary symmetry
\eq{
\label{eq:TR_eff_AB_K_valley}
\cc c^\dagger_{\eta,l,\bsl{r}} \cc^{-1} = c^\dagger_{\eta,\bar{l},\bsl{r}}
}
which flips the layers and acts like time-reversal. This symmetry can be anticipated since the opposite layers have opposite spins in the twisted AB stacking. 
Adding displacement field would induce an energy difference between the two layers, \ie,
\eq{
H_{\eta,\varepsilon} =  \int d^2 r \sum_{l} c^\dagger_{\eta,l,\bsl{r}}c_{\eta,l,\bsl{r}} (-)^l \frac{\varepsilon}{2}\ ,
}
which breaks the effective TR symmetry, as well as the $C_{2x}$ symmetry.
Including higher harmonics for the intralayer and interlatyer terms may break the effective TR symmetry.

Similar to the discussion in \appref{app:Kvalley_AA}, we can express the Hamiltonian in the momentum space based on the Fourier transformation in \eqnref{eq:Kvalley_FT}, leading to 
\eq{
\label{eq:Kvalley_H_momentum_space_general_AB}
H_{\eta,0}^{AB} = \sum_{\bsl{k}} \sum_{\bsl{Q},\bsl{Q}'\in \Q} c^\dagger_{\eta,\bsl{k},\bsl{Q}} h_{\eta,\bsl{Q}\bsl{Q}'}^{AB}(\bsl{k})c_{\eta,\bsl{k},\bsl{Q}'}\ ,
}
where $c^\dagger_{\eta,\bsl{k},\bsl{Q}}$ is defined in \eqnref{eq:Kvalley_moire_basis_momentum_space}, and $\Q$ is defined right below \eqnref{eq:Kvalley_H_momentum_space_general_AA}.
Specifically, $h_{\eta,\bsl{Q}\bsl{Q}'}^{AB}(\bsl{k})$ reads
\eqa{
\label{eq:Kvalley_model_AB}
h_{\eta,\bsl{Q}\bsl{Q}'}^{AB}(\bsl{k}) 
&   =  \delta_{\bsl{Q}\bsl{Q}'} \left( \frac{- \hbar^2 \left( \bsl{k} -\bsl{Q} \right)^2}{ 2 m^*} + \frac{\varepsilon}{2} \eta (-)^{\bsl{Q}} \right) \\
& \quad + V \sum_{i=1}^3 \sum_{\alpha=\pm} \left[  e^{- \alpha \ii \psi} \delta_{\bsl{Q}+\alpha \bsl{g}_i,\bsl{Q}'}  \right]  \\
& \quad + w \sum_{i=1}^{3}  \sum_{\alpha=\pm} \left[  e^{\ii   \alpha \frac{2\pi (i-1)}{3} }\delta_{\bsl{Q},\bsl{Q}'+  \alpha \bsl{q}_i} \right] \ ,
}
where we have included the displacement field, and $w$ is chosen to be real by picking the relative phase between the two layers, and $(-)^{\bsl{Q}} =\pm1$ for $\bsl{Q}\in\{ \bsl{G}_M \pm \bsl{q}_1 \}$.
We can see that without the displacement field, $h_{\eta,\bsl{Q}\bsl{Q}'}^{AB}(\bsl{k})$ is the same for two valleys due to the effective time-reversal symmetry.

\subsection{AB-Stacking: $\Gamma$ Valley}

In this part, we discuss the continuum model for the $\Gamma$ valley.
Thus, the basis of the moire model at $\Gamma$ valley for {\ABtmt} is the same as that for {\AAtmt} which is labelled as $\psi^\dagger_{\bsl{r},l,s}$ with $l$ being the layer index and $s=\uparrow/\downarrow$ labeling the spin. The important difference is that, in the $\Gamma$ valley, the momentum space lattice is triangular instead of a honeycomb at $\pm \K$ points.
The reps of the symmetries of the moir\'e model read
\eqa{
\label{eq:sym_rep_Gamma_AB}
& C_3 \psi^\dagger_{\bsl{r},l} C_3^{-1} = \psi^\dagger_{C_3 \bsl{r},l} e^{-\ii s_z \frac{\pi}{3} }\\
& C_{2x} \psi^\dagger_{\bsl{r},l} (C_{2x})^{-1} = \psi^\dagger_{C_{2x} \bsl{r},\bar{l}} (-\ii s_x) \\
& \TR \psi^\dagger_{\bsl{r},l} \TR^{-1} = \psi^\dagger_{ \bsl{r},l} \ii s_y\\
& T_{\bsl{R}_M} \psi^\dagger_{\bsl{r},l} T_{\bsl{R}_M}^{-1} = \psi^\dagger_{ \bsl{r}+\bsl{R}_M,l} \ ,
}
where $s_x,s_y,s_z$ are the Pauli matrices for the spin index, and $\bar{l} = t,b$ if $l=b,t$.

After choosing the kinetic term in the continuum model as the intralayer spin-independent $\nabla^2$ term similar to the AA-stacking case, the form of the continuum model reads
\eq{
\label{eq:Gamma_model_general_AB}
H_{\Gamma} = \int d^2 r \mat{ \psi^\dagger_{\bsl{r},b} &  \psi^\dagger_{\bsl{r},t} } h_{\Gamma}^{AB}(\bsl{r})
\mat{ \psi^\dagger_{\bsl{r},b}  \\  \psi^\dagger_{\bsl{r},t} }\ ,
}
where 
\eq{
\label{eq:hAB_Gamma_r}
h_{\Gamma}^{AB}(\bsl{r})=
\mat{
\frac{\hbar^2 \nabla^2}{2 m_\Gamma^*} s_0 + V_{\Gamma,b}(\bsl{r}) & t_{\Gamma}(\bsl{r})\\
t_{\Gamma}^\dagger (\bsl{r}) & \frac{\hbar^2 \nabla^2 }{2 m_\Gamma^*} s_0 + V_{\Gamma,t}(\bsl{r}) 
}+E_\Gamma\ ,
}
\eq{
\psi^\dagger_{\bsl{r},l} = ( \psi^\dagger_{\bsl{r},l,\uparrow} , \psi^\dagger_{\bsl{r},l,\downarrow} )\ ,
}
$t$ and $b$ correspond to the top and bottom layers, respectively, $E_\Gamma$ accounts for the energy difference between the $\Gamma$-valley and $\pm\K$-valley bands, and $V_{\Gamma,l}(\bsl{r})$ and $t_{\Gamma}(\bsl{r})$ are $2\times 2$ matrix functions.
Similar to the discussion for {\AAtmt}, we only include the terms up to the first harmonics in $V_{\Gamma,l}(\bsl{r})$, and the zeroth harmonics in the $t_{\Gamma}(\bsl{r})$.
Then, combined with the symmetry reps in \eqnref{eq:sym_rep_Gamma_AB}, we arrive at
\eqa{
\label{eq:Vt_AB_Gamma_Valley_sim}
& V_{\Gamma,l}(\bsl{r}) =  \left[\sum_{i=1}^3 V_{\Gamma} e^{\ii  \psi_{\Gamma}}  e^{\ii \bsl{g}_i\cdot\bsl{r} } +  \sum_{i=1}^3 V_{\Gamma} e^{-\ii  \psi_{\Gamma}}  e^{-\ii \bsl{g}_i\cdot\bsl{r} } \right] s_0\\
& t_{\Gamma}(\bsl{r}) = w_{\Gamma} s_0 \ .
}

With the simplification in \eqnref{eq:hAB_Gamma_r} and \eqnref{eq:Vt_AB_Gamma_Valley_sim}, the $\Gamma$-valley continuum model in \eqnref{eq:Gamma_model_general_AB} has an effective TR symmetry
\eq{
\label{eq:TR_eff_AB_Gamma_valley}
\cc \psi^\dagger_{\bsl{r},l} \cc^{-1} = \psi^\dagger_{\bsl{r}, \bar{l} }
}
arising because of the spin-layer locking in the AB stacked structure. Similar to the $\pm\K$-valley case, adding displacement breaks the effective inversion symmetry, as well as the $C_{2x}$ symmetry, where the displacement field term reads
\eq{
H_{\Gamma,\varepsilon} =  \int d^2 r \sum_{l} \psi^\dagger_{\bsl{r},l}\psi_{\bsl{r},l} (-)^l \frac{\varepsilon}{2}\ .
}
Including the first harmonics in $t_{\Gamma}(\bsl{r})$ and the higher harmonics in $V_{\Gamma,l}(\bsl{r})$ can also break the effective TR symmetry.

To express the Hamiltonian in the momentum space, we use the Fourier transformation of the basis, which reads
\eq{
\label{eq:Gammavalley_FT}
\psi^\dagger_{\bsl{r},l,s} = \frac{1}{\sqrt{\V}} \sum_{\bsl{k}} \sum_{\bsl{G}_M } e^{-\ii (\bsl{k} - \bsl{G}_M) \cdot \bsl{r} } \psi^\dagger_{\bsl{k}-\bsl{G}_M,l,s}\ .
}
As shown in \eqnref{eq:Vt_AB_Gamma_Valley_sim}, the Hamiltonian has the same form in the two spin subspaces.
As a result, $H_{\eta,0}$ reads
\eq{
H_{\Gamma,0}^{AB} = \sum_{\bsl{k},\bsl{G}_M,\bsl{G}_M'} \sum_{l,l',s} \psi^\dagger_{\bsl{k}-\bsl{G}_M,l,s} h_{ll',\bsl{G}_M\bsl{G}_M'}^{AB}(\bsl{k})\psi_{\bsl{k}-\bsl{G}_M',l',s}\ ,
}
where 
\eqa{
\label{eq:Gammavalley_model_AB}
& h_{ll,\bsl{G}_M\bsl{G}_M'}^{AB}(\bsl{k}) =  \left[ -\frac{\hbar^2 (\bsl{k}-\bsl{G})^2 }{2 m_\Gamma^*} + (-)^l \frac{\varepsilon}{2}  +E_\Gamma\right] \delta_{\bsl{G}_M, \bsl{G}_M'} \\
& \qquad + V_{\Gamma}  \sum_{i=1}^3 \left[ e^{\ii  \psi_{\Gamma}}  \delta_{\bsl{G}_M, \bsl{G}_M'- \bsl{g}_i }  +  e^{-\ii  \psi_{\Gamma}}   \delta_{\bsl{G}_M, \bsl{G}_M' + \bsl{g}_i } \right] \\
& h_{l\bar{l},\bsl{G}_M\bsl{G}_M'}^{AB}(\bsl{k}) = w_{\Gamma}  \delta_{\bsl{G}_M, \bsl{G}_M'}\ ,
}
and we have included the displacement field.

\subsection{AB-Stacking: Fitting to the DFT Data}
\label{app:ABFitting}
As shown in \figref{fig:ABrelaxevolution}, both the $\pm\K$-valley and the $\Gamma$-valley valence bands are close to the charge neutrality.
Therefore, we only use both the $\pm \K$-valley model (\eqnref{eq:Kvalley_model_AB}) and the $\Gamma$-valley model (\eqnref{eq:Gammavalley_model_AB}) in the fitting.
We mainly fit to the top 4 valence bands (2 in each valley) for the $\pm \K$-valley model (\eqnref{eq:Kvalley_model_AB}) and top 6 valence bands in the $\Gamma$ valley; similar to the case of AA stacking, we fit to the DFT bands at $\theta=3.89^\circ$.
The good match between those DFT bands and those from the models is shown in \figref{fig:fitting_main} (b) and \figref{fig:fitting_BZ} (b), where the model parameter values obtained from the fitting are shown in \tabref{tab:AB_parameters_DFT}.
In the following, we provide more details for the fitting for the $\pm \K$-valley model (\eqnref{eq:Kvalley_model_AB}) and the $\Gamma$-valley model, separately.

\subsubsection{$\pm \K$-valley}

We choose the value of $m^*$ to be the effective mass ($\sim 0.62 m_e$) of the top valence band for the untwisted AB-stacking bilayer structure around $\K$, as shown in \figref{fig:bs-mono-AA-2H}.
The value of effective mass around $\K$ is similar for the monolayer {\mt} and the untwisted AB-stacking bilayer {\mt}.

Owing to the effective TR symmetry \eqnref{eq:TR_eff_AB_K_valley} and the true microscopic TR symmetry, the $-\K$ valley has the same bands as the $+\K$ valley, leading to at least double degeneracy of each band from the $\pm\K$-valley continuum model.
Along the $\Gamma_M-M_M$ line, the bands from the $\pm\K$-valley continuum model have 4-fold degeneracy because we set the interlayer coupling to be zero ($w=0$), which is justified by the nearly 4-fold degeneracy for the top 4 DFT valence bands along $\Gamma_M-M_M$, as shown in \figref{fig:fitting_main}.
The small $w=0$ can be understood as the follows.
The two layers in one valley now have opposite spin; owing to the spin U(1) symmetry for the low-energy states near $\pm \K$ valleys in monolayer {\mt}, we expect the spin $U(1)$ symmetry is approximately preserved in {\tmt}, which means the interlayer coupling is very small for the AB-stacking.
The zero interlayer coupling makes the eigenstates have well-defined valley and layer/spin.
As a result, the two states with the same spin are degenerate at $\K_M$, since the combination of the effective TR symmetry and the TR symmetry leaves the spin invariant.
This is consistent with the DFT result in \figref{fig:irreps-2H-vbs}.
In \figref{fig:C3eigenvalues_highsymline} (b), we show the $C_3$ eigenvalues of the bands from the $\pm\K$-valley model, which is consistent with \figref{fig:irreps-2H-vbs} and which shows that each of the top four valence bands is in the $A_1@1a$ atomic limit, representing an atomic $s$ orbital at the origin of the center of moir\'e Wigner-Seitz unit cell. 

\subsubsection{$\Gamma$-valley}

In our fitting, we choose the values of $m_\Gamma^*$ to be the effective mass of the top valence band for the untwisted AB-stacking bilayer structure around $\Gamma$ in \figref{fig:bs-mono-AA-2H}. (There is some anisotropy, so in practice we round the effective mass to $10m_e$. The resulting bands show little dependence on the precise value of the mass.)
Although the value of effective mass around $\K$ is similar for the monolayer {\mt} and the untwisted AB-stacking bilayer {\mt}, it differs a lot around $\Gamma$ as shown in \figref{fig:bs-mono-AA-2H}.
The change of the $\Gamma$ effective mass from the monolayer {\mt} to the untwisted AB-stacking bilayer {\mt} comes from the interlayer coupling between the monolayer conduction and valence bands.
This effect can only be taken into account in the $\Gamma$-valley continuum model by changing the value of $m^*$, since the the monolayer conduction bands are not explicitly included in the $\Gamma$-valley continuum model.
Therefore, we do not take the values of $m_\Gamma^*$ from the monolayer; instead, we round it to the $\Gamma$ effective mass in the untwisted AB-stacking bilayer {\mt}.
Furthermore, we determine the interlayer coupling $w_\Gamma$ for the $\Gamma$-valley model based on the interlayer coupling at $\Gamma$ for the untwisted AB-stacking bilayer structure, as shown in \figref{fig:bs-mono-AA-2H}.

The $\Gamma$-valley continuum model in \eqnref{eq:Gamma_model_general_AB} preserves spin $\U(1)$. 
Owing to the effective TR symmetry \eqnref{eq:TR_eff_AB_K_valley} and the TR symmetry, the spin-down bands are the same as the spin-up bands, leading to at least double degeneracy of each band from the $\Gamma$-valley continuum model.
As shown in \figref{fig:C3eigenvalues_highsymline} (c), the $C_3$ eigenvalues of the bands from the $\pm\K$-valley model are consistent with the DFT calculation shown in \figref{fig:irreps-2H-vbs}.

The $\Gamma$-valley bands are extremely flat as shown in \figref{fig:fitting_main}.
To understand it, first we note that the interlayer coupling $w_{\Gamma}$ is very large (shown in \tabref{tab:AB_parameters_DFT}), allowing us to project the model into $\psi^\dagger_{\bsl{r},s} = \frac{1}{\sqrt{2}}(\psi^\dagger_{t,\bsl{r},s}+\psi^\dagger_{b,\bsl{r},s})$ basis.
(Note that we have rotated the relative phase between $\psi^\dagger_{t,\bsl{r},s}$ and $\psi^\dagger_{b,\bsl{r},s}$ to make $w_\Gamma$ positive.)
Then, the $\Gamma$-valley continuum model in \eqnref{eq:hAB_Gamma_r} with zero displacement field becomes
\eq{
\label{eq:htilde_AB_r}
\tilde{h}^{AB}(\bsl{r}) =  \frac{\hbar^2 \nabla^2 }{2 m_\Gamma^*}  + w_\Gamma +E_\Gamma + V_{\Gamma}(\bsl{r}) \ ,
}
where the spin degeneracy is implicit, and
\eqa{
V_{\Gamma}(\bsl{r}) & =  \sum_{i=1}^3 V_{\Gamma} e^{\ii  \psi_{\Gamma}}  e^{\ii \bsl{g}_i\cdot\bsl{r} } +  \sum_{i=1}^3 V_{\Gamma} e^{-\ii  \psi_{\Gamma}}  e^{-\ii \bsl{g}_i\cdot\bsl{r} } \\
& =  2 V_{\Gamma} \sum_{i=1}^3 \cos( \bsl{g}_i\cdot\bsl{r} + \psi_{\Gamma} ) \ .
}
For convenience of the discussion, let us consider the hole Hamiltonian 
\eq{
-\tilde{h}^{AB}(\bsl{r}) =  -\frac{\hbar^2 \nabla^2 }{2 m_\Gamma^*}  - w_\Gamma - E_\Gamma - V_{\Gamma}(\bsl{r}) \ ;
}
the highest bands of $\tilde{h}^{AB}(\bsl{r})$ are lowest bands of $-\tilde{h}^{AB}(\bsl{r})$.
With $\psi=0$ and $V_\Gamma >0$ in \tabref{tab:AB_parameters_DFT}, the minimum of the potential $-V_{\Gamma}(\bsl{r})$ is at the moir\'e lattice points $\bsl{r}=\bsl{R}_M$, which has value $-V_{\Gamma}(\bsl{R}_M)=-6V_{\Gamma}$.
To go from one moir\'e lattice point $\bsl{R}_M$ to its nearest neighbors $\bsl{R}_M$ to $\bsl{R}_M + \bsl{a}_{M,1}$, the minimum energy barrier is $8 V_{\Gamma} = 576$meV.
The energy barrier is much larger than the kinetic energy term estimated for $m^*_{\Gamma} = 10 m_e$: $\frac{\hbar^2 |\bsl{q}_1|^2}{2 m^*_{\Gamma}} = 2.49$meV.
As a result, the lowest-energy states of $-\tilde{h}^{AB}(\bsl{r})$ are bounded around $\bsl{R}_M$; then we can expand $- V_{\Gamma}(\bsl{r})$ around $\bsl{R}_M$ to the second order of $\bsl{r}-\bsl{R}_M$ (the first order of $\bsl{r}-\bsl{R}_M$ is forbidden by the $C_3$ symmetry):
\eqa{
- V_{\Gamma}(\bsl{r}) & =   V_\Gamma  \sum_{i} (\bsl{g}_i\cdot\bsl{r})^2 + O(|\bsl{r}|^4) \\
& =  \frac{3}{2} V_\Gamma |b_{M,1}|^2 \left|\bsl{r}\right|^2 + O(|\bsl{r}|^4)\ .
}
Then, the effective Hamiltonian for those lowest-energy states can be approximated by an array of decoupled harmonic oscillators centered around the $\bsl{R}_M$.
Harmonic oscillators centered around different $\bsl{R}_M$'s are related by the moir\'e lattice translations, and thus have the same spectrum.
As a result, we have a set of flat bands with energies given by solving
\eq{
h_{\Gamma,harmonic}^{AB}(\bsl{r}) = -\frac{\hbar^2 \nabla^2 }{2 m_\Gamma^*}  + w_\Gamma - E_\Gamma + \frac{3}{2} V_\Gamma |b_{M,1}|^2 \left|\bsl{r}\right|^2
}
which yields 
\eq{
\label{eq:E_Gamma_harmonics_AB}
E_{\Gamma,harmonic}^{AB}(n_x , n_y) =   w_\Gamma  E_\Gamma - \hbar \omega - (n_x + n_y) \frac{\hbar \omega }{2}\ ,
}
where $n_x,n_y=0,1,2,3,...$ and 
\eq{
\hbar\omega = \hbar \sqrt{\frac{3 V_\Gamma |b_{M,1}|^2 }{m_\Gamma^*}} = 56.76\text{meV}\ .
}
This spectrum tells us (i) after including spin, the highest valence band $(n_x,n_y)=(0,0)$ is doubly degenerate and the second highest valence band $(n_x,n_y)=(1,0),(0,1)$ is 4-fold degenerate, (ii) the highest and second valence bands have spinless $C_3$ eigenvalues $1$ and $e^{\pm \ii 2\pi/3}$, respectively,  and (iii) the energy difference between the highest valence band and the second highest valence is $\hbar\omega= 56.76\text{meV}$.
All of the features are consistent with the DFT calculation.
In \figref{fig:fitting_BZ} (b), the flat bands from the $\Gamma$ valley around $-30$meV and $-86$meV have near 2-fold degeneracy (spinless $C_3$ eigenvalue $1$) and near 4-fold degeneracy (spinless $C_3$ eigenvalue $e^{\pm \ii 2\pi/3}$), and the energy difference between them is $56.74$meV. 

\begin{figure}[t]
    \centering
    \includegraphics[width=0.9\columnwidth]{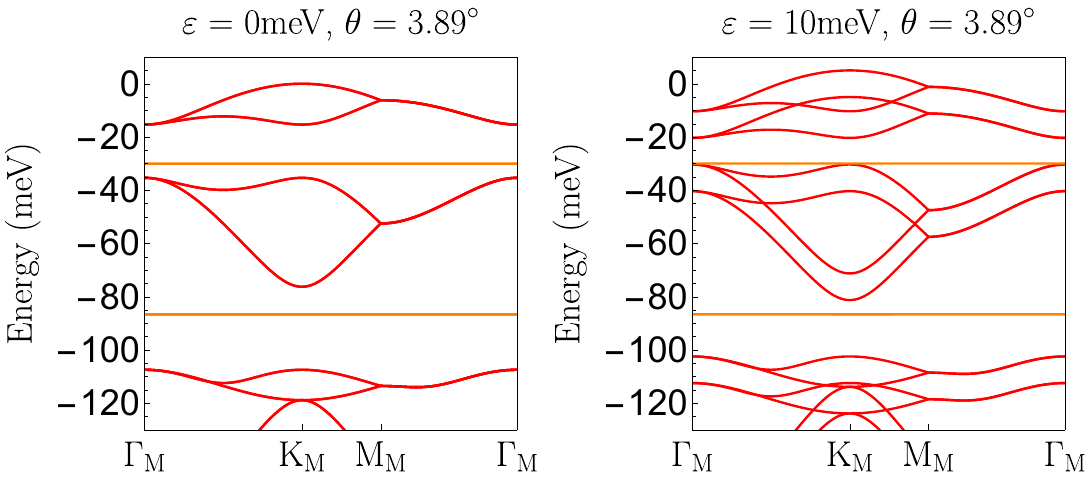}
    \caption{The evolution of the bands of the AB-stacking {\tmt} based on the continuum model with parameter values in \tabref{tab:AB_parameters_DFT}.
    The red and orange lines come from the $\pm\K$ valley and the $\Gamma$ valley, respectively.
    }
    \label{fig:AB_Efield}
\end{figure}

At last, we discuss the effect of the displacement field.
As shown in \figref{fig:AB_Efield}, the effect of the displacement field on the $\pm K$-valley bands are just to shift the bands from different layers relative to each other, since the valley is a good quantum number due to the zero interlayer coupling (see \tabref{tab:AB_parameters_DFT}).
On the other hand, the effect of the displacement field on the low-energy $\Gamma$-valley bands is negligible, which is consistent with the fact that the very large interlayer (see \tabref{tab:AB_parameters_DFT}) coupling make the eigenstates equally distributed between the two layers.

\subsection{AB-Stacking Electron Bands}
\label{app:electronpocket}

For completeness, we also briefly discuss the conduction bands although they are not the focus of this work. The Fermi surface of the electron bands in the untwisted structure is more complex than the valence bands (see \figref{fig:bs-mono-AA-2H}), exhibiting an electron-like pocket along the $\Gamma \K$ line which is below the $\pm \K$-valley states in both bilayer configurations. Let us denote the center of the pocket by $s \mbf{K}$, $s \sim .7$, along with its $C_3$- and $\mathcal{T}$-related partners. Thus in total there are 6 pockets in the untwisted BZ. \figref{fig:bs-mono-AA-2H} shows that there are only two low-energy states (spin $\u$ and spin $\d$) at $s \mbf{K}$, unlike the four total states (spin $\u$ and spin $\d$ in both layers) at $\mbf{K}$. This can be understood from the simple model $\frac{\mbf{k}^2}{2m_p} \tau_0 - w_p \tau_1$ where $\tau$ is a layer Pauli matrix, $\mbf{k}$ is the momentum measured from the Fermi pockets with mass $m_p$, and $w_p$ is the effective inter-layer coupling, which splits the two layers into bonding/anti-bonding states with energy $\mbf{k}^2/2m_p \mp w_p$. 

Comparing the monolayer and bilayer band structures in \figref{fig:bs-mono-AA-2H}, we estimate $w_p \sim 200$meV. Since this is a large hybridization that removes the anti-bonding state from the low-energy bands, an effective moir\'e model on the bonding state is
\bea
\label{eq:uglybandsmodel}
H_{\mbf{p}_i} &= \frac{(- i \nabla + \mbf{p}_i)^2}{2 m_p} + V_p(\mbf{r})
\eea
where $V_p(\mbf{r})$ is a moir\'e-periodic potential and $\mbf{p}_j = R(\frac{2\pi j}{6}) s \mbf{K}$ is the center of the pocket. Note that the momentum-space origin of $H_{\mbf{p}_i}$, $\mbf{p}_i$ mod $\mbf{G}_M$, depends sensitively on the value of $s$ and $\th$ and is generically off the moir\'e high-symmetry points. The resulting shifting and folding of the $\mbf{p}_i$ pocket onto the moir\'e BZ explains the non-uniform evolution of the conduction bands in \figref{fig:AArelaxevolution} and \ref{fig:ABrelaxevolution} as a function of twist angle. In comparison, the valence bands which originate at the K points converge smoothly.

\begin{figure}
    \centering
    \includegraphics[width=1\linewidth]{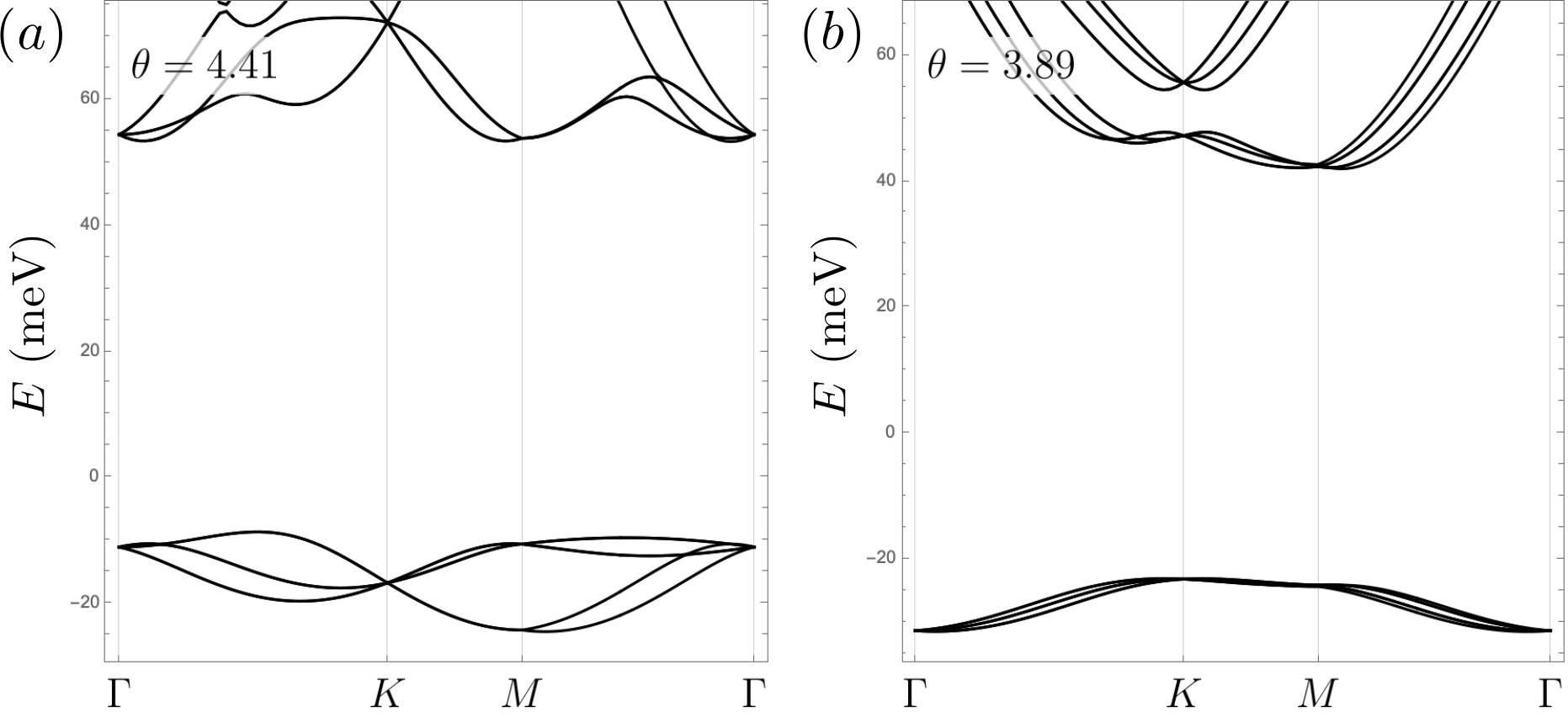}
    \caption{Conduction bands in the AB-stacked structure using $s = .68$ for $\th = 4.41^\circ$ (a) and $\th = 3.89^\circ$ (b). Note that there are 6 low energy bands (per valley) in each case, although in (a) they are nearly degenerate. 
    The features of the band structures are well-matched by the \emph{lowest} 12 bands in ab-initio in Fig. \ref{fig:ABrelaxevolution} between both angles. One should note that the ab-initio band structures also show bands from the $K$ valley at slightly higher energies.}
    \label{fig:ABmodelelectron}
\end{figure}

We show band structures at $\th = 4.41^\circ, 3.89^\circ$ in Fig. \ref{fig:ABmodelelectron} using $s = .68$ as an example to illustrate the strong effect of $\mbf{p}_i$. Here we chose similar parameters to the $K$-valley model in Tab. \ref{tab:AB_parameters_DFT}, namely $m_p = .62 m_e$, $V = 26.5$meV (half the $K$ valley value), and $\psi = -52^\circ$. We see that Fig. \ref{fig:ABmodelelectron} qualitatively matches the lowest bands in Fig. \ref{fig:ABrelaxevolution}. Since the only symmetry preserved at $\mbf{p}_i$ is $C_{2x}$, symmetry does not strongly constrain the form of Eq. \ref{eq:uglybandsmodel}. Nevertheless, the minimal model here successfully reproduces many features of the bands. 

\bibliography{refs}

\begin{thebibliography}{85}%
\makeatletter
\providecommand \@ifxundefined [1]{%
 \@ifx{#1\undefined}
}%
\providecommand \@ifnum [1]{%
 \ifnum #1\expandafter \@firstoftwo
 \else \expandafter \@secondoftwo
 \fi
}%
\providecommand \@ifx [1]{%
 \ifx #1\expandafter \@firstoftwo
 \else \expandafter \@secondoftwo
 \fi
}%
\providecommand \natexlab [1]{#1}%
\providecommand \enquote  [1]{``#1''}%
\providecommand \bibnamefont  [1]{#1}%
\providecommand \bibfnamefont [1]{#1}%
\providecommand \citenamefont [1]{#1}%
\providecommand \href@noop [0]{\@secondoftwo}%
\providecommand \href [0]{\begingroup \@sanitize@url \@href}%
\providecommand \@href[1]{\@@startlink{#1}\@@href}%
\providecommand \@@href[1]{\endgroup#1\@@endlink}%
\providecommand \@sanitize@url [0]{\catcode `\\12\catcode `\$12\catcode `\&12\catcode `\#12\catcode `\^12\catcode `\_12\catcode `\%12\relax}%
\providecommand \@@startlink[1]{}%
\providecommand \@@endlink[0]{}%
\providecommand \url  [0]{\begingroup\@sanitize@url \@url }%
\providecommand \@url [1]{\endgroup\@href {#1}{\urlprefix }}%
\providecommand \urlprefix  [0]{URL }%
\providecommand \Eprint [0]{\href }%
\providecommand \doibase [0]{http://dx.doi.org/}%
\providecommand \selectlanguage [0]{\@gobble}%
\providecommand \bibinfo  [0]{\@secondoftwo}%
\providecommand \bibfield  [0]{\@secondoftwo}%
\providecommand \translation [1]{[#1]}%
\providecommand \BibitemOpen [0]{}%
\providecommand \bibitemStop [0]{}%
\providecommand \bibitemNoStop [0]{.\EOS\space}%
\providecommand \EOS [0]{\spacefactor3000\relax}%
\providecommand \BibitemShut  [1]{\csname bibitem#1\endcsname}%
\let\auto@bib@innerbib\@empty
\bibitem [{\citenamefont {Neupert}\ \emph {et~al.}(2011)\citenamefont {Neupert}, \citenamefont {Santos}, \citenamefont {Chamon},\ and\ \citenamefont {Mudry}}]{neupert}%
  \BibitemOpen
  \bibfield  {author} {\bibinfo {author} {\bibfnamefont {T.}~\bibnamefont {Neupert}}, \bibinfo {author} {\bibfnamefont {L.}~\bibnamefont {Santos}}, \bibinfo {author} {\bibfnamefont {C.}~\bibnamefont {Chamon}}, \ and\ \bibinfo {author} {\bibfnamefont {C.}~\bibnamefont {Mudry}},\ }\href {\doibase 10.1103/PhysRevLett.106.236804} {\bibfield  {journal} {\bibinfo  {journal} {Phys. Rev. Lett.}\ }\textbf {\bibinfo {volume} {106}},\ \bibinfo {pages} {236804} (\bibinfo {year} {2011})}\BibitemShut {NoStop}%
\bibitem [{\citenamefont {{Sheng}}\ \emph {et~al.}(2011)\citenamefont {{Sheng}}, \citenamefont {{Gu}}, \citenamefont {{Sun}},\ and\ \citenamefont {{Sheng}}}]{sheng}%
  \BibitemOpen
  \bibfield  {author} {\bibinfo {author} {\bibfnamefont {D.~N.}\ \bibnamefont {{Sheng}}}, \bibinfo {author} {\bibfnamefont {Z.-C.}\ \bibnamefont {{Gu}}}, \bibinfo {author} {\bibfnamefont {K.}~\bibnamefont {{Sun}}}, \ and\ \bibinfo {author} {\bibfnamefont {L.}~\bibnamefont {{Sheng}}},\ }\href {\doibase 10.1038/ncomms1380} {\bibfield  {journal} {\bibinfo  {journal} {Nature Communications}\ }\textbf {\bibinfo {volume} {2}},\ \bibinfo {eid} {389} (\bibinfo {year} {2011})},\ \Eprint {http://arxiv.org/abs/1102.2658} {arXiv:1102.2658 [cond-mat.str-el]} \BibitemShut {NoStop}%
\bibitem [{\citenamefont {Regnault}\ and\ \citenamefont {Bernevig}(2011)}]{regnault}%
  \BibitemOpen
  \bibfield  {author} {\bibinfo {author} {\bibfnamefont {N.}~\bibnamefont {Regnault}}\ and\ \bibinfo {author} {\bibfnamefont {B.~A.}\ \bibnamefont {Bernevig}},\ }\href {\doibase 10.1103/PhysRevX.1.021014} {\bibfield  {journal} {\bibinfo  {journal} {Phys. Rev. X}\ }\textbf {\bibinfo {volume} {1}},\ \bibinfo {pages} {021014} (\bibinfo {year} {2011})}\BibitemShut {NoStop}%
\bibitem [{\citenamefont {Tang}\ \emph {et~al.}(2011)\citenamefont {Tang}, \citenamefont {Mei},\ and\ \citenamefont {Wen}}]{Tang11}%
  \BibitemOpen
  \bibfield  {author} {\bibinfo {author} {\bibfnamefont {E.}~\bibnamefont {Tang}}, \bibinfo {author} {\bibfnamefont {J.-W.}\ \bibnamefont {Mei}}, \ and\ \bibinfo {author} {\bibfnamefont {X.-G.}\ \bibnamefont {Wen}},\ }\href {\doibase 10.1103/PhysRevLett.106.236802} {\bibfield  {journal} {\bibinfo  {journal} {Phys. Rev. Lett.}\ }\textbf {\bibinfo {volume} {106}},\ \bibinfo {pages} {236802} (\bibinfo {year} {2011})}\BibitemShut {NoStop}%
\bibitem [{\citenamefont {Sun}\ \emph {et~al.}(2011)\citenamefont {Sun}, \citenamefont {Gu}, \citenamefont {Katsura},\ and\ \citenamefont {Das~Sarma}}]{Sun2011}%
  \BibitemOpen
  \bibfield  {author} {\bibinfo {author} {\bibfnamefont {K.}~\bibnamefont {Sun}}, \bibinfo {author} {\bibfnamefont {Z.}~\bibnamefont {Gu}}, \bibinfo {author} {\bibfnamefont {H.}~\bibnamefont {Katsura}}, \ and\ \bibinfo {author} {\bibfnamefont {S.}~\bibnamefont {Das~Sarma}},\ }\href {\doibase 10.1103/PhysRevLett.106.236803} {\bibfield  {journal} {\bibinfo  {journal} {Phys. Rev. Lett.}\ }\textbf {\bibinfo {volume} {106}},\ \bibinfo {pages} {236803} (\bibinfo {year} {2011})}\BibitemShut {NoStop}%
\bibitem [{\citenamefont {Bergholtz}\ and\ \citenamefont {Liu}(2013)}]{Bergholtz13}%
  \BibitemOpen
  \bibfield  {author} {\bibinfo {author} {\bibfnamefont {E.~J.}\ \bibnamefont {Bergholtz}}\ and\ \bibinfo {author} {\bibfnamefont {Z.}~\bibnamefont {Liu}},\ }\href {\doibase 10.1142/S021797921330017X} {\bibfield  {journal} {\bibinfo  {journal} {International Journal of Modern Physics B}\ }\textbf {\bibinfo {volume} {27}},\ \bibinfo {pages} {1330017} (\bibinfo {year} {2013})},\ \Eprint {http://arxiv.org/abs/https://doi.org/10.1142/S021797921330017X} {https://doi.org/10.1142/S021797921330017X} \BibitemShut {NoStop}%
\bibitem [{\citenamefont {Parameswaran}\ \emph {et~al.}(2013)\citenamefont {Parameswaran}, \citenamefont {Roy},\ and\ \citenamefont {Sondhi}}]{Parameswaran13}%
  \BibitemOpen
  \bibfield  {author} {\bibinfo {author} {\bibfnamefont {S.~A.}\ \bibnamefont {Parameswaran}}, \bibinfo {author} {\bibfnamefont {R.}~\bibnamefont {Roy}}, \ and\ \bibinfo {author} {\bibfnamefont {S.~L.}\ \bibnamefont {Sondhi}},\ }\href {\doibase https://doi.org/10.1016/j.crhy.2013.04.003} {\bibfield  {journal} {\bibinfo  {journal} {Comptes Rendus Physique}\ }\textbf {\bibinfo {volume} {14}},\ \bibinfo {pages} {816} (\bibinfo {year} {2013})},\ \bibinfo {note} {topological insulators / Isolants topologiques}\BibitemShut {NoStop}%
\bibitem [{\citenamefont {Abouelkomsan}\ \emph {et~al.}(2020)\citenamefont {Abouelkomsan}, \citenamefont {Liu},\ and\ \citenamefont {Bergholtz}}]{Abouelkomsan2020FCIMoire}%
  \BibitemOpen
  \bibfield  {author} {\bibinfo {author} {\bibfnamefont {A.}~\bibnamefont {Abouelkomsan}}, \bibinfo {author} {\bibfnamefont {Z.}~\bibnamefont {Liu}}, \ and\ \bibinfo {author} {\bibfnamefont {E.~J.}\ \bibnamefont {Bergholtz}},\ }\href {\doibase 10.1103/PhysRevLett.124.106803} {\bibfield  {journal} {\bibinfo  {journal} {Phys. Rev. Lett.}\ }\textbf {\bibinfo {volume} {124}},\ \bibinfo {pages} {106803} (\bibinfo {year} {2020})}\BibitemShut {NoStop}%
\bibitem [{\citenamefont {Ledwith}\ \emph {et~al.}(2020)\citenamefont {Ledwith}, \citenamefont {Tarnopolsky}, \citenamefont {Khalaf},\ and\ \citenamefont {Vishwanath}}]{Vishwanath2020FCITBG}%
  \BibitemOpen
  \bibfield  {author} {\bibinfo {author} {\bibfnamefont {P.~J.}\ \bibnamefont {Ledwith}}, \bibinfo {author} {\bibfnamefont {G.}~\bibnamefont {Tarnopolsky}}, \bibinfo {author} {\bibfnamefont {E.}~\bibnamefont {Khalaf}}, \ and\ \bibinfo {author} {\bibfnamefont {A.}~\bibnamefont {Vishwanath}},\ }\href {\doibase 10.1103/PhysRevResearch.2.023237} {\bibfield  {journal} {\bibinfo  {journal} {Phys. Rev. Res.}\ }\textbf {\bibinfo {volume} {2}},\ \bibinfo {pages} {023237} (\bibinfo {year} {2020})}\BibitemShut {NoStop}%
\bibitem [{\citenamefont {Repellin}\ and\ \citenamefont {Senthil}(2020)}]{RepellinFCITBG}%
  \BibitemOpen
  \bibfield  {author} {\bibinfo {author} {\bibfnamefont {C.}~\bibnamefont {Repellin}}\ and\ \bibinfo {author} {\bibfnamefont {T.}~\bibnamefont {Senthil}},\ }\href {\doibase 10.1103/PhysRevResearch.2.023238} {\bibfield  {journal} {\bibinfo  {journal} {Phys. Rev. Res.}\ }\textbf {\bibinfo {volume} {2}},\ \bibinfo {pages} {023238} (\bibinfo {year} {2020})}\BibitemShut {NoStop}%
\bibitem [{\citenamefont {Parker}\ \emph {et~al.}(2021)\citenamefont {Parker}, \citenamefont {Ledwith}, \citenamefont {Khalaf}, \citenamefont {Soejima}, \citenamefont {Hauschild}, \citenamefont {Xie}, \citenamefont {Pierce}, \citenamefont {Zaletel}, \citenamefont {Yacoby},\ and\ \citenamefont {Vishwanath}}]{Parker2021fieldtunedFCITBG}%
  \BibitemOpen
  \bibfield  {author} {\bibinfo {author} {\bibfnamefont {D.}~\bibnamefont {Parker}}, \bibinfo {author} {\bibfnamefont {P.}~\bibnamefont {Ledwith}}, \bibinfo {author} {\bibfnamefont {E.}~\bibnamefont {Khalaf}}, \bibinfo {author} {\bibfnamefont {T.}~\bibnamefont {Soejima}}, \bibinfo {author} {\bibfnamefont {J.}~\bibnamefont {Hauschild}}, \bibinfo {author} {\bibfnamefont {Y.}~\bibnamefont {Xie}}, \bibinfo {author} {\bibfnamefont {A.}~\bibnamefont {Pierce}}, \bibinfo {author} {\bibfnamefont {M.~P.}\ \bibnamefont {Zaletel}}, \bibinfo {author} {\bibfnamefont {A.}~\bibnamefont {Yacoby}}, \ and\ \bibinfo {author} {\bibfnamefont {A.}~\bibnamefont {Vishwanath}},\ }\href@noop {} {\enquote {\bibinfo {title} {Field-tuned and zero-field fractional chern insulators in magic angle graphene},}\ } (\bibinfo {year} {2021}),\ \Eprint {http://arxiv.org/abs/2112.13837} {arXiv:2112.13837 [cond-mat.str-el]} \BibitemShut {NoStop}%
\bibitem [{\citenamefont {Wilhelm}\ \emph {et~al.}(2021)\citenamefont {Wilhelm}, \citenamefont {Lang},\ and\ \citenamefont {L\"auchli}}]{Wilhelm2021FCITBG}%
  \BibitemOpen
  \bibfield  {author} {\bibinfo {author} {\bibfnamefont {P.}~\bibnamefont {Wilhelm}}, \bibinfo {author} {\bibfnamefont {T.~C.}\ \bibnamefont {Lang}}, \ and\ \bibinfo {author} {\bibfnamefont {A.~M.}\ \bibnamefont {L\"auchli}},\ }\href {\doibase 10.1103/PhysRevB.103.125406} {\bibfield  {journal} {\bibinfo  {journal} {Phys. Rev. B}\ }\textbf {\bibinfo {volume} {103}},\ \bibinfo {pages} {125406} (\bibinfo {year} {2021})}\BibitemShut {NoStop}%
\bibitem [{\citenamefont {Sheffer}\ and\ \citenamefont {Stern}(2021)}]{Stern2021FCITBG}%
  \BibitemOpen
  \bibfield  {author} {\bibinfo {author} {\bibfnamefont {Y.}~\bibnamefont {Sheffer}}\ and\ \bibinfo {author} {\bibfnamefont {A.}~\bibnamefont {Stern}},\ }\href {\doibase 10.1103/PhysRevB.104.L121405} {\bibfield  {journal} {\bibinfo  {journal} {Phys. Rev. B}\ }\textbf {\bibinfo {volume} {104}},\ \bibinfo {pages} {L121405} (\bibinfo {year} {2021})}\BibitemShut {NoStop}%
\bibitem [{\citenamefont {Li}\ \emph {et~al.}(2021)\citenamefont {Li}, \citenamefont {Kumar}, \citenamefont {Sun},\ and\ \citenamefont {Lin}}]{Li2021FCItTMD}%
  \BibitemOpen
  \bibfield  {author} {\bibinfo {author} {\bibfnamefont {H.}~\bibnamefont {Li}}, \bibinfo {author} {\bibfnamefont {U.}~\bibnamefont {Kumar}}, \bibinfo {author} {\bibfnamefont {K.}~\bibnamefont {Sun}}, \ and\ \bibinfo {author} {\bibfnamefont {S.-Z.}\ \bibnamefont {Lin}},\ }\href {\doibase 10.1103/PhysRevResearch.3.L032070} {\bibfield  {journal} {\bibinfo  {journal} {Phys. Rev. Res.}\ }\textbf {\bibinfo {volume} {3}},\ \bibinfo {pages} {L032070} (\bibinfo {year} {2021})}\BibitemShut {NoStop}%
\bibitem [{\citenamefont {Cr\'epel}\ and\ \citenamefont {Fu}(2023)}]{Crepel2022FCITMD}%
  \BibitemOpen
  \bibfield  {author} {\bibinfo {author} {\bibfnamefont {V.}~\bibnamefont {Cr\'epel}}\ and\ \bibinfo {author} {\bibfnamefont {L.}~\bibnamefont {Fu}},\ }\href {\doibase 10.1103/PhysRevB.107.L201109} {\bibfield  {journal} {\bibinfo  {journal} {Phys. Rev. B}\ }\textbf {\bibinfo {volume} {107}},\ \bibinfo {pages} {L201109} (\bibinfo {year} {2023})}\BibitemShut {NoStop}%
\bibitem [{\citenamefont {Dong}\ \emph {et~al.}(2022)\citenamefont {Dong}, \citenamefont {Wang},\ and\ \citenamefont {Fu}}]{Fu2022FCIDirac}%
  \BibitemOpen
  \bibfield  {author} {\bibinfo {author} {\bibfnamefont {J.}~\bibnamefont {Dong}}, \bibinfo {author} {\bibfnamefont {J.}~\bibnamefont {Wang}}, \ and\ \bibinfo {author} {\bibfnamefont {L.}~\bibnamefont {Fu}},\ }\href@noop {} {\enquote {\bibinfo {title} {Dirac electron under periodic magnetic field: Platform for fractional chern insulator and generalized wigner crystal},}\ } (\bibinfo {year} {2022}),\ \Eprint {http://arxiv.org/abs/2208.10516} {arXiv:2208.10516 [cond-mat.mes-hall]} \BibitemShut {NoStop}%
\bibitem [{\citenamefont {Abouelkomsan}\ \emph {et~al.}(2023{\natexlab{a}})\citenamefont {Abouelkomsan}, \citenamefont {Yang},\ and\ \citenamefont {Bergholtz}}]{Abouelkomsan2023FCITBG}%
  \BibitemOpen
  \bibfield  {author} {\bibinfo {author} {\bibfnamefont {A.}~\bibnamefont {Abouelkomsan}}, \bibinfo {author} {\bibfnamefont {K.}~\bibnamefont {Yang}}, \ and\ \bibinfo {author} {\bibfnamefont {E.~J.}\ \bibnamefont {Bergholtz}},\ }\href {\doibase 10.1103/PhysRevResearch.5.L012015} {\bibfield  {journal} {\bibinfo  {journal} {Phys. Rev. Res.}\ }\textbf {\bibinfo {volume} {5}},\ \bibinfo {pages} {L012015} (\bibinfo {year} {2023}{\natexlab{a}})}\BibitemShut {NoStop}%
\bibitem [{\citenamefont {{Morales-Dur{\'a}n}}\ \emph {et~al.}(2023{\natexlab{a}})\citenamefont {{Morales-Dur{\'a}n}}, \citenamefont {{Wang}}, \citenamefont {{Schleder}}, \citenamefont {{Angeli}}, \citenamefont {{Zhu}}, \citenamefont {{Kaxiras}}, \citenamefont {{Repellin}},\ and\ \citenamefont {{Cano}}}]{Cano2023FCItTMD}%
  \BibitemOpen
  \bibfield  {author} {\bibinfo {author} {\bibfnamefont {N.}~\bibnamefont {{Morales-Dur{\'a}n}}}, \bibinfo {author} {\bibfnamefont {J.}~\bibnamefont {{Wang}}}, \bibinfo {author} {\bibfnamefont {G.~R.}\ \bibnamefont {{Schleder}}}, \bibinfo {author} {\bibfnamefont {M.}~\bibnamefont {{Angeli}}}, \bibinfo {author} {\bibfnamefont {Z.}~\bibnamefont {{Zhu}}}, \bibinfo {author} {\bibfnamefont {E.}~\bibnamefont {{Kaxiras}}}, \bibinfo {author} {\bibfnamefont {C.}~\bibnamefont {{Repellin}}}, \ and\ \bibinfo {author} {\bibfnamefont {J.}~\bibnamefont {{Cano}}},\ }\href {\doibase 10.1103/PhysRevResearch.5.L032022} {\bibfield  {journal} {\bibinfo  {journal} {Physical Review Research}\ }\textbf {\bibinfo {volume} {5}},\ \bibinfo {eid} {L032022} (\bibinfo {year} {2023}{\natexlab{a}})},\ \Eprint {http://arxiv.org/abs/2304.06669} {arXiv:2304.06669 [cond-mat.str-el]} \BibitemShut {NoStop}%
\bibitem [{\citenamefont {{Wu}}\ \emph {et~al.}(2023)\citenamefont {{Wu}}, \citenamefont {{Shaffer}}, \citenamefont {{Wu}},\ and\ \citenamefont {{Santos}}}]{2023arXiv230907222W}%
  \BibitemOpen
  \bibfield  {author} {\bibinfo {author} {\bibfnamefont {Y.-M.}\ \bibnamefont {{Wu}}}, \bibinfo {author} {\bibfnamefont {D.}~\bibnamefont {{Shaffer}}}, \bibinfo {author} {\bibfnamefont {Z.}~\bibnamefont {{Wu}}}, \ and\ \bibinfo {author} {\bibfnamefont {L.~H.}\ \bibnamefont {{Santos}}},\ }\href@noop {} {\bibfield  {journal} {\bibinfo  {journal} {arXiv e-prints}\ ,\ \bibinfo {eid} {arXiv:2309.07222}} (\bibinfo {year} {2023})},\ \Eprint {http://arxiv.org/abs/2309.07222} {arXiv:2309.07222 [cond-mat.mes-hall]} \BibitemShut {NoStop}%
\bibitem [{\citenamefont {Reddy}\ \emph {et~al.}(2023)\citenamefont {Reddy}, \citenamefont {Alsallom}, \citenamefont {Zhang}, \citenamefont {Devakul},\ and\ \citenamefont {Fu}}]{reddy2023fractional}%
  \BibitemOpen
  \bibfield  {author} {\bibinfo {author} {\bibfnamefont {A.~P.}\ \bibnamefont {Reddy}}, \bibinfo {author} {\bibfnamefont {F.~F.}\ \bibnamefont {Alsallom}}, \bibinfo {author} {\bibfnamefont {Y.}~\bibnamefont {Zhang}}, \bibinfo {author} {\bibfnamefont {T.}~\bibnamefont {Devakul}}, \ and\ \bibinfo {author} {\bibfnamefont {L.}~\bibnamefont {Fu}},\ }\href@noop {} {\enquote {\bibinfo {title} {Fractional quantum anomalous hall states in twisted bilayer mote$_2$ and wse$_2$},}\ } (\bibinfo {year} {2023}),\ \Eprint {http://arxiv.org/abs/2304.12261} {arXiv:2304.12261 [cond-mat.mes-hall]} \BibitemShut {NoStop}%
\bibitem [{\citenamefont {Wang}\ \emph {et~al.}(2023)\citenamefont {Wang}, \citenamefont {Zhang}, \citenamefont {Liu}, \citenamefont {He}, \citenamefont {Xu}, \citenamefont {Ran}, \citenamefont {Cao},\ and\ \citenamefont {Xiao}}]{wang2023fractional}%
  \BibitemOpen
  \bibfield  {author} {\bibinfo {author} {\bibfnamefont {C.}~\bibnamefont {Wang}}, \bibinfo {author} {\bibfnamefont {X.-W.}\ \bibnamefont {Zhang}}, \bibinfo {author} {\bibfnamefont {X.}~\bibnamefont {Liu}}, \bibinfo {author} {\bibfnamefont {Y.}~\bibnamefont {He}}, \bibinfo {author} {\bibfnamefont {X.}~\bibnamefont {Xu}}, \bibinfo {author} {\bibfnamefont {Y.}~\bibnamefont {Ran}}, \bibinfo {author} {\bibfnamefont {T.}~\bibnamefont {Cao}}, \ and\ \bibinfo {author} {\bibfnamefont {D.}~\bibnamefont {Xiao}},\ }\href@noop {} {\enquote {\bibinfo {title} {Fractional chern insulator in twisted bilayer mote$_2$},}\ } (\bibinfo {year} {2023}),\ \Eprint {http://arxiv.org/abs/2304.11864} {arXiv:2304.11864 [cond-mat.str-el]} \BibitemShut {NoStop}%
\bibitem [{\citenamefont {{Qiu}}\ \emph {et~al.}(2023)\citenamefont {{Qiu}}, \citenamefont {{Li}}, \citenamefont {{Luo}},\ and\ \citenamefont {{Wu}}}]{Wu2023IntegerFillingstMoTe2}%
  \BibitemOpen
  \bibfield  {author} {\bibinfo {author} {\bibfnamefont {W.-X.}\ \bibnamefont {{Qiu}}}, \bibinfo {author} {\bibfnamefont {B.}~\bibnamefont {{Li}}}, \bibinfo {author} {\bibfnamefont {X.-J.}\ \bibnamefont {{Luo}}}, \ and\ \bibinfo {author} {\bibfnamefont {F.}~\bibnamefont {{Wu}}},\ }\href {\doibase 10.48550/arXiv.2305.01006} {\bibfield  {journal} {\bibinfo  {journal} {arXiv e-prints}\ ,\ \bibinfo {eid} {arXiv:2305.01006}} (\bibinfo {year} {2023})},\ \Eprint {http://arxiv.org/abs/2305.01006} {arXiv:2305.01006 [cond-mat.mes-hall]} \BibitemShut {NoStop}%
\bibitem [{\citenamefont {Dong}\ \emph {et~al.}(2023)\citenamefont {Dong}, \citenamefont {Wang}, \citenamefont {Ledwith}, \citenamefont {Vishwanath},\ and\ \citenamefont {Parker}}]{Dong2023CFLtMoTe2}%
  \BibitemOpen
  \bibfield  {author} {\bibinfo {author} {\bibfnamefont {J.}~\bibnamefont {Dong}}, \bibinfo {author} {\bibfnamefont {J.}~\bibnamefont {Wang}}, \bibinfo {author} {\bibfnamefont {P.~J.}\ \bibnamefont {Ledwith}}, \bibinfo {author} {\bibfnamefont {A.}~\bibnamefont {Vishwanath}}, \ and\ \bibinfo {author} {\bibfnamefont {D.~E.}\ \bibnamefont {Parker}},\ }\href@noop {} {\bibfield  {journal} {\bibinfo  {journal} {arXiv preprint arXiv:2306.01719}\ } (\bibinfo {year} {2023})}\BibitemShut {NoStop}%
\bibitem [{\citenamefont {Goldman}\ \emph {et~al.}(2023)\citenamefont {Goldman}, \citenamefont {Reddy}, \citenamefont {Paul},\ and\ \citenamefont {Fu}}]{Goldman2023CFLtMoTe2}%
  \BibitemOpen
  \bibfield  {author} {\bibinfo {author} {\bibfnamefont {H.}~\bibnamefont {Goldman}}, \bibinfo {author} {\bibfnamefont {A.~P.}\ \bibnamefont {Reddy}}, \bibinfo {author} {\bibfnamefont {N.}~\bibnamefont {Paul}}, \ and\ \bibinfo {author} {\bibfnamefont {L.}~\bibnamefont {Fu}},\ }\href@noop {} {\bibfield  {journal} {\bibinfo  {journal} {arXiv preprint arXiv:2306.02513}\ } (\bibinfo {year} {2023})}\BibitemShut {NoStop}%
\bibitem [{\citenamefont {{Morales-Dur{\'a}n}}\ \emph {et~al.}(2023{\natexlab{b}})\citenamefont {{Morales-Dur{\'a}n}}, \citenamefont {{Wei}},\ and\ \citenamefont {{MacDonald}}}]{MacDonald2023MagicAngletTMD}%
  \BibitemOpen
  \bibfield  {author} {\bibinfo {author} {\bibfnamefont {N.}~\bibnamefont {{Morales-Dur{\'a}n}}}, \bibinfo {author} {\bibfnamefont {N.}~\bibnamefont {{Wei}}}, \ and\ \bibinfo {author} {\bibfnamefont {A.~H.}\ \bibnamefont {{MacDonald}}},\ }\href {\doibase 10.48550/arXiv.2308.03143} {\bibfield  {journal} {\bibinfo  {journal} {arXiv e-prints}\ ,\ \bibinfo {eid} {arXiv:2308.03143}} (\bibinfo {year} {2023}{\natexlab{b}})},\ \Eprint {http://arxiv.org/abs/2308.03143} {arXiv:2308.03143 [cond-mat.str-el]} \BibitemShut {NoStop}%
\bibitem [{\citenamefont {{Reddy}}\ and\ \citenamefont {{Fu}}(2023)}]{Reddy2023GlobalPDFCI}%
  \BibitemOpen
  \bibfield  {author} {\bibinfo {author} {\bibfnamefont {A.~P.}\ \bibnamefont {{Reddy}}}\ and\ \bibinfo {author} {\bibfnamefont {L.}~\bibnamefont {{Fu}}},\ }\href {\doibase 10.48550/arXiv.2308.10406} {\bibfield  {journal} {\bibinfo  {journal} {arXiv e-prints}\ ,\ \bibinfo {eid} {arXiv:2308.10406}} (\bibinfo {year} {2023})},\ \Eprint {http://arxiv.org/abs/2308.10406} {arXiv:2308.10406 [cond-mat.mes-hall]} \BibitemShut {NoStop}%
\bibitem [{\citenamefont {{Song}}\ \emph {et~al.}(2023)\citenamefont {{Song}}, \citenamefont {{Zhang}},\ and\ \citenamefont {{Senthil}}}]{Song2023tTMDFCI}%
  \BibitemOpen
  \bibfield  {author} {\bibinfo {author} {\bibfnamefont {X.-Y.}\ \bibnamefont {{Song}}}, \bibinfo {author} {\bibfnamefont {Y.-H.}\ \bibnamefont {{Zhang}}}, \ and\ \bibinfo {author} {\bibfnamefont {T.}~\bibnamefont {{Senthil}}},\ }\href {\doibase 10.48550/arXiv.2308.10903} {\bibfield  {journal} {\bibinfo  {journal} {arXiv e-prints}\ ,\ \bibinfo {eid} {arXiv:2308.10903}} (\bibinfo {year} {2023})},\ \Eprint {http://arxiv.org/abs/2308.10903} {arXiv:2308.10903 [cond-mat.str-el]} \BibitemShut {NoStop}%
\bibitem [{\citenamefont {{Xu}}\ \emph {et~al.}(2023{\natexlab{a}})\citenamefont {{Xu}}, \citenamefont {{Li}}, \citenamefont {{Xu}}, \citenamefont {{Bi}},\ and\ \citenamefont {{Zhang}}}]{Xu2023MLWOFCItTMD}%
  \BibitemOpen
  \bibfield  {author} {\bibinfo {author} {\bibfnamefont {C.}~\bibnamefont {{Xu}}}, \bibinfo {author} {\bibfnamefont {J.}~\bibnamefont {{Li}}}, \bibinfo {author} {\bibfnamefont {Y.}~\bibnamefont {{Xu}}}, \bibinfo {author} {\bibfnamefont {Z.}~\bibnamefont {{Bi}}}, \ and\ \bibinfo {author} {\bibfnamefont {Y.}~\bibnamefont {{Zhang}}},\ }\href {\doibase 10.48550/arXiv.2308.09697} {\bibfield  {journal} {\bibinfo  {journal} {arXiv e-prints}\ ,\ \bibinfo {eid} {arXiv:2308.09697}} (\bibinfo {year} {2023}{\natexlab{a}})},\ \Eprint {http://arxiv.org/abs/2308.09697} {arXiv:2308.09697 [cond-mat.str-el]} \BibitemShut {NoStop}%
\bibitem [{\citenamefont {{Wang}}\ \emph {et~al.}(2023)\citenamefont {{Wang}}, \citenamefont {{Devakul}}, \citenamefont {{Zaletel}},\ and\ \citenamefont {{Fu}}}]{Zaletel2023tMoTe2FCI}%
  \BibitemOpen
  \bibfield  {author} {\bibinfo {author} {\bibfnamefont {T.}~\bibnamefont {{Wang}}}, \bibinfo {author} {\bibfnamefont {T.}~\bibnamefont {{Devakul}}}, \bibinfo {author} {\bibfnamefont {M.~P.}\ \bibnamefont {{Zaletel}}}, \ and\ \bibinfo {author} {\bibfnamefont {L.}~\bibnamefont {{Fu}}},\ }\href {\doibase 10.48550/arXiv.2306.02501} {\bibfield  {journal} {\bibinfo  {journal} {arXiv e-prints}\ ,\ \bibinfo {eid} {arXiv:2306.02501}} (\bibinfo {year} {2023})},\ \Eprint {http://arxiv.org/abs/2306.02501} {arXiv:2306.02501 [cond-mat.str-el]} \BibitemShut {NoStop}%
\bibitem [{\citenamefont {{Liu}}\ \emph {et~al.}(2023)\citenamefont {{Liu}}, \citenamefont {{Wang}}, \citenamefont {{Zhang}}, \citenamefont {{Cao}},\ and\ \citenamefont {{Xiao}}}]{Liu2023HFtMoTe2num2}%
  \BibitemOpen
  \bibfield  {author} {\bibinfo {author} {\bibfnamefont {X.}~\bibnamefont {{Liu}}}, \bibinfo {author} {\bibfnamefont {C.}~\bibnamefont {{Wang}}}, \bibinfo {author} {\bibfnamefont {X.-W.}\ \bibnamefont {{Zhang}}}, \bibinfo {author} {\bibfnamefont {T.}~\bibnamefont {{Cao}}}, \ and\ \bibinfo {author} {\bibfnamefont {D.}~\bibnamefont {{Xiao}}},\ }\href {\doibase 10.48550/arXiv.2308.07488} {\bibfield  {journal} {\bibinfo  {journal} {arXiv e-prints}\ ,\ \bibinfo {eid} {arXiv:2308.07488}} (\bibinfo {year} {2023})},\ \Eprint {http://arxiv.org/abs/2308.07488} {arXiv:2308.07488 [cond-mat.mes-hall]} \BibitemShut {NoStop}%
\bibitem [{\citenamefont {Yu}\ \emph {et~al.}(2023)\citenamefont {Yu}, \citenamefont {Herzog-Arbeitman}, \citenamefont {Wang}, \citenamefont {Vafek}, \citenamefont {Bernevig},\ and\ \citenamefont {Regnault}}]{Yu2023FCI}%
  \BibitemOpen
  \bibfield  {author} {\bibinfo {author} {\bibfnamefont {J.}~\bibnamefont {Yu}}, \bibinfo {author} {\bibfnamefont {J.}~\bibnamefont {Herzog-Arbeitman}}, \bibinfo {author} {\bibfnamefont {M.}~\bibnamefont {Wang}}, \bibinfo {author} {\bibfnamefont {O.}~\bibnamefont {Vafek}}, \bibinfo {author} {\bibfnamefont {B.~A.}\ \bibnamefont {Bernevig}}, \ and\ \bibinfo {author} {\bibfnamefont {N.}~\bibnamefont {Regnault}},\ }\href@noop {} {\enquote {\bibinfo {title} {Fractional chern insulators vs. non-magnetic states in twisted bilayer mote$_2$},}\ } (\bibinfo {year} {2023}),\ \Eprint {http://arxiv.org/abs/2309.14429} {arXiv:2309.14429 [cond-mat.mes-hall]} \BibitemShut {NoStop}%
\bibitem [{\citenamefont {Abouelkomsan}\ \emph {et~al.}(2023{\natexlab{b}})\citenamefont {Abouelkomsan}, \citenamefont {Reddy}, \citenamefont {Fu},\ and\ \citenamefont {Bergholtz}}]{Fu2023BandMixingFCItMoTe2}%
  \BibitemOpen
  \bibfield  {author} {\bibinfo {author} {\bibfnamefont {A.}~\bibnamefont {Abouelkomsan}}, \bibinfo {author} {\bibfnamefont {A.~P.}\ \bibnamefont {Reddy}}, \bibinfo {author} {\bibfnamefont {L.}~\bibnamefont {Fu}}, \ and\ \bibinfo {author} {\bibfnamefont {E.~J.}\ \bibnamefont {Bergholtz}},\ }\href@noop {} {\bibfield  {journal} {\bibinfo  {journal} {arXiv:2309.16548}\ } (\bibinfo {year} {2023}{\natexlab{b}})}\BibitemShut {NoStop}%
\bibitem [{\citenamefont {Cai}\ \emph {et~al.}(2023)\citenamefont {Cai}, \citenamefont {Anderson}, \citenamefont {Wang}, \citenamefont {Zhang}, \citenamefont {Liu}, \citenamefont {Holtzmann}, \citenamefont {Zhang}, \citenamefont {Fan}, \citenamefont {Taniguchi}, \citenamefont {Watanabe}, \citenamefont {Ran}, \citenamefont {Cao}, \citenamefont {Fu}, \citenamefont {Xiao}, \citenamefont {Yao},\ and\ \citenamefont {Xu}}]{cai2023signatures}%
  \BibitemOpen
  \bibfield  {author} {\bibinfo {author} {\bibfnamefont {J.}~\bibnamefont {Cai}}, \bibinfo {author} {\bibfnamefont {E.}~\bibnamefont {Anderson}}, \bibinfo {author} {\bibfnamefont {C.}~\bibnamefont {Wang}}, \bibinfo {author} {\bibfnamefont {X.}~\bibnamefont {Zhang}}, \bibinfo {author} {\bibfnamefont {X.}~\bibnamefont {Liu}}, \bibinfo {author} {\bibfnamefont {W.}~\bibnamefont {Holtzmann}}, \bibinfo {author} {\bibfnamefont {Y.}~\bibnamefont {Zhang}}, \bibinfo {author} {\bibfnamefont {F.}~\bibnamefont {Fan}}, \bibinfo {author} {\bibfnamefont {T.}~\bibnamefont {Taniguchi}}, \bibinfo {author} {\bibfnamefont {K.}~\bibnamefont {Watanabe}}, \bibinfo {author} {\bibfnamefont {Y.}~\bibnamefont {Ran}}, \bibinfo {author} {\bibfnamefont {T.}~\bibnamefont {Cao}}, \bibinfo {author} {\bibfnamefont {L.}~\bibnamefont {Fu}}, \bibinfo {author} {\bibfnamefont {D.}~\bibnamefont {Xiao}}, \bibinfo {author} {\bibfnamefont {W.}~\bibnamefont {Yao}}, \ and\ \bibinfo {author} {\bibfnamefont {X.}~\bibnamefont {Xu}},\ }\href {\doibase
  10.1038/s41586-023-06289-w} {\bibfield  {journal} {\bibinfo  {journal} {Nature}\ } (\bibinfo {year} {2023}),\ 10.1038/s41586-023-06289-w}\BibitemShut {NoStop}%
\bibitem [{\citenamefont {Zeng}\ \emph {et~al.}(2023)\citenamefont {Zeng}, \citenamefont {Xia}, \citenamefont {Kang}, \citenamefont {Zhu}, \citenamefont {Kn{\"u}ppel}, \citenamefont {Vaswani}, \citenamefont {Watanabe}, \citenamefont {Taniguchi}, \citenamefont {Mak},\ and\ \citenamefont {Shan}}]{zeng2023integer}%
  \BibitemOpen
  \bibfield  {author} {\bibinfo {author} {\bibfnamefont {Y.}~\bibnamefont {Zeng}}, \bibinfo {author} {\bibfnamefont {Z.}~\bibnamefont {Xia}}, \bibinfo {author} {\bibfnamefont {K.}~\bibnamefont {Kang}}, \bibinfo {author} {\bibfnamefont {J.}~\bibnamefont {Zhu}}, \bibinfo {author} {\bibfnamefont {P.}~\bibnamefont {Kn{\"u}ppel}}, \bibinfo {author} {\bibfnamefont {C.}~\bibnamefont {Vaswani}}, \bibinfo {author} {\bibfnamefont {K.}~\bibnamefont {Watanabe}}, \bibinfo {author} {\bibfnamefont {T.}~\bibnamefont {Taniguchi}}, \bibinfo {author} {\bibfnamefont {K.~F.}\ \bibnamefont {Mak}}, \ and\ \bibinfo {author} {\bibfnamefont {J.}~\bibnamefont {Shan}},\ }\href {\doibase 10.1038/s41586-023-06452-3} {\bibfield  {journal} {\bibinfo  {journal} {Nature}\ } (\bibinfo {year} {2023}),\ 10.1038/s41586-023-06452-3}\BibitemShut {NoStop}%
\bibitem [{\citenamefont {Park}\ \emph {et~al.}(2023)\citenamefont {Park}, \citenamefont {Cai}, \citenamefont {Anderson}, \citenamefont {Zhang}, \citenamefont {Zhu}, \citenamefont {Liu}, \citenamefont {Wang}, \citenamefont {Holtzmann}, \citenamefont {Hu}, \citenamefont {Liu} \emph {et~al.}}]{park2023observation}%
  \BibitemOpen
  \bibfield  {author} {\bibinfo {author} {\bibfnamefont {H.}~\bibnamefont {Park}}, \bibinfo {author} {\bibfnamefont {J.}~\bibnamefont {Cai}}, \bibinfo {author} {\bibfnamefont {E.}~\bibnamefont {Anderson}}, \bibinfo {author} {\bibfnamefont {Y.}~\bibnamefont {Zhang}}, \bibinfo {author} {\bibfnamefont {J.}~\bibnamefont {Zhu}}, \bibinfo {author} {\bibfnamefont {X.}~\bibnamefont {Liu}}, \bibinfo {author} {\bibfnamefont {C.}~\bibnamefont {Wang}}, \bibinfo {author} {\bibfnamefont {W.}~\bibnamefont {Holtzmann}}, \bibinfo {author} {\bibfnamefont {C.}~\bibnamefont {Hu}}, \bibinfo {author} {\bibfnamefont {Z.}~\bibnamefont {Liu}},  \emph {et~al.},\ }\href@noop {} {\bibfield  {journal} {\bibinfo  {journal} {arXiv preprint arXiv:2308.02657}\ } (\bibinfo {year} {2023})}\BibitemShut {NoStop}%
\bibitem [{\citenamefont {{Xu}}\ \emph {et~al.}(2023{\natexlab{b}})\citenamefont {{Xu}}, \citenamefont {{Sun}}, \citenamefont {{Jia}}, \citenamefont {{Liu}}, \citenamefont {{Xu}}, \citenamefont {{Li}}, \citenamefont {{Gu}}, \citenamefont {{Watanabe}}, \citenamefont {{Taniguchi}}, \citenamefont {{Tong}}, \citenamefont {{Jia}}, \citenamefont {{Shi}}, \citenamefont {{Jiang}}, \citenamefont {{Zhang}}, \citenamefont {{Liu}},\ and\ \citenamefont {{Li}}}]{Xu2023FCItMoTe2}%
  \BibitemOpen
  \bibfield  {author} {\bibinfo {author} {\bibfnamefont {F.}~\bibnamefont {{Xu}}}, \bibinfo {author} {\bibfnamefont {Z.}~\bibnamefont {{Sun}}}, \bibinfo {author} {\bibfnamefont {T.}~\bibnamefont {{Jia}}}, \bibinfo {author} {\bibfnamefont {C.}~\bibnamefont {{Liu}}}, \bibinfo {author} {\bibfnamefont {C.}~\bibnamefont {{Xu}}}, \bibinfo {author} {\bibfnamefont {C.}~\bibnamefont {{Li}}}, \bibinfo {author} {\bibfnamefont {Y.}~\bibnamefont {{Gu}}}, \bibinfo {author} {\bibfnamefont {K.}~\bibnamefont {{Watanabe}}}, \bibinfo {author} {\bibfnamefont {T.}~\bibnamefont {{Taniguchi}}}, \bibinfo {author} {\bibfnamefont {B.}~\bibnamefont {{Tong}}}, \bibinfo {author} {\bibfnamefont {J.}~\bibnamefont {{Jia}}}, \bibinfo {author} {\bibfnamefont {Z.}~\bibnamefont {{Shi}}}, \bibinfo {author} {\bibfnamefont {S.}~\bibnamefont {{Jiang}}}, \bibinfo {author} {\bibfnamefont {Y.}~\bibnamefont {{Zhang}}}, \bibinfo {author} {\bibfnamefont {X.}~\bibnamefont {{Liu}}}, \ and\ \bibinfo {author} {\bibfnamefont {T.}~\bibnamefont {{Li}}},\ }\href
  {\doibase 10.48550/arXiv.2308.06177} {\bibfield  {journal} {\bibinfo  {journal} {arXiv e-prints}\ ,\ \bibinfo {eid} {arXiv:2308.06177}} (\bibinfo {year} {2023}{\natexlab{b}})},\ \Eprint {http://arxiv.org/abs/2308.06177} {arXiv:2308.06177 [cond-mat.mes-hall]} \BibitemShut {NoStop}%
\bibitem [{\citenamefont {Lu}\ \emph {et~al.}(2023{\natexlab{a}})\citenamefont {Lu}, \citenamefont {Han}, \citenamefont {Yao}, \citenamefont {Reddy}, \citenamefont {Yang}, \citenamefont {Seo}, \citenamefont {Watanabe}, \citenamefont {Taniguchi}, \citenamefont {Fu},\ and\ \citenamefont {Ju}}]{Lu2023FCI}%
  \BibitemOpen
  \bibfield  {author} {\bibinfo {author} {\bibfnamefont {Z.}~\bibnamefont {Lu}}, \bibinfo {author} {\bibfnamefont {T.}~\bibnamefont {Han}}, \bibinfo {author} {\bibfnamefont {Y.}~\bibnamefont {Yao}}, \bibinfo {author} {\bibfnamefont {A.~P.}\ \bibnamefont {Reddy}}, \bibinfo {author} {\bibfnamefont {J.}~\bibnamefont {Yang}}, \bibinfo {author} {\bibfnamefont {J.}~\bibnamefont {Seo}}, \bibinfo {author} {\bibfnamefont {K.}~\bibnamefont {Watanabe}}, \bibinfo {author} {\bibfnamefont {T.}~\bibnamefont {Taniguchi}}, \bibinfo {author} {\bibfnamefont {L.}~\bibnamefont {Fu}}, \ and\ \bibinfo {author} {\bibfnamefont {L.}~\bibnamefont {Ju}},\ }\href@noop {} {\  (\bibinfo {year} {2023}{\natexlab{a}})},\ \Eprint {http://arxiv.org/abs/2309.17436} {arXiv:2309.17436 [cond-mat.mes-hall]} \BibitemShut {NoStop}%
\bibitem [{\citenamefont {Cao}\ \emph {et~al.}(2018{\natexlab{a}})\citenamefont {Cao}, \citenamefont {Fatemi}, \citenamefont {Demir}, \citenamefont {Fang}, \citenamefont {Tomarken}, \citenamefont {Luo}, \citenamefont {Sanchez-Yamagishi}, \citenamefont {Watanabe}, \citenamefont {Taniguchi}, \citenamefont {Kaxiras} \emph {et~al.}}]{Cao2018TBGMott}%
  \BibitemOpen
  \bibfield  {author} {\bibinfo {author} {\bibfnamefont {Y.}~\bibnamefont {Cao}}, \bibinfo {author} {\bibfnamefont {V.}~\bibnamefont {Fatemi}}, \bibinfo {author} {\bibfnamefont {A.}~\bibnamefont {Demir}}, \bibinfo {author} {\bibfnamefont {S.}~\bibnamefont {Fang}}, \bibinfo {author} {\bibfnamefont {S.~L.}\ \bibnamefont {Tomarken}}, \bibinfo {author} {\bibfnamefont {J.~Y.}\ \bibnamefont {Luo}}, \bibinfo {author} {\bibfnamefont {J.~D.}\ \bibnamefont {Sanchez-Yamagishi}}, \bibinfo {author} {\bibfnamefont {K.}~\bibnamefont {Watanabe}}, \bibinfo {author} {\bibfnamefont {T.}~\bibnamefont {Taniguchi}}, \bibinfo {author} {\bibfnamefont {E.}~\bibnamefont {Kaxiras}},  \emph {et~al.},\ }\href {https://doi.org/10.1038/nature26154} {\bibfield  {journal} {\bibinfo  {journal} {Nature}\ }\textbf {\bibinfo {volume} {556}},\ \bibinfo {pages} {80} (\bibinfo {year} {2018}{\natexlab{a}})}\BibitemShut {NoStop}%
\bibitem [{\citenamefont {Cao}\ \emph {et~al.}(2018{\natexlab{b}})\citenamefont {Cao}, \citenamefont {Fatemi}, \citenamefont {Fang}, \citenamefont {Watanabe}, \citenamefont {Taniguchi}, \citenamefont {Kaxiras},\ and\ \citenamefont {Jarillo-Herrero}}]{Cao2018TBGSC}%
  \BibitemOpen
  \bibfield  {author} {\bibinfo {author} {\bibfnamefont {Y.}~\bibnamefont {Cao}}, \bibinfo {author} {\bibfnamefont {V.}~\bibnamefont {Fatemi}}, \bibinfo {author} {\bibfnamefont {S.}~\bibnamefont {Fang}}, \bibinfo {author} {\bibfnamefont {K.}~\bibnamefont {Watanabe}}, \bibinfo {author} {\bibfnamefont {T.}~\bibnamefont {Taniguchi}}, \bibinfo {author} {\bibfnamefont {E.}~\bibnamefont {Kaxiras}}, \ and\ \bibinfo {author} {\bibfnamefont {P.}~\bibnamefont {Jarillo-Herrero}},\ }\href {https://doi.org/10.1038/nature26160} {\bibfield  {journal} {\bibinfo  {journal} {Nature}\ }\textbf {\bibinfo {volume} {556}},\ \bibinfo {pages} {43} (\bibinfo {year} {2018}{\natexlab{b}})}\BibitemShut {NoStop}%
\bibitem [{\citenamefont {Spanton}\ \emph {et~al.}(2018)\citenamefont {Spanton}, \citenamefont {Zibrov}, \citenamefont {Zhou}, \citenamefont {Taniguchi}, \citenamefont {Watanabe}, \citenamefont {Zaletel},\ and\ \citenamefont {Young}}]{Young2018FCIBLGMoire}%
  \BibitemOpen
  \bibfield  {author} {\bibinfo {author} {\bibfnamefont {E.~M.}\ \bibnamefont {Spanton}}, \bibinfo {author} {\bibfnamefont {A.~A.}\ \bibnamefont {Zibrov}}, \bibinfo {author} {\bibfnamefont {H.}~\bibnamefont {Zhou}}, \bibinfo {author} {\bibfnamefont {T.}~\bibnamefont {Taniguchi}}, \bibinfo {author} {\bibfnamefont {K.}~\bibnamefont {Watanabe}}, \bibinfo {author} {\bibfnamefont {M.~P.}\ \bibnamefont {Zaletel}}, \ and\ \bibinfo {author} {\bibfnamefont {A.~F.}\ \bibnamefont {Young}},\ }\href {\doibase 10.1126/science.aan8458} {\bibfield  {journal} {\bibinfo  {journal} {Science}\ }\textbf {\bibinfo {volume} {360}},\ \bibinfo {pages} {62} (\bibinfo {year} {2018})},\ \Eprint {http://arxiv.org/abs/https://www.science.org/doi/pdf/10.1126/science.aan8458} {https://www.science.org/doi/pdf/10.1126/science.aan8458} \BibitemShut {NoStop}%
\bibitem [{\citenamefont {Xie}\ \emph {et~al.}(2021)\citenamefont {Xie}, \citenamefont {Pierce}, \citenamefont {Park}, \citenamefont {Parker}, \citenamefont {Khalaf}, \citenamefont {Ledwith}, \citenamefont {Cao}, \citenamefont {Lee}, \citenamefont {Chen}, \citenamefont {Forrester}, \citenamefont {Watanabe}, \citenamefont {Taniguchi}, \citenamefont {Vishwanath}, \citenamefont {Jarillo-Herrero},\ and\ \citenamefont {Yacoby}}]{Xie2021TBGFCIfiniteB}%
  \BibitemOpen
  \bibfield  {author} {\bibinfo {author} {\bibfnamefont {Y.}~\bibnamefont {Xie}}, \bibinfo {author} {\bibfnamefont {A.~T.}\ \bibnamefont {Pierce}}, \bibinfo {author} {\bibfnamefont {J.~M.}\ \bibnamefont {Park}}, \bibinfo {author} {\bibfnamefont {D.~E.}\ \bibnamefont {Parker}}, \bibinfo {author} {\bibfnamefont {E.}~\bibnamefont {Khalaf}}, \bibinfo {author} {\bibfnamefont {P.}~\bibnamefont {Ledwith}}, \bibinfo {author} {\bibfnamefont {Y.}~\bibnamefont {Cao}}, \bibinfo {author} {\bibfnamefont {S.~H.}\ \bibnamefont {Lee}}, \bibinfo {author} {\bibfnamefont {S.}~\bibnamefont {Chen}}, \bibinfo {author} {\bibfnamefont {P.~R.}\ \bibnamefont {Forrester}}, \bibinfo {author} {\bibfnamefont {K.}~\bibnamefont {Watanabe}}, \bibinfo {author} {\bibfnamefont {T.}~\bibnamefont {Taniguchi}}, \bibinfo {author} {\bibfnamefont {A.}~\bibnamefont {Vishwanath}}, \bibinfo {author} {\bibfnamefont {P.}~\bibnamefont {Jarillo-Herrero}}, \ and\ \bibinfo {author} {\bibfnamefont {A.}~\bibnamefont {Yacoby}},\ }\href {\doibase
  10.1038/s41586-021-04002-3} {\bibfield  {journal} {\bibinfo  {journal} {Nature}\ }\textbf {\bibinfo {volume} {600}},\ \bibinfo {pages} {439} (\bibinfo {year} {2021})}\BibitemShut {NoStop}%
\bibitem [{\citenamefont {Lu}\ \emph {et~al.}(2023{\natexlab{b}})\citenamefont {Lu}, \citenamefont {Han}, \citenamefont {Yao}, \citenamefont {Reddy}, \citenamefont {Yang}, \citenamefont {Seo}, \citenamefont {Watanabe}, \citenamefont {Taniguchi}, \citenamefont {Fu},\ and\ \citenamefont {Ju}}]{LongJu2023FCIPentalayerGraphenehBN}%
  \BibitemOpen
  \bibfield  {author} {\bibinfo {author} {\bibfnamefont {Z.}~\bibnamefont {Lu}}, \bibinfo {author} {\bibfnamefont {T.}~\bibnamefont {Han}}, \bibinfo {author} {\bibfnamefont {Y.}~\bibnamefont {Yao}}, \bibinfo {author} {\bibfnamefont {A.~P.}\ \bibnamefont {Reddy}}, \bibinfo {author} {\bibfnamefont {J.}~\bibnamefont {Yang}}, \bibinfo {author} {\bibfnamefont {J.}~\bibnamefont {Seo}}, \bibinfo {author} {\bibfnamefont {K.}~\bibnamefont {Watanabe}}, \bibinfo {author} {\bibfnamefont {T.}~\bibnamefont {Taniguchi}}, \bibinfo {author} {\bibfnamefont {L.}~\bibnamefont {Fu}}, \ and\ \bibinfo {author} {\bibfnamefont {L.}~\bibnamefont {Ju}},\ }\href {https://arxiv.org/abs/2309.17436} {\bibfield  {journal} {\bibinfo  {journal} {arXiv:2309.17436}\ } (\bibinfo {year} {2023}{\natexlab{b}})}\BibitemShut {NoStop}%
\bibitem [{\citenamefont {Wu}\ \emph {et~al.}(2019)\citenamefont {Wu}, \citenamefont {Lovorn}, \citenamefont {Tutuc}, \citenamefont {Martin},\ and\ \citenamefont {MacDonald}}]{Wu2019TIintTMD}%
  \BibitemOpen
  \bibfield  {author} {\bibinfo {author} {\bibfnamefont {F.}~\bibnamefont {Wu}}, \bibinfo {author} {\bibfnamefont {T.}~\bibnamefont {Lovorn}}, \bibinfo {author} {\bibfnamefont {E.}~\bibnamefont {Tutuc}}, \bibinfo {author} {\bibfnamefont {I.}~\bibnamefont {Martin}}, \ and\ \bibinfo {author} {\bibfnamefont {A.~H.}\ \bibnamefont {MacDonald}},\ }\href {\doibase 10.1103/PhysRevLett.122.086402} {\bibfield  {journal} {\bibinfo  {journal} {Phys. Rev. Lett.}\ }\textbf {\bibinfo {volume} {122}},\ \bibinfo {pages} {086402} (\bibinfo {year} {2019})}\BibitemShut {NoStop}%
\bibitem [{\citenamefont {Li}\ \emph {et~al.}(2023)\citenamefont {Li}, \citenamefont {Qiu},\ and\ \citenamefont {Wu}}]{Fengcheng2023tMoTe2HFnum1}%
  \BibitemOpen
  \bibfield  {author} {\bibinfo {author} {\bibfnamefont {B.}~\bibnamefont {Li}}, \bibinfo {author} {\bibfnamefont {W.-X.}\ \bibnamefont {Qiu}}, \ and\ \bibinfo {author} {\bibfnamefont {F.}~\bibnamefont {Wu}},\ }\href@noop {} {\bibfield  {journal} {\bibinfo  {journal} {arXiv preprint arXiv:2310.02217}\ } (\bibinfo {year} {2023})}\BibitemShut {NoStop}%
\bibitem [{vas()}]{vaspmanual}%
  \BibitemOpen
  \href {https://www.vasp.at/wiki/index.php/ISMEAR} {\enquote {\bibinfo {title} {Vasp manual https://www.vasp.at/wiki/index.php/ismear},}\ }\BibitemShut {NoStop}%
\bibitem [{\citenamefont {{Li}}\ \emph {et~al.}(2021)\citenamefont {{Li}}, \citenamefont {{Jiang}}, \citenamefont {{Shen}}, \citenamefont {{Zhang}}, \citenamefont {{Li}}, \citenamefont {{Tao}}, \citenamefont {{Devakul}}, \citenamefont {{Watanabe}}, \citenamefont {{Taniguchi}}, \citenamefont {{Fu}}, \citenamefont {{Shan}},\ and\ \citenamefont {{Mak}}}]{2021Natur.600..641L}%
  \BibitemOpen
  \bibfield  {author} {\bibinfo {author} {\bibfnamefont {T.}~\bibnamefont {{Li}}}, \bibinfo {author} {\bibfnamefont {S.}~\bibnamefont {{Jiang}}}, \bibinfo {author} {\bibfnamefont {B.}~\bibnamefont {{Shen}}}, \bibinfo {author} {\bibfnamefont {Y.}~\bibnamefont {{Zhang}}}, \bibinfo {author} {\bibfnamefont {L.}~\bibnamefont {{Li}}}, \bibinfo {author} {\bibfnamefont {Z.}~\bibnamefont {{Tao}}}, \bibinfo {author} {\bibfnamefont {T.}~\bibnamefont {{Devakul}}}, \bibinfo {author} {\bibfnamefont {K.}~\bibnamefont {{Watanabe}}}, \bibinfo {author} {\bibfnamefont {T.}~\bibnamefont {{Taniguchi}}}, \bibinfo {author} {\bibfnamefont {L.}~\bibnamefont {{Fu}}}, \bibinfo {author} {\bibfnamefont {J.}~\bibnamefont {{Shan}}}, \ and\ \bibinfo {author} {\bibfnamefont {K.~F.}\ \bibnamefont {{Mak}}},\ }\href {\doibase 10.1038/s41586-021-04171-1} {\bibfield  {journal} {\bibinfo  {journal} {\nat}\ }\textbf {\bibinfo {volume} {600}},\ \bibinfo {pages} {641} (\bibinfo {year} {2021})},\ \Eprint {http://arxiv.org/abs/2107.01796}
  {arXiv:2107.01796 [cond-mat.mes-hall]} \BibitemShut {NoStop}%
\bibitem [{\citenamefont {Yankowitz}\ and\ \citenamefont {Mak}(2022)}]{yankowitz2022moire}%
  \BibitemOpen
  \bibfield  {author} {\bibinfo {author} {\bibfnamefont {M.}~\bibnamefont {Yankowitz}}\ and\ \bibinfo {author} {\bibfnamefont {K.~F.}\ \bibnamefont {Mak}},\ }\href@noop {} {\bibfield  {journal} {\bibinfo  {journal} {APL Materials}\ }\textbf {\bibinfo {volume} {10}} (\bibinfo {year} {2022})}\BibitemShut {NoStop}%
\bibitem [{\citenamefont {{Zhao}}\ \emph {et~al.}(2022)\citenamefont {{Zhao}}, \citenamefont {{Kang}}, \citenamefont {{Li}}, \citenamefont {{Tschirhart}}, \citenamefont {{Redekop}}, \citenamefont {{Watanabe}}, \citenamefont {{Taniguchi}}, \citenamefont {{Young}}, \citenamefont {{Shan}},\ and\ \citenamefont {{Mak}}}]{2022arXiv220702312Z}%
  \BibitemOpen
  \bibfield  {author} {\bibinfo {author} {\bibfnamefont {W.}~\bibnamefont {{Zhao}}}, \bibinfo {author} {\bibfnamefont {K.}~\bibnamefont {{Kang}}}, \bibinfo {author} {\bibfnamefont {L.}~\bibnamefont {{Li}}}, \bibinfo {author} {\bibfnamefont {C.}~\bibnamefont {{Tschirhart}}}, \bibinfo {author} {\bibfnamefont {E.}~\bibnamefont {{Redekop}}}, \bibinfo {author} {\bibfnamefont {K.}~\bibnamefont {{Watanabe}}}, \bibinfo {author} {\bibfnamefont {T.}~\bibnamefont {{Taniguchi}}}, \bibinfo {author} {\bibfnamefont {A.}~\bibnamefont {{Young}}}, \bibinfo {author} {\bibfnamefont {J.}~\bibnamefont {{Shan}}}, \ and\ \bibinfo {author} {\bibfnamefont {K.~F.}\ \bibnamefont {{Mak}}},\ }\href {\doibase 10.48550/arXiv.2207.02312} {\bibfield  {journal} {\bibinfo  {journal} {arXiv e-prints}\ ,\ \bibinfo {eid} {arXiv:2207.02312}} (\bibinfo {year} {2022})},\ \Eprint {http://arxiv.org/abs/2207.02312} {arXiv:2207.02312 [cond-mat.mes-hall]} \BibitemShut {NoStop}%
\bibitem [{\citenamefont {Mak}\ and\ \citenamefont {Shan}(2022)}]{mak2022semiconductor}%
  \BibitemOpen
  \bibfield  {author} {\bibinfo {author} {\bibfnamefont {K.~F.}\ \bibnamefont {Mak}}\ and\ \bibinfo {author} {\bibfnamefont {J.}~\bibnamefont {Shan}},\ }\href@noop {} {\bibfield  {journal} {\bibinfo  {journal} {Nature Nanotechnology}\ }\textbf {\bibinfo {volume} {17}},\ \bibinfo {pages} {686} (\bibinfo {year} {2022})}\BibitemShut {NoStop}%
\bibitem [{\citenamefont {Mai}\ \emph {et~al.}(2023{\natexlab{a}})\citenamefont {Mai}, \citenamefont {Huang}, \citenamefont {Yu}, \citenamefont {Feldman},\ and\ \citenamefont {Phillips}}]{mai2023interaction}%
  \BibitemOpen
  \bibfield  {author} {\bibinfo {author} {\bibfnamefont {P.}~\bibnamefont {Mai}}, \bibinfo {author} {\bibfnamefont {E.~W.}\ \bibnamefont {Huang}}, \bibinfo {author} {\bibfnamefont {J.}~\bibnamefont {Yu}}, \bibinfo {author} {\bibfnamefont {B.~E.}\ \bibnamefont {Feldman}}, \ and\ \bibinfo {author} {\bibfnamefont {P.~W.}\ \bibnamefont {Phillips}},\ }\href@noop {} {\bibfield  {journal} {\bibinfo  {journal} {npj Quantum Materials}\ }\textbf {\bibinfo {volume} {8}},\ \bibinfo {pages} {14} (\bibinfo {year} {2023}{\natexlab{a}})}\BibitemShut {NoStop}%
\bibitem [{\citenamefont {Uchida}\ \emph {et~al.}(2014)\citenamefont {Uchida}, \citenamefont {Furuya}, \citenamefont {Iwata},\ and\ \citenamefont {Oshiyama}}]{coincidence1}%
  \BibitemOpen
  \bibfield  {author} {\bibinfo {author} {\bibfnamefont {K.}~\bibnamefont {Uchida}}, \bibinfo {author} {\bibfnamefont {S.}~\bibnamefont {Furuya}}, \bibinfo {author} {\bibfnamefont {J.-I.}\ \bibnamefont {Iwata}}, \ and\ \bibinfo {author} {\bibfnamefont {A.}~\bibnamefont {Oshiyama}},\ }\href@noop {} {\bibfield  {journal} {\bibinfo  {journal} {Physical Review B}\ }\textbf {\bibinfo {volume} {90}},\ \bibinfo {pages} {155451} (\bibinfo {year} {2014})}\BibitemShut {NoStop}%
\bibitem [{\citenamefont {Koda}\ \emph {et~al.}(2016)\citenamefont {Koda}, \citenamefont {Bechstedt}, \citenamefont {Marques},\ and\ \citenamefont {Teles}}]{coincidence2}%
  \BibitemOpen
  \bibfield  {author} {\bibinfo {author} {\bibfnamefont {D.~S.}\ \bibnamefont {Koda}}, \bibinfo {author} {\bibfnamefont {F.}~\bibnamefont {Bechstedt}}, \bibinfo {author} {\bibfnamefont {M.}~\bibnamefont {Marques}}, \ and\ \bibinfo {author} {\bibfnamefont {L.~K.}\ \bibnamefont {Teles}},\ }\href@noop {} {\bibfield  {journal} {\bibinfo  {journal} {The Journal of Physical Chemistry C}\ }\textbf {\bibinfo {volume} {120}},\ \bibinfo {pages} {10895} (\bibinfo {year} {2016})}\BibitemShut {NoStop}%
\bibitem [{\citenamefont {Kresse}\ and\ \citenamefont {Hafner}(1993{\natexlab{a}})}]{vasp1}%
  \BibitemOpen
  \bibfield  {author} {\bibinfo {author} {\bibfnamefont {G.}~\bibnamefont {Kresse}}\ and\ \bibinfo {author} {\bibfnamefont {J.}~\bibnamefont {Hafner}},\ }\href {\doibase 10.1103/PhysRevB.47.558} {\bibfield  {journal} {\bibinfo  {journal} {Phys. Rev. B}\ }\textbf {\bibinfo {volume} {47}},\ \bibinfo {pages} {558} (\bibinfo {year} {1993}{\natexlab{a}})}\BibitemShut {NoStop}%
\bibitem [{\citenamefont {Kresse}\ and\ \citenamefont {Hafner}(1993{\natexlab{b}})}]{vasp2}%
  \BibitemOpen
  \bibfield  {author} {\bibinfo {author} {\bibfnamefont {G.}~\bibnamefont {Kresse}}\ and\ \bibinfo {author} {\bibfnamefont {J.}~\bibnamefont {Hafner}},\ }\href@noop {} {\bibfield  {journal} {\bibinfo  {journal} {Physical review B}\ }\textbf {\bibinfo {volume} {47}},\ \bibinfo {pages} {558} (\bibinfo {year} {1993}{\natexlab{b}})}\BibitemShut {NoStop}%
\bibitem [{\citenamefont {Kresse}\ \emph {et~al.}(1996)\citenamefont {Kresse} \emph {et~al.}}]{vasp3}%
  \BibitemOpen
  \bibfield  {author} {\bibinfo {author} {\bibfnamefont {G.}~\bibnamefont {Kresse}} \emph {et~al.},\ }\href@noop {} {\bibfield  {journal} {\bibinfo  {journal} {Phys. Rev. B}\ }\textbf {\bibinfo {volume} {54}},\ \bibinfo {pages} {169} (\bibinfo {year} {1996})}\BibitemShut {NoStop}%
\bibitem [{\citenamefont {Kresse}\ and\ \citenamefont {Hafner}(1994)}]{vasp4}%
  \BibitemOpen
  \bibfield  {author} {\bibinfo {author} {\bibfnamefont {G.}~\bibnamefont {Kresse}}\ and\ \bibinfo {author} {\bibfnamefont {J.}~\bibnamefont {Hafner}},\ }\href@noop {} {\bibfield  {journal} {\bibinfo  {journal} {Journal of Physics: Condensed Matter}\ }\textbf {\bibinfo {volume} {6}},\ \bibinfo {pages} {8245} (\bibinfo {year} {1994})}\BibitemShut {NoStop}%
\bibitem [{\citenamefont {Kresse}\ and\ \citenamefont {Joubert}(1999)}]{vasp5}%
  \BibitemOpen
  \bibfield  {author} {\bibinfo {author} {\bibfnamefont {G.}~\bibnamefont {Kresse}}\ and\ \bibinfo {author} {\bibfnamefont {D.}~\bibnamefont {Joubert}},\ }\href@noop {} {\bibfield  {journal} {\bibinfo  {journal} {Physical review b}\ }\textbf {\bibinfo {volume} {59}},\ \bibinfo {pages} {1758} (\bibinfo {year} {1999})}\BibitemShut {NoStop}%
\bibitem [{\citenamefont {Puotinen}\ and\ \citenamefont {Newnham}(1961)}]{1961_D.Puotinen}%
  \BibitemOpen
  \bibfield  {author} {\bibinfo {author} {\bibfnamefont {D.}~\bibnamefont {Puotinen}}\ and\ \bibinfo {author} {\bibfnamefont {R.~E.}\ \bibnamefont {Newnham}},\ }\href {\doibase https://doi.org/10.1107/S0365110X61002084} {\bibfield  {journal} {\bibinfo  {journal} {Acta Crystallographica}\ }\textbf {\bibinfo {volume} {14}},\ \bibinfo {pages} {691} (\bibinfo {year} {1961})}\BibitemShut {NoStop}%
\bibitem [{\citenamefont {Fang}\ \emph {et~al.}(2012)\citenamefont {Fang}, \citenamefont {Gilbert},\ and\ \citenamefont {Bernevig}}]{PhysRevB.86.115112}%
  \BibitemOpen
  \bibfield  {author} {\bibinfo {author} {\bibfnamefont {C.}~\bibnamefont {Fang}}, \bibinfo {author} {\bibfnamefont {M.~J.}\ \bibnamefont {Gilbert}}, \ and\ \bibinfo {author} {\bibfnamefont {B.~A.}\ \bibnamefont {Bernevig}},\ }\href {\doibase 10.1103/PhysRevB.86.115112} {\bibfield  {journal} {\bibinfo  {journal} {Phys. Rev. B}\ }\textbf {\bibinfo {volume} {86}},\ \bibinfo {pages} {115112} (\bibinfo {year} {2012})}\BibitemShut {NoStop}%
\bibitem [{\citenamefont {{Xu}}\ \emph {et~al.}(2022)\citenamefont {{Xu}}, \citenamefont {{Kang}}, \citenamefont {{Watanabe}}, \citenamefont {{Taniguchi}}, \citenamefont {{Mak}},\ and\ \citenamefont {{Shan}}}]{2022NatNa..17..934X}%
  \BibitemOpen
  \bibfield  {author} {\bibinfo {author} {\bibfnamefont {Y.}~\bibnamefont {{Xu}}}, \bibinfo {author} {\bibfnamefont {K.}~\bibnamefont {{Kang}}}, \bibinfo {author} {\bibfnamefont {K.}~\bibnamefont {{Watanabe}}}, \bibinfo {author} {\bibfnamefont {T.}~\bibnamefont {{Taniguchi}}}, \bibinfo {author} {\bibfnamefont {K.~F.}\ \bibnamefont {{Mak}}}, \ and\ \bibinfo {author} {\bibfnamefont {J.}~\bibnamefont {{Shan}}},\ }\href {\doibase 10.1038/s41565-022-01180-7} {\bibfield  {journal} {\bibinfo  {journal} {Nature Nanotechnology}\ }\textbf {\bibinfo {volume} {17}},\ \bibinfo {pages} {934} (\bibinfo {year} {2022})},\ \Eprint {http://arxiv.org/abs/2202.02055} {arXiv:2202.02055 [cond-mat.str-el]} \BibitemShut {NoStop}%
\bibitem [{\citenamefont {Xu}\ \emph {et~al.}(2020)\citenamefont {Xu}, \citenamefont {Liu}, \citenamefont {Rhodes}, \citenamefont {Watanabe}, \citenamefont {Taniguchi}, \citenamefont {Hone}, \citenamefont {Elser}, \citenamefont {Mak},\ and\ \citenamefont {Shan}}]{xu2020correlated}%
  \BibitemOpen
  \bibfield  {author} {\bibinfo {author} {\bibfnamefont {Y.}~\bibnamefont {Xu}}, \bibinfo {author} {\bibfnamefont {S.}~\bibnamefont {Liu}}, \bibinfo {author} {\bibfnamefont {D.~A.}\ \bibnamefont {Rhodes}}, \bibinfo {author} {\bibfnamefont {K.}~\bibnamefont {Watanabe}}, \bibinfo {author} {\bibfnamefont {T.}~\bibnamefont {Taniguchi}}, \bibinfo {author} {\bibfnamefont {J.}~\bibnamefont {Hone}}, \bibinfo {author} {\bibfnamefont {V.}~\bibnamefont {Elser}}, \bibinfo {author} {\bibfnamefont {K.~F.}\ \bibnamefont {Mak}}, \ and\ \bibinfo {author} {\bibfnamefont {J.}~\bibnamefont {Shan}},\ }\href@noop {} {\bibfield  {journal} {\bibinfo  {journal} {Nature}\ }\textbf {\bibinfo {volume} {587}},\ \bibinfo {pages} {214} (\bibinfo {year} {2020})}\BibitemShut {NoStop}%
\bibitem [{\citenamefont {Kometter}\ \emph {et~al.}(2023)\citenamefont {Kometter}, \citenamefont {Yu}, \citenamefont {Devakul}, \citenamefont {Reddy}, \citenamefont {Zhang}, \citenamefont {Foutty}, \citenamefont {Watanabe}, \citenamefont {Taniguchi}, \citenamefont {Fu},\ and\ \citenamefont {Feldman}}]{kometter2023hofstadter}%
  \BibitemOpen
  \bibfield  {author} {\bibinfo {author} {\bibfnamefont {C.~R.}\ \bibnamefont {Kometter}}, \bibinfo {author} {\bibfnamefont {J.}~\bibnamefont {Yu}}, \bibinfo {author} {\bibfnamefont {T.}~\bibnamefont {Devakul}}, \bibinfo {author} {\bibfnamefont {A.~P.}\ \bibnamefont {Reddy}}, \bibinfo {author} {\bibfnamefont {Y.}~\bibnamefont {Zhang}}, \bibinfo {author} {\bibfnamefont {B.~A.}\ \bibnamefont {Foutty}}, \bibinfo {author} {\bibfnamefont {K.}~\bibnamefont {Watanabe}}, \bibinfo {author} {\bibfnamefont {T.}~\bibnamefont {Taniguchi}}, \bibinfo {author} {\bibfnamefont {L.}~\bibnamefont {Fu}}, \ and\ \bibinfo {author} {\bibfnamefont {B.~E.}\ \bibnamefont {Feldman}},\ }\href@noop {} {\bibfield  {journal} {\bibinfo  {journal} {Nature Physics}\ ,\ \bibinfo {pages} {1}} (\bibinfo {year} {2023})}\BibitemShut {NoStop}%
\bibitem [{\citenamefont {Morales-Dur\'an}\ \emph {et~al.}(2023)\citenamefont {Morales-Dur\'an}, \citenamefont {Potasz},\ and\ \citenamefont {MacDonald}}]{PhysRevB.107.235131}%
  \BibitemOpen
  \bibfield  {author} {\bibinfo {author} {\bibfnamefont {N.}~\bibnamefont {Morales-Dur\'an}}, \bibinfo {author} {\bibfnamefont {P.}~\bibnamefont {Potasz}}, \ and\ \bibinfo {author} {\bibfnamefont {A.~H.}\ \bibnamefont {MacDonald}},\ }\href {\doibase 10.1103/PhysRevB.107.235131} {\bibfield  {journal} {\bibinfo  {journal} {Phys. Rev. B}\ }\textbf {\bibinfo {volume} {107}},\ \bibinfo {pages} {235131} (\bibinfo {year} {2023})}\BibitemShut {NoStop}%
\bibitem [{\citenamefont {Tan}\ \emph {et~al.}(2023)\citenamefont {Tan}, \citenamefont {Rasmita}, \citenamefont {Zhang}, \citenamefont {Cai}, \citenamefont {Cai}, \citenamefont {Dai}, \citenamefont {Watanabe}, \citenamefont {Taniguchi}, \citenamefont {MacDonald},\ and\ \citenamefont {Gao}}]{tan2023layer}%
  \BibitemOpen
  \bibfield  {author} {\bibinfo {author} {\bibfnamefont {Q.}~\bibnamefont {Tan}}, \bibinfo {author} {\bibfnamefont {A.}~\bibnamefont {Rasmita}}, \bibinfo {author} {\bibfnamefont {Z.}~\bibnamefont {Zhang}}, \bibinfo {author} {\bibfnamefont {H.}~\bibnamefont {Cai}}, \bibinfo {author} {\bibfnamefont {X.}~\bibnamefont {Cai}}, \bibinfo {author} {\bibfnamefont {X.}~\bibnamefont {Dai}}, \bibinfo {author} {\bibfnamefont {K.}~\bibnamefont {Watanabe}}, \bibinfo {author} {\bibfnamefont {T.}~\bibnamefont {Taniguchi}}, \bibinfo {author} {\bibfnamefont {A.~H.}\ \bibnamefont {MacDonald}}, \ and\ \bibinfo {author} {\bibfnamefont {W.}~\bibnamefont {Gao}},\ }\href@noop {} {\bibfield  {journal} {\bibinfo  {journal} {Nature Materials}\ ,\ \bibinfo {pages} {1}} (\bibinfo {year} {2023})}\BibitemShut {NoStop}%
\bibitem [{\citenamefont {Morales-Dur\'an}\ \emph {et~al.}(2022)\citenamefont {Morales-Dur\'an}, \citenamefont {Hu}, \citenamefont {Potasz},\ and\ \citenamefont {MacDonald}}]{PhysRevLett.128.217202}%
  \BibitemOpen
  \bibfield  {author} {\bibinfo {author} {\bibfnamefont {N.}~\bibnamefont {Morales-Dur\'an}}, \bibinfo {author} {\bibfnamefont {N.~C.}\ \bibnamefont {Hu}}, \bibinfo {author} {\bibfnamefont {P.}~\bibnamefont {Potasz}}, \ and\ \bibinfo {author} {\bibfnamefont {A.~H.}\ \bibnamefont {MacDonald}},\ }\href {\doibase 10.1103/PhysRevLett.128.217202} {\bibfield  {journal} {\bibinfo  {journal} {Phys. Rev. Lett.}\ }\textbf {\bibinfo {volume} {128}},\ \bibinfo {pages} {217202} (\bibinfo {year} {2022})}\BibitemShut {NoStop}%
\bibitem [{\citenamefont {{Foutty}}\ \emph {et~al.}(2023)\citenamefont {{Foutty}}, \citenamefont {{Kometter}}, \citenamefont {{Devakul}}, \citenamefont {{Reddy}}, \citenamefont {{Watanabe}}, \citenamefont {{Taniguchi}}, \citenamefont {{Fu}},\ and\ \citenamefont {{Feldman}}}]{2023arXiv230409808F}%
  \BibitemOpen
  \bibfield  {author} {\bibinfo {author} {\bibfnamefont {B.~A.}\ \bibnamefont {{Foutty}}}, \bibinfo {author} {\bibfnamefont {C.~R.}\ \bibnamefont {{Kometter}}}, \bibinfo {author} {\bibfnamefont {T.}~\bibnamefont {{Devakul}}}, \bibinfo {author} {\bibfnamefont {A.~P.}\ \bibnamefont {{Reddy}}}, \bibinfo {author} {\bibfnamefont {K.}~\bibnamefont {{Watanabe}}}, \bibinfo {author} {\bibfnamefont {T.}~\bibnamefont {{Taniguchi}}}, \bibinfo {author} {\bibfnamefont {L.}~\bibnamefont {{Fu}}}, \ and\ \bibinfo {author} {\bibfnamefont {B.~E.}\ \bibnamefont {{Feldman}}},\ }\href {\doibase 10.48550/arXiv.2304.09808} {\bibfield  {journal} {\bibinfo  {journal} {arXiv e-prints}\ ,\ \bibinfo {eid} {arXiv:2304.09808}} (\bibinfo {year} {2023})},\ \Eprint {http://arxiv.org/abs/2304.09808} {arXiv:2304.09808 [cond-mat.mes-hall]} \BibitemShut {NoStop}%
\bibitem [{\citenamefont {{Setty}}\ \emph {et~al.}(2023)\citenamefont {{Setty}}, \citenamefont {{Xie}}, \citenamefont {{Sur}}, \citenamefont {{Chen}}, \citenamefont {{Vergniory}},\ and\ \citenamefont {{Si}}}]{2023arXiv230914340S}%
  \BibitemOpen
  \bibfield  {author} {\bibinfo {author} {\bibfnamefont {C.}~\bibnamefont {{Setty}}}, \bibinfo {author} {\bibfnamefont {F.}~\bibnamefont {{Xie}}}, \bibinfo {author} {\bibfnamefont {S.}~\bibnamefont {{Sur}}}, \bibinfo {author} {\bibfnamefont {L.}~\bibnamefont {{Chen}}}, \bibinfo {author} {\bibfnamefont {M.~G.}\ \bibnamefont {{Vergniory}}}, \ and\ \bibinfo {author} {\bibfnamefont {Q.}~\bibnamefont {{Si}}},\ }\href {\doibase 10.48550/arXiv.2309.14340} {\bibfield  {journal} {\bibinfo  {journal} {arXiv e-prints}\ ,\ \bibinfo {eid} {arXiv:2309.14340}} (\bibinfo {year} {2023})},\ \Eprint {http://arxiv.org/abs/2309.14340} {arXiv:2309.14340 [cond-mat.str-el]} \BibitemShut {NoStop}%
\bibitem [{\citenamefont {Bernevig}\ \emph {et~al.}(2021)\citenamefont {Bernevig}, \citenamefont {Lian}, \citenamefont {Cowsik}, \citenamefont {Xie}, \citenamefont {Regnault},\ and\ \citenamefont {Song}}]{PhysRevB.103.205415}%
  \BibitemOpen
  \bibfield  {author} {\bibinfo {author} {\bibfnamefont {B.~A.}\ \bibnamefont {Bernevig}}, \bibinfo {author} {\bibfnamefont {B.}~\bibnamefont {Lian}}, \bibinfo {author} {\bibfnamefont {A.}~\bibnamefont {Cowsik}}, \bibinfo {author} {\bibfnamefont {F.}~\bibnamefont {Xie}}, \bibinfo {author} {\bibfnamefont {N.}~\bibnamefont {Regnault}}, \ and\ \bibinfo {author} {\bibfnamefont {Z.-D.}\ \bibnamefont {Song}},\ }\href {\doibase 10.1103/PhysRevB.103.205415} {\bibfield  {journal} {\bibinfo  {journal} {Phys. Rev. B}\ }\textbf {\bibinfo {volume} {103}},\ \bibinfo {pages} {205415} (\bibinfo {year} {2021})}\BibitemShut {NoStop}%
\bibitem [{\citenamefont {{Song}}\ and\ \citenamefont {{Bernevig}}(2022)}]{2022PhRvL.129d7601S}%
  \BibitemOpen
  \bibfield  {author} {\bibinfo {author} {\bibfnamefont {Z.-D.}\ \bibnamefont {{Song}}}\ and\ \bibinfo {author} {\bibfnamefont {B.~A.}\ \bibnamefont {{Bernevig}}},\ }\href {\doibase 10.1103/PhysRevLett.129.047601} {\bibfield  {journal} {\bibinfo  {journal} {\prl}\ }\textbf {\bibinfo {volume} {129}},\ \bibinfo {eid} {047601} (\bibinfo {year} {2022})},\ \Eprint {http://arxiv.org/abs/2111.05865} {arXiv:2111.05865 [cond-mat.str-el]} \BibitemShut {NoStop}%
\bibitem [{\citenamefont {Bultinck}\ \emph {et~al.}(2019)\citenamefont {Bultinck}, \citenamefont {Khalaf}, \citenamefont {Liu}, \citenamefont {Chatterjee}, \citenamefont {Vishwanath},\ and\ \citenamefont {Zaletel}}]{Bultinck2019GroundSA}%
  \BibitemOpen
  \bibfield  {author} {\bibinfo {author} {\bibfnamefont {N.}~\bibnamefont {Bultinck}}, \bibinfo {author} {\bibfnamefont {E.}~\bibnamefont {Khalaf}}, \bibinfo {author} {\bibfnamefont {S.}~\bibnamefont {Liu}}, \bibinfo {author} {\bibfnamefont {S.}~\bibnamefont {Chatterjee}}, \bibinfo {author} {\bibfnamefont {A.}~\bibnamefont {Vishwanath}}, \ and\ \bibinfo {author} {\bibfnamefont {M.~P.}\ \bibnamefont {Zaletel}},\ }\href {https://api.semanticscholar.org/CorpusID:207880424} {\bibfield  {journal} {\bibinfo  {journal} {arXiv: Strongly Correlated Electrons}\ } (\bibinfo {year} {2019})}\BibitemShut {NoStop}%
\bibitem [{\citenamefont {{Soldini}}\ \emph {et~al.}(2023)\citenamefont {{Soldini}}, \citenamefont {{Astrakhantsev}}, \citenamefont {{Iraola}}, \citenamefont {{Tiwari}}, \citenamefont {{Fischer}}, \citenamefont {{Valent{\'\i}}}, \citenamefont {{Vergniory}}, \citenamefont {{Wagner}},\ and\ \citenamefont {{Neupert}}}]{2023PhRvB.107x5145S}%
  \BibitemOpen
  \bibfield  {author} {\bibinfo {author} {\bibfnamefont {M.~O.}\ \bibnamefont {{Soldini}}}, \bibinfo {author} {\bibfnamefont {N.}~\bibnamefont {{Astrakhantsev}}}, \bibinfo {author} {\bibfnamefont {M.}~\bibnamefont {{Iraola}}}, \bibinfo {author} {\bibfnamefont {A.}~\bibnamefont {{Tiwari}}}, \bibinfo {author} {\bibfnamefont {M.~H.}\ \bibnamefont {{Fischer}}}, \bibinfo {author} {\bibfnamefont {R.}~\bibnamefont {{Valent{\'\i}}}}, \bibinfo {author} {\bibfnamefont {M.~G.}\ \bibnamefont {{Vergniory}}}, \bibinfo {author} {\bibfnamefont {G.}~\bibnamefont {{Wagner}}}, \ and\ \bibinfo {author} {\bibfnamefont {T.}~\bibnamefont {{Neupert}}},\ }\href {\doibase 10.1103/PhysRevB.107.245145} {\bibfield  {journal} {\bibinfo  {journal} {\prb}\ }\textbf {\bibinfo {volume} {107}},\ \bibinfo {eid} {245145} (\bibinfo {year} {2023})},\ \Eprint {http://arxiv.org/abs/2209.10556} {arXiv:2209.10556 [cond-mat.str-el]} \BibitemShut {NoStop}%
\bibitem [{\citenamefont {{Herzog-Arbeitman}}\ \emph {et~al.}(2022{\natexlab{a}})\citenamefont {{Herzog-Arbeitman}}, \citenamefont {{Bernevig}},\ and\ \citenamefont {{Song}}}]{2022arXiv221200030H}%
  \BibitemOpen
  \bibfield  {author} {\bibinfo {author} {\bibfnamefont {J.}~\bibnamefont {{Herzog-Arbeitman}}}, \bibinfo {author} {\bibfnamefont {B.~A.}\ \bibnamefont {{Bernevig}}}, \ and\ \bibinfo {author} {\bibfnamefont {Z.-D.}\ \bibnamefont {{Song}}},\ }\href {\doibase 10.48550/arXiv.2212.00030} {\bibfield  {journal} {\bibinfo  {journal} {arXiv e-prints}\ ,\ \bibinfo {eid} {arXiv:2212.00030}} (\bibinfo {year} {2022}{\natexlab{a}})},\ \Eprint {http://arxiv.org/abs/2212.00030} {arXiv:2212.00030 [cond-mat.str-el]} \BibitemShut {NoStop}%
\bibitem [{\citenamefont {Mai}\ \emph {et~al.}(2023{\natexlab{b}})\citenamefont {Mai}, \citenamefont {Feldman},\ and\ \citenamefont {Phillips}}]{Phillips2023DQMCHaldaneModel}%
  \BibitemOpen
  \bibfield  {author} {\bibinfo {author} {\bibfnamefont {P.}~\bibnamefont {Mai}}, \bibinfo {author} {\bibfnamefont {B.~E.}\ \bibnamefont {Feldman}}, \ and\ \bibinfo {author} {\bibfnamefont {P.~W.}\ \bibnamefont {Phillips}},\ }\href {\doibase 10.1103/PhysRevResearch.5.013162} {\bibfield  {journal} {\bibinfo  {journal} {Phys. Rev. Res.}\ }\textbf {\bibinfo {volume} {5}},\ \bibinfo {pages} {013162} (\bibinfo {year} {2023}{\natexlab{b}})}\BibitemShut {NoStop}%
\bibitem [{\citenamefont {{Wagner}}\ \emph {et~al.}(2023)\citenamefont {{Wagner}}, \citenamefont {{Crippa}}, \citenamefont {{Amaricci}}, \citenamefont {{Hansmann}}, \citenamefont {{Klett}}, \citenamefont {{K{\"o}nig}}, \citenamefont {{Sch{\"a}fer}}, \citenamefont {{Di Sante}}, \citenamefont {{Cano}}, \citenamefont {{Millis}}, \citenamefont {{Georges}},\ and\ \citenamefont {{Sangiovanni}}}]{2023arXiv230105588W}%
  \BibitemOpen
  \bibfield  {author} {\bibinfo {author} {\bibfnamefont {N.}~\bibnamefont {{Wagner}}}, \bibinfo {author} {\bibfnamefont {L.}~\bibnamefont {{Crippa}}}, \bibinfo {author} {\bibfnamefont {A.}~\bibnamefont {{Amaricci}}}, \bibinfo {author} {\bibfnamefont {P.}~\bibnamefont {{Hansmann}}}, \bibinfo {author} {\bibfnamefont {M.}~\bibnamefont {{Klett}}}, \bibinfo {author} {\bibfnamefont {E.}~\bibnamefont {{K{\"o}nig}}}, \bibinfo {author} {\bibfnamefont {T.}~\bibnamefont {{Sch{\"a}fer}}}, \bibinfo {author} {\bibfnamefont {D.}~\bibnamefont {{Di Sante}}}, \bibinfo {author} {\bibfnamefont {J.}~\bibnamefont {{Cano}}}, \bibinfo {author} {\bibfnamefont {A.}~\bibnamefont {{Millis}}}, \bibinfo {author} {\bibfnamefont {A.}~\bibnamefont {{Georges}}}, \ and\ \bibinfo {author} {\bibfnamefont {G.}~\bibnamefont {{Sangiovanni}}},\ }\href {\doibase 10.48550/arXiv.2301.05588} {\bibfield  {journal} {\bibinfo  {journal} {arXiv e-prints}\ ,\ \bibinfo {eid} {arXiv:2301.05588}} (\bibinfo {year} {2023})},\ \Eprint
  {http://arxiv.org/abs/2301.05588} {arXiv:2301.05588 [cond-mat.str-el]} \BibitemShut {NoStop}%
\bibitem [{\citenamefont {Mai}\ \emph {et~al.}(2023{\natexlab{c}})\citenamefont {Mai}, \citenamefont {Zhao}, \citenamefont {Feldman},\ and\ \citenamefont {Phillips}}]{Phillips2023DQMC}%
  \BibitemOpen
  \bibfield  {author} {\bibinfo {author} {\bibfnamefont {P.}~\bibnamefont {Mai}}, \bibinfo {author} {\bibfnamefont {J.}~\bibnamefont {Zhao}}, \bibinfo {author} {\bibfnamefont {B.~E.}\ \bibnamefont {Feldman}}, \ and\ \bibinfo {author} {\bibfnamefont {P.~W.}\ \bibnamefont {Phillips}},\ }\href {\doibase 10.1038/s41467-023-41465-6} {\bibfield  {journal} {\bibinfo  {journal} {Nature Communications}\ }\textbf {\bibinfo {volume} {14}},\ \bibinfo {pages} {5999} (\bibinfo {year} {2023}{\natexlab{c}})}\BibitemShut {NoStop}%
\bibitem [{\citenamefont {Ding}\ \emph {et~al.}(2023)\citenamefont {Ding}, \citenamefont {Yang}, \citenamefont {Wang}, \citenamefont {Zhu}, \citenamefont {Peng}, \citenamefont {Mai}, \citenamefont {Huang}, \citenamefont {Moritz}, \citenamefont {Phillips}, \citenamefont {Feldman} \emph {et~al.}}]{ding2023particle}%
  \BibitemOpen
  \bibfield  {author} {\bibinfo {author} {\bibfnamefont {J.~K.}\ \bibnamefont {Ding}}, \bibinfo {author} {\bibfnamefont {L.}~\bibnamefont {Yang}}, \bibinfo {author} {\bibfnamefont {W.~O.}\ \bibnamefont {Wang}}, \bibinfo {author} {\bibfnamefont {Z.}~\bibnamefont {Zhu}}, \bibinfo {author} {\bibfnamefont {C.}~\bibnamefont {Peng}}, \bibinfo {author} {\bibfnamefont {P.}~\bibnamefont {Mai}}, \bibinfo {author} {\bibfnamefont {E.~W.}\ \bibnamefont {Huang}}, \bibinfo {author} {\bibfnamefont {B.}~\bibnamefont {Moritz}}, \bibinfo {author} {\bibfnamefont {P.~W.}\ \bibnamefont {Phillips}}, \bibinfo {author} {\bibfnamefont {B.~E.}\ \bibnamefont {Feldman}},  \emph {et~al.},\ }\href@noop {} {\bibfield  {journal} {\bibinfo  {journal} {arXiv preprint arXiv:2309.07876}\ } (\bibinfo {year} {2023})}\BibitemShut {NoStop}%
\bibitem [{\citenamefont {Olin}\ \emph {et~al.}(2023)\citenamefont {Olin}, \citenamefont {Jmukhadze}, \citenamefont {MacDonald},\ and\ \citenamefont {Lee}}]{olin2023ab}%
  \BibitemOpen
  \bibfield  {author} {\bibinfo {author} {\bibfnamefont {S.}~\bibnamefont {Olin}}, \bibinfo {author} {\bibfnamefont {E.}~\bibnamefont {Jmukhadze}}, \bibinfo {author} {\bibfnamefont {A.~H.}\ \bibnamefont {MacDonald}}, \ and\ \bibinfo {author} {\bibfnamefont {W.-C.}\ \bibnamefont {Lee}},\ }\href@noop {} {\bibfield  {journal} {\bibinfo  {journal} {arXiv preprint arXiv:2310.17824}\ } (\bibinfo {year} {2023})}\BibitemShut {NoStop}%
\bibitem [{\citenamefont {Jinnouchi}\ \emph {et~al.}(2019)\citenamefont {Jinnouchi}, \citenamefont {Karsai},\ and\ \citenamefont {Kresse}}]{jinnouchi_--fly_2019}%
  \BibitemOpen
  \bibfield  {author} {\bibinfo {author} {\bibfnamefont {R.}~\bibnamefont {Jinnouchi}}, \bibinfo {author} {\bibfnamefont {F.}~\bibnamefont {Karsai}}, \ and\ \bibinfo {author} {\bibfnamefont {G.}~\bibnamefont {Kresse}},\ }\href {\doibase 10.1103/PhysRevB.100.014105} {\bibfield  {journal} {\bibinfo  {journal} {Physical Review B}\ }\textbf {\bibinfo {volume} {100}},\ \bibinfo {pages} {014105} (\bibinfo {year} {2019})}\BibitemShut {NoStop}%
\bibitem [{\citenamefont {Batzner}\ \emph {et~al.}(2022)\citenamefont {Batzner}, \citenamefont {Musaelian}, \citenamefont {Sun}, \citenamefont {Geiger}, \citenamefont {Mailoa}, \citenamefont {Kornbluth}, \citenamefont {Molinari}, \citenamefont {Smidt},\ and\ \citenamefont {Kozinsky}}]{batzner_e3-equivariant_2022}%
  \BibitemOpen
  \bibfield  {author} {\bibinfo {author} {\bibfnamefont {S.}~\bibnamefont {Batzner}}, \bibinfo {author} {\bibfnamefont {A.}~\bibnamefont {Musaelian}}, \bibinfo {author} {\bibfnamefont {L.}~\bibnamefont {Sun}}, \bibinfo {author} {\bibfnamefont {M.}~\bibnamefont {Geiger}}, \bibinfo {author} {\bibfnamefont {J.~P.}\ \bibnamefont {Mailoa}}, \bibinfo {author} {\bibfnamefont {M.}~\bibnamefont {Kornbluth}}, \bibinfo {author} {\bibfnamefont {N.}~\bibnamefont {Molinari}}, \bibinfo {author} {\bibfnamefont {T.~E.}\ \bibnamefont {Smidt}}, \ and\ \bibinfo {author} {\bibfnamefont {B.}~\bibnamefont {Kozinsky}},\ }\href {\doibase 10.1038/s41467-022-29939-5} {\bibfield  {journal} {\bibinfo  {journal} {Nature Communications}\ }\textbf {\bibinfo {volume} {13}},\ \bibinfo {pages} {2453} (\bibinfo {year} {2022})}\BibitemShut {NoStop}%
\bibitem [{\citenamefont {Larsen}\ \emph {et~al.}(2017)\citenamefont {Larsen}, \citenamefont {Mortensen}, \citenamefont {Blomqvist}, \citenamefont {Castelli}, \citenamefont {Christensen}, \citenamefont {Dulak}, \citenamefont {Friis}, \citenamefont {Groves}, \citenamefont {Hammer}, \citenamefont {Hargus}, \citenamefont {Hermes}, \citenamefont {Jennings}, \citenamefont {Jensen}, \citenamefont {Kermode}, \citenamefont {Kitchin}, \citenamefont {Kolsbjerg}, \citenamefont {Kubal}, \citenamefont {Kaasbjerg}, \citenamefont {Lysgaard}, \citenamefont {Maronsson}, \citenamefont {Maxson}, \citenamefont {Olsen}, \citenamefont {Pastewka}, \citenamefont {Peterson}, \citenamefont {Rostgaard}, \citenamefont {Schiøtz}, \citenamefont {Schütt}, \citenamefont {Strange}, \citenamefont {Thygesen}, \citenamefont {Vegge}, \citenamefont {Vilhelmsen}, \citenamefont {Walter}, \citenamefont {Zeng},\ and\ \citenamefont {Jacobsen}}]{HjorthLarsen2017}%
  \BibitemOpen
  \bibfield  {author} {\bibinfo {author} {\bibfnamefont {A.~H.}\ \bibnamefont {Larsen}}, \bibinfo {author} {\bibfnamefont {J.~J.}\ \bibnamefont {Mortensen}}, \bibinfo {author} {\bibfnamefont {J.}~\bibnamefont {Blomqvist}}, \bibinfo {author} {\bibfnamefont {I.~E.}\ \bibnamefont {Castelli}}, \bibinfo {author} {\bibfnamefont {R.}~\bibnamefont {Christensen}}, \bibinfo {author} {\bibfnamefont {M.}~\bibnamefont {Dulak}}, \bibinfo {author} {\bibfnamefont {J.}~\bibnamefont {Friis}}, \bibinfo {author} {\bibfnamefont {M.~N.}\ \bibnamefont {Groves}}, \bibinfo {author} {\bibfnamefont {B.}~\bibnamefont {Hammer}}, \bibinfo {author} {\bibfnamefont {C.}~\bibnamefont {Hargus}}, \bibinfo {author} {\bibfnamefont {E.~D.}\ \bibnamefont {Hermes}}, \bibinfo {author} {\bibfnamefont {P.~C.}\ \bibnamefont {Jennings}}, \bibinfo {author} {\bibfnamefont {P.~B.}\ \bibnamefont {Jensen}}, \bibinfo {author} {\bibfnamefont {J.}~\bibnamefont {Kermode}}, \bibinfo {author} {\bibfnamefont {J.~R.}\ \bibnamefont {Kitchin}}, \bibinfo {author}
  {\bibfnamefont {E.~L.}\ \bibnamefont {Kolsbjerg}}, \bibinfo {author} {\bibfnamefont {J.}~\bibnamefont {Kubal}}, \bibinfo {author} {\bibfnamefont {K.}~\bibnamefont {Kaasbjerg}}, \bibinfo {author} {\bibfnamefont {S.}~\bibnamefont {Lysgaard}}, \bibinfo {author} {\bibfnamefont {J.~B.}\ \bibnamefont {Maronsson}}, \bibinfo {author} {\bibfnamefont {T.}~\bibnamefont {Maxson}}, \bibinfo {author} {\bibfnamefont {T.}~\bibnamefont {Olsen}}, \bibinfo {author} {\bibfnamefont {L.}~\bibnamefont {Pastewka}}, \bibinfo {author} {\bibfnamefont {A.}~\bibnamefont {Peterson}}, \bibinfo {author} {\bibfnamefont {C.}~\bibnamefont {Rostgaard}}, \bibinfo {author} {\bibfnamefont {J.}~\bibnamefont {Schiøtz}}, \bibinfo {author} {\bibfnamefont {O.}~\bibnamefont {Schütt}}, \bibinfo {author} {\bibfnamefont {M.}~\bibnamefont {Strange}}, \bibinfo {author} {\bibfnamefont {K.~S.}\ \bibnamefont {Thygesen}}, \bibinfo {author} {\bibfnamefont {T.}~\bibnamefont {Vegge}}, \bibinfo {author} {\bibfnamefont {L.}~\bibnamefont {Vilhelmsen}}, \bibinfo
  {author} {\bibfnamefont {M.}~\bibnamefont {Walter}}, \bibinfo {author} {\bibfnamefont {Z.}~\bibnamefont {Zeng}}, \ and\ \bibinfo {author} {\bibfnamefont {K.~W.}\ \bibnamefont {Jacobsen}},\ }\href {\doibase 10.1088/1361-648X/aa680e} {\bibfield  {journal} {\bibinfo  {journal} {Journal of Physics: Condensed Matter}\ }\textbf {\bibinfo {volume} {29}},\ \bibinfo {pages} {273002} (\bibinfo {year} {2017})}\BibitemShut {NoStop}%
\bibitem [{\citenamefont {Herath}\ \emph {et~al.}(2020)\citenamefont {Herath}, \citenamefont {Tavadze}, \citenamefont {He}, \citenamefont {Bousquet}, \citenamefont {Singh}, \citenamefont {Muñoz},\ and\ \citenamefont {Romero}}]{pyprocar}%
  \BibitemOpen
  \bibfield  {author} {\bibinfo {author} {\bibfnamefont {U.}~\bibnamefont {Herath}}, \bibinfo {author} {\bibfnamefont {P.}~\bibnamefont {Tavadze}}, \bibinfo {author} {\bibfnamefont {X.}~\bibnamefont {He}}, \bibinfo {author} {\bibfnamefont {E.}~\bibnamefont {Bousquet}}, \bibinfo {author} {\bibfnamefont {S.}~\bibnamefont {Singh}}, \bibinfo {author} {\bibfnamefont {F.}~\bibnamefont {Muñoz}}, \ and\ \bibinfo {author} {\bibfnamefont {A.~H.}\ \bibnamefont {Romero}},\ }\href {\doibase https://doi.org/10.1016/j.cpc.2019.107080} {\bibfield  {journal} {\bibinfo  {journal} {Computer Physics Communications}\ }\textbf {\bibinfo {volume} {251}},\ \bibinfo {pages} {107080} (\bibinfo {year} {2020})}\BibitemShut {NoStop}%
\bibitem [{\citenamefont {{Bistritzer}}\ and\ \citenamefont {{MacDonald}}(2011)}]{2011PNAS..10812233B}%
  \BibitemOpen
  \bibfield  {author} {\bibinfo {author} {\bibfnamefont {R.}~\bibnamefont {{Bistritzer}}}\ and\ \bibinfo {author} {\bibfnamefont {A.~H.}\ \bibnamefont {{MacDonald}}},\ }\href {\doibase 10.1073/pnas.1108174108} {\bibfield  {journal} {\bibinfo  {journal} {Proceedings of the National Academy of Science}\ }\textbf {\bibinfo {volume} {108}},\ \bibinfo {pages} {12233} (\bibinfo {year} {2011})},\ \Eprint {http://arxiv.org/abs/1009.4203} {arXiv:1009.4203 [cond-mat.mes-hall]} \BibitemShut {NoStop}%
\bibitem [{\citenamefont {{Song}}\ \emph {et~al.}(2019)\citenamefont {{Song}}, \citenamefont {{Wang}}, \citenamefont {{Shi}}, \citenamefont {{Li}}, \citenamefont {{Fang}},\ and\ \citenamefont {{Bernevig}}}]{2018arXiv180710676S}%
  \BibitemOpen
  \bibfield  {author} {\bibinfo {author} {\bibfnamefont {Z.-D.}\ \bibnamefont {{Song}}}, \bibinfo {author} {\bibfnamefont {Z.}~\bibnamefont {{Wang}}}, \bibinfo {author} {\bibfnamefont {W.}~\bibnamefont {{Shi}}}, \bibinfo {author} {\bibfnamefont {G.}~\bibnamefont {{Li}}}, \bibinfo {author} {\bibfnamefont {C.}~\bibnamefont {{Fang}}}, \ and\ \bibinfo {author} {\bibfnamefont {B.~A.}\ \bibnamefont {{Bernevig}}},\ }\href {\doibase 10.1103/PhysRevLett.123.036401} {\bibfield  {journal} {\bibinfo  {journal} {\prl}\ }\textbf {\bibinfo {volume} {123}},\ \bibinfo {eid} {036401} (\bibinfo {year} {2019})},\ \Eprint {http://arxiv.org/abs/1807.10676} {arXiv:1807.10676 [cond-mat.mes-hall]} \BibitemShut {NoStop}%
\bibitem [{\citenamefont {{Herzog-Arbeitman}}\ \emph {et~al.}(2022{\natexlab{b}})\citenamefont {{Herzog-Arbeitman}}, \citenamefont {{Chew}},\ and\ \citenamefont {{Bernevig}}}]{2022PhRvB.106h5140H}%
  \BibitemOpen
  \bibfield  {author} {\bibinfo {author} {\bibfnamefont {J.}~\bibnamefont {{Herzog-Arbeitman}}}, \bibinfo {author} {\bibfnamefont {A.}~\bibnamefont {{Chew}}}, \ and\ \bibinfo {author} {\bibfnamefont {B.~A.}\ \bibnamefont {{Bernevig}}},\ }\href {\doibase 10.1103/PhysRevB.106.085140} {\bibfield  {journal} {\bibinfo  {journal} {\prb}\ }\textbf {\bibinfo {volume} {106}},\ \bibinfo {eid} {085140} (\bibinfo {year} {2022}{\natexlab{b}})},\ \Eprint {http://arxiv.org/abs/2206.07717} {arXiv:2206.07717 [cond-mat.mes-hall]} \BibitemShut {NoStop}%
\bibitem [{\citenamefont {Xiao}\ \emph {et~al.}(2012)\citenamefont {Xiao}, \citenamefont {Liu}, \citenamefont {Feng}, \citenamefont {Xu},\ and\ \citenamefont {Yao}}]{Xiao2012TMD}%
  \BibitemOpen
  \bibfield  {author} {\bibinfo {author} {\bibfnamefont {D.}~\bibnamefont {Xiao}}, \bibinfo {author} {\bibfnamefont {G.-B.}\ \bibnamefont {Liu}}, \bibinfo {author} {\bibfnamefont {W.}~\bibnamefont {Feng}}, \bibinfo {author} {\bibfnamefont {X.}~\bibnamefont {Xu}}, \ and\ \bibinfo {author} {\bibfnamefont {W.}~\bibnamefont {Yao}},\ }\href {\doibase 10.1103/PhysRevLett.108.196802} {\bibfield  {journal} {\bibinfo  {journal} {Phys. Rev. Lett.}\ }\textbf {\bibinfo {volume} {108}},\ \bibinfo {pages} {196802} (\bibinfo {year} {2012})}\BibitemShut {NoStop}%
\end{thebibliography}%
\bibliographystyle{apsrev4-1}

\end{document}